\documentclass[prx,aps,10pt,nofootinbib,superscriptaddress]{revtex4-2} %superscriptaddress
\usepackage{scalerel}
\usepackage{graphicx}
\usepackage[nointegrals]{wasysym} %
\usepackage[export]{adjustbox}
\usepackage{amsmath,amsfonts,amssymb,latexsym}
\usepackage{hhline}
\usepackage{bm}
\usepackage{verbatim}
\usepackage{enumitem}
\usepackage{mathrsfs}
\usepackage{slashed}
\usepackage{empheq}
\usepackage{physics}
\usepackage[most]{tcolorbox}

\newtcbtheorem{boxedtheorem}{Theorem}%
{colback=gray!10,
 colframe=black,
 fonttitle=\bfseries,
 coltitle=black,
 breakable,
 enhanced,
 boxrule=0.8pt,
 arc=2mm,
 left=2mm,
 right=2mm,
 top=1mm,
 bottom=1mm}{th}

\usepackage{graphicx}
\usepackage{dcolumn}
\usepackage{bm}
\usepackage{multirow}

\usepackage[normalem]{ulem}

\usepackage{amsmath}
\usepackage{amsthm}
\usepackage{amstext}
\usepackage{amssymb}
\usepackage{mathrsfs}
\usepackage{amsfonts}
\usepackage{amsbsy} 

\usepackage{braket}

\usepackage[all,matrix,cmtip]{xy}
\usepackage{mathtools}

\usepackage{xcolor}
\definecolor{red}{rgb}{1,0,0}
\definecolor{blue}{rgb}{0,0,1}
\definecolor{dblue}{rgb}{0,0,0.4}
\definecolor{green}{rgb}{0,1,0}
\definecolor{black}{rgb}{0,0,0}
\definecolor{white}{rgb}{1,1,1}
\definecolor{pastelblue}{RGB}{20,93,160}

\definecolor{brn}{rgb}{.8,.4,.0}
\definecolor{redo}{rgb}{1,.5,.0}
\definecolor{ddgrn}{rgb}{0,0.4,0}
\definecolor{dgrn}{rgb}{0,0.55,0}
\definecolor{dbl}{rgb}{0,0,0.5}

\usepackage[colorlinks,citecolor=pastelblue,linkcolor=pastelblue,urlcolor=pastelblue]{hyperref}

\renewcommand{\Im}{{\rm Im}} 
\renewcommand{\Re}{{\rm Re}}

\newcommand{\sgn}{{\rm sgn}}

\newcommand{\bpm}{\begin{pmatrix}}
	\newcommand{\epm}{\end{pmatrix}}
\newcommand{\bmm}{\begin{matrix}}
	\newcommand{\emm}{\end{matrix}}
\newcommand{\bvm}{\begin{vmatrix}}
	\newcommand{\evm}{\end{vmatrix}}

\usepackage{euscript}

\makeatletter
\newsavebox{\@brx}
\newcommand{\llangle}[1][]{\savebox{\@brx}{\(\m@th{#1\langle}\)}%
	\mathopen{\copy\@brx\kern-0.5\wd\@brx\usebox{\@brx}}}
\newcommand{\rrangle}[1][]{\savebox{\@brx}{\(\m@th{#1\rangle}\)}%
	\mathclose{\copy\@brx\kern-0.5\wd\@brx\usebox{\@brx}}}
\makeatother

\newcommand{\ZS}[1]{\textcolor{blue}{\it [ZS: #1]}}
\newcommand{\bs}{\boldsymbol}

\allowdisplaybreaks

\usepackage{tikz}
\definecolor{myRed}{RGB}{188,0,4} % #BC0004
\definecolor{myGray}{RGB}{146,146,146} % #929292
\definecolor{myBlue}{RGB}{0,0,133} % #000085

\begin{document}

%\preprint{APS/123-QED}

\title{Lattice composite Fermi liquid with broken inversion symmetry}%

\author{Pavel A. Nosov}
\email{pnosov@fas.harvard.edu}
 %\homepage{}
\affiliation{Department of Physics, Harvard University, Cambridge, MA 02138, USA
}%

\author{Zhengyan Darius Shi}%
\email{zhengyanshi@stanford.edu}
\affiliation{
Leinweber Institute for Theoretical Physics, Stanford University, Stanford, California 94305, USA}

\date{\today}

\begin{abstract}
    We study transport in lattice composite Fermi liquids realized in half-filled Chern bands with broken inversion symmetry. We show that reduced crystalline symmetry exposes intrinsic singular dynamical responses of composite fermions that are otherwise hidden in the conventional Landau-level setting. At zero wave vector, inversion breaking allows gauge-field fluctuations to generate a non-analytic longitudinal optical resistivity, with $\operatorname{Re}\rho^{xx}(\omega)\sim |\omega|^{4/3}$ for gate-screened Coulomb interactions. At finite wave vector $\bm{q}$, inversion breaking leads to nonreciprocal transport and a non-analytic $\sim |\bm{q}|$  dependence of the Hall conductivity, both of which can be probed through surface acoustic wave propagation. We also discuss a distinct mechanism for singular DC transport in lattice composite Fermi liquids: renormalization of $2k_F$ scattering at the composite Fermi surface enhances Umklapp relaxation and can lead to a non-analytic temperature dependence of the resistivity. Taken together, our results identify transport signatures of lattice composite Fermi liquids that are absent in their continuum quantum Hall counterparts and can be directly tested in ongoing experiments on twisted MoTe$_2$ and rhombohedral graphene, where evidence for zero-field composite Fermi liquids has recently been reported.
\end{abstract}

\maketitle
% \tableofcontents

% \newpage

\section{Introduction}

The composite Fermi liquid (CFL) is a strongly interacting metal that emerges when electrons half-fill the lowest Landau level. Unlike an ordinary electronic Fermi liquid, the elementary low energy excitations of a CFL are built out of composite fermion (CF)s, which can be viewed as electrons bound to two units of magnetic flux quanta~\cite{Jain1989_CFframework,Jain1990_CFdetails,Lopez1991_CSGL,Halperin1993_HLRtheory}. Precisely at half-filling of the lowest Landau level, the attached flux cancels the background magnetic flux such that each CF sees zero magnetic field on average. As a result, CFs can fill a Fermi sea at nonzero density, leading to a compressible metallic phase of matter~\cite{Halperin1993_HLRtheory,Son2015_DiracCFL}. This elegant conceptual framework provides a unified understanding of many neighboring topological phases in the quantum Hall phase diagram, including the Jain states at $\nu = p/(2p+1)$ with Abelian topological order~\cite{Jain1989_CFframework,Jain1990_CFdetails} as well as the Pfaffian state at $\nu = 5/2$ with non-Abelian topological order~\cite{Moore1991_nonAbelian,Read1999_pair}.

Recent developments in moire materials suggest that the validity of the CF framework extends beyond the conventional Landau level context. Central to this story is the discovery of zero-field fractional quantum anomalous Hall (FQAH) insulators~\cite{Cai2023_FQAHTMD,Xu2023_FQAHTMD,Zeng2023_FQAHTMD,Park2023_FQAH_TMD,Lu2023_FQAHPenta,Lu2025_EQAH}. Following early thermodynamic signatures of FQAH at certain Jain fillings in twisted MoTe$_2$~\cite{Cai2023_FQAHTMD,Zeng2023_FQAHTMD}, numerical studies~\cite{Dong2023_ACFL,Goldman2023_ACFL} revealed that these Jain states may arise as daughter states of an anomalous composite Fermi liquid (ACFL) at half filling of a parent Chern band. The numerical prediction of an ACFL phase has received experimental support from transport measurements in both twisted MoTe$_2$ and rhombohedral multilayer graphene~\cite{Park2023_FQAH_TMD,Xu2023_FQAHTMD,Lu2023_FQAHPenta} as well as trion sensing measurements in twisted MoTe$_2$~\cite{Anderson2024_trionCFL}, though a smoking-gun detection of the composite Fermi surface remains an important future target. 

The observation of ACFLs (and more generally lattice CFLs) in half-filled Chern bands raises an important theoretical question: does the zero-field/lattice setting reveal new physics of CFs beyond what is possible in a Landau level? In this paper, we approach this question from the perspective of symmetries. It is well known that the conventional CFL is strongly constrained by symmetries of the half-filled Landau level, including magnetic translation symmetry, continuous rotation symmetry, and Galilean boost symmetry. While these symmetries provide powerful constraints on physical observables, they are additional structures imposed on top of the essential CF construction~\cite{Jain1989_CFframework,Jain1990_CFdetails,Lopez1991_CSGL,Halperin1993_HLRtheory}. A central aim of this work is to demonstrate how certain Landau level symmetries conceal striking dynamical signatures of CFs, which only reappear when the corresponding symmetries are removed by an appropriate lattice potential. As a result, the phenomenology of ACFLs/lattice CFLs can be qualitatively distinct from the conventional CFL, despite their resemblance in many basic thermodynamic/transport probes. 

The most important Landau level symmetry that we will remove is the spatial inversion symmetry that relates momenta $\bs{k} \leftrightarrow - \bs{k}$. In the Landau level setting, the CF dispersion satisfies $\xi(\bs{k}) = \frac{k^2}{2m} - \mu$, where $m$ is the CF effective mass and $\mu$ is the chemical potential. This dispersion is guaranteed by Galilean invariance and obviously respects the inversion symmetry. Even on a lattice, toy models of half-filled Chern bands often assume a $C_6$ symmetric lattice potential which breaks the Galilean symmetry but continues to respect spatial inversion~\cite{Goldman2023_ACFL,Dong2023_ACFL}. However, in realistic models of twisted MoTe$_2$ and rhombohedral multilayer graphene, where signatures of lattice CFLs have been reported, trigonal warping effects generically reduce the $C_6$ symmetry further down to $C_3$, which no longer contains spatial inversion as a subgroup~\cite{Wu2019_TMD_foundational,Dong2023_FQAHpenta_MIT,Dong2023_FQAHpenta_harvard,Zhou2023_FQAHpenta}. Motivated by this experimental input, we explore the consequences of a CF Fermi surface that is invariant under $C_3$ rotations but not under spatial inversion. Our main qualitative conclusions rely only on this symmetry reduction and therefore apply to generic inversion-asymmetric CF dispersions. For explicit calculations, however, we often specialize to a weakly trigonally warped dispersion of the form
\begin{equation}\label{eq:trigonal_disp}
    \xi(\bs{k}) = \frac{k^2}{2m} \left(1 + \lambda \cos 3 \theta_{\bs{k}}\right) - \mu \,, 
\end{equation}
where $\theta_{\bs{k}}$ is the angle of the CF momentum $\bs{k}$ relative to the $x$-axis, and $\lambda \ll 1$ is a parameter that controls the strength of trigonal warping. Previously, we showed that with attractive density-density interactions, the natural paired state of CFs with this kind of inversion-asymmetric dispersion is a novel phase of matter with gapless Bogoliubov Fermi pockets of CFs. We dubbed this phase the Composite Bogoliubov Fermi liquid (CBFL) and analyzed some of its striking properties that depart from the CFL and gapped quantum Hall states that can exist at the same lattice filling~\cite{Shi2026_CBFL}. In this paper, our focus will instead be on the normal state of the lattice CFL. 

%  Motivated by this experimental input, we will work throughout the paper with a weakly trigonally warped CF dispersion of the form 
%Though many of our results apply to more general CF dispersions without inversion symmetry, we will often present explicit formulae using the dispersion in Eq.~\eqref{eq:trigonal_disp} for concreteness.  

The first key result of this paper is that lattice CFLs with broken inversion symmetry can exhibit a singular homogeneous (i.e. $\bs{q} = 0$) optical resistivity with a scaling form 
\begin{equation}\label{eq:main_transport_optical}
    \Re \,\rho^{xx}(\omega, T = 0) \approx C(z) \,\lambda^2 \, |\omega|^{4/z} + \mathcal{O}(\lambda^4) \,, \quad C(3) = \frac{1.926 \pi^2}{\mu^{4/3} \left(4/3+mv_0/2\pi\right)^{4/3}} \,,
\end{equation}
 where $z$ is the dynamical exponent of gauge fluctuations in the lattice CFL, $C(z)$ is a positive  dispersion-specific constant. Physically, $z = 2$ corresponds to a microscopic model with long-range Coulomb interactions $v(\bm{q})\propto 1/|\bm{q}|$, while $z = 3$ corresponds to a screened short-range interaction $v(\bm{q})=v_0$, for which we were able to analytically compute $C(z=3)$ as a function of $m, \mu, \lambda$ in Eq.~\eqref{eq:trigonal_disp} as well as the interaction strength $v_0$.

The non-analytic frequency scaling of the optical resistivity in Eq.~\eqref{eq:main_transport_optical} is striking. In the half-filled Landau level, magnetic Galilean invariance forces the longitudinal optical resistivity to have the form $\rho^{xx}(\omega, T = 0) = - i \omega m_e/n$ %\textcolor{magenta}{[PN: shouldn't this be $\rho^{xx}(\omega, T = 0) = - i \omega m_e/n$ ?]} 
where $n$ is the charge density and $m_e$ is the electron mass. This non-perturbative result, sometimes known as Kohn's theorem \cite{Kohn1961}, forbids any non-analytic corrections to $\rho^{xx}(\omega, T = 0)$ of the form presented in Eq.~\eqref{eq:main_transport_optical}. Furthermore, even when magnetic Galilean invariance is explicitly broken, a non-analytic correction to $\rho^{xx}(\omega, T = 0)$ remains forbidden for any convex Fermi surface with inversion symmetry,  due to non-perturbative constraints from infrared symmetries and 't Hooft anomalies~\cite{Else2020_EFL,Shi2022_gifts} (the same result has been verified using perturbative diagrammatic calculations in certain controlled expansions~\cite{Shi2022_loopcurrent,Guo2022_YukawaSYK2d,Li2023_NFLoptical,Shi2023_controlled,Guo2024_fluc_criticalFS,Gindikin2024_optical_NFL,Gindikin2025_NFLstability}). An important observation of this paper is that in the absence of inversion symmetry, these non-perturbative constraints are relaxed so that fluctuations of the CF Fermi surface directly lead to non-analytic low energy optical transport. Physically, the leading singular optical response comes from gauge-mediated small-momentum-transfer collisions between CFs on tangential patches of the Fermi surface. 
For a convex inversion-symmetric Fermi surface, the two selected patches are related by $\bs{k}\leftrightarrow-\bs{k}$, so the net CF velocity change vanishes. 
Once inversion symmetry is broken, the selected patches are no longer antipodal, making the scattering kinematics generic, much like for a concave Fermi surface~\cite{Maslov2011_NFL_nonGalileo}, and allowing the low-energy processes to produce a nonzero change in the CF current.

The second set of results concerns the transport behavior of lattice inversion-asymmetric CFLs at nonzero wave vector $\bs{q}$. In the longitudinal (L) - transverse (T) basis defined by the vector $\bs{q}$, we find the conductivity matrix 
\begin{equation}\label{eq:main_transport_finiteQ}
\begin{aligned}
\sigma^{LL}(\bs q,\omega)
&\propto
|\bs q|
\left[
1-iA_1\lambda\cos\left(3\phi_{\bs q}\right)
\right]
+\mathcal O(\lambda^2|\bm{q}|, q^2),
\\
\sigma^{LT/TL}(\bs q,\omega)
&=
\mp\frac{e^2}{2h}
+iB_1\lambda|\bs q|
\sin\left(3\phi_{\bs q}\right)
-B_2\lambda^2|\bs q|
\sin\left(6\phi_{\bs q}\right)
+\mathcal O(\lambda^3|\bs q|,q^2),
\\
\sigma^{TT}(\bs q,\omega)
&\propto \frac{i A_3 q^2}{\omega} + 
|\bs q|
\left[
1+iA_2\lambda\cos\left(3\phi_{\bs q}\right)
\right]
+\mathcal O(\lambda^2|\bm{q}|,q^2).
\end{aligned}
\end{equation}
where $\bs{q} = |\bs{q}| (\cos \phi_{\bs{q}}, \sin \phi_{\bs{q}})$ and $A_{1,2,3}, B_{1,2}$ are positive constants independent of $\lambda$.  We assumed a kinematic regime 
$(q/k_F)^{z-1} \ll \omega/(v_F q) \ll 1$, which is relevant for long-wavelength surface acoustic wave probes. Results for other kinematic regimes are more complicated and can be found in Sec.~\ref{sec:finiteQ_transport}. While a leading $|\bs{q}|$ dependence of the longitudinal conductivity is a standard feature of composite Fermi liquids \cite{Mirlin1997_SAW0,Simon1996_SAWderivation}, we find that broken inversion symmetry produces a qualitatively new effect in the Hall channel: $\sigma^{LT}$ also acquires an angle-dependent non-analytic contribution linear in $|\bs{q}|$, which is absent in inversion-symmetric CFLs. In addition, we find that inversion breaking generates a non-analytic correction to the longitudinal conductivity $\sigma^{LL}$, which has the same $|\bs{q}|$ scaling as the leading inversion-symmetric contribution. These new results lead to modifications of the velocity shift $\Delta v(\bs{q})/v$ and attenuation rate $\kappa(\bs{q})$ in surface acoustic wave propagation (see Sec.~\ref{sec:discussion_Q} for discussion), which is of direct relevance to near-term experiments in moire realizations of lattice CFLs.

In addition to these inversion-breaking responses, we also discuss a distinct mechanism for singular temperature dependence of the DC resistivity $\rho^{xx}(\omega=0,T)$ in lattice CFLs with or without inversion, building on the insights of Ref.~\cite{Lee2024_powerlaw_resistivity}. This mechanism relies on the singularity of CF density correlations near the $2k_F$ scattering wave vectors, which is generated by gauge-field fluctuations. When the CF Fermi surface is sufficiently large that Umklapp processes can access these $2k_F$ singularities, the momentum relaxation rate is enhanced, leading to the scaling form
\begin{equation}\label{eq:main_transport_DC}
    \Re \,\rho^{xx}(\omega=0, T) = A \, T^{\frac{z+4}{z} - 2 \sigma} \,, \quad \gamma_{2k_F}(\omega) \sim \omega^{-2 \sigma} \,,
\end{equation}
where $A$ is a non-universal positive constant and $\gamma_{2k_F}$ is the frequency-dependent $2k_F$ vertex function (to be defined more carefully in Sec.~\ref{subsec:zeroQ_DC}). When the scaling exponent $\sigma > \frac{2}{z} - \frac{1}{2}$, this $2k_F$-assisted DC resistivity dominates over the conventional Umklapp contribution, which scales as $T^2$~\cite{Lawrence1973_FLUmklapp,MacDonald1981_FLUmklapp}. Within a large $N$ expansion, we estimate the exponent to be $\sigma = \frac{1}{2N}$, which extrapolates to $\sigma = \frac{1}{2}$ in the physical $N = 1$ limit, leading to $\Re \,\rho^{xx}(\omega = 0, T) \sim T^{4/z}$.

Taken together, these results show that zero-field/lattice CFLs do more than reproduce familiar composite fermion physics in a new platform. They expose essential dynamical signatures of CFLs that are hidden in the continuum quantum Hall setting by symmetry. The same topological mechanism of flux attachment now operates in an environment with reduced spatial symmetry, allowing new singular contributions to both homogeneous and finite-momentum transport. 

The rest of this paper will be organized as follows. In Sec.~\ref{sec:review_latticeCFL}, we review the effective field theory of lattice CFLs and derive various basic observables within the standard random phase approximation (RPA). In Sec.~\ref{sec:zeroQ_transport}, we study electrical transport in a lattice CFL with broken inversion symmetry and derive Eq.~\eqref{eq:main_transport_optical}. In Sec.~\ref{subsec:zeroQ_DC}, we also discuss the gauge-fluctuation-enhanced Umklapp scattering in lattice CFLs, with or without inversion symmetry, and derive the DC resistivity scaling in Eq.~\eqref{eq:main_transport_DC}. These calculations go beyond the random phase approximation and requires the inclusion of a subset of higher-loop Feynman diagrams, which we justify within a large $N$ expansion. In Sec.~\ref{sec:finiteQ_transport}, we move on to transport at nonzero $\bs{q}$ and derive the form of the conductivity matrix as stated in Eq.~\ref{eq:main_transport_finiteQ}.  Finally, we conclude in Sec.~\ref{sec:discussion} with a discussion of possible experimental probes of our theoretical predictions and outline some interesting directions for future research in composite fermion physics.

\section{Lattice CFL in an inversion-asymmetric Chern band}\label{sec:review_latticeCFL}

\subsection{Parton construction}\label{subsec:review_parton}

To set the stage, we review the effective field theory for a lattice CFL realized in a half-filled Chern band. The microscopic electronic model takes the general form
\begin{equation}
    H = H_0 + H_{\rm int} \,, \quad H_0 = \sum_{\bs{k}} \left[\epsilon_c(\bs{k}) - \mu\right] \, c^{\dagger}_{\bs{k}} c_{\bs{k}} \,, \quad H_{\rm int} = \sum_{\bs{k}, \bs{p}, \bs{q}} V_{\bs{q}}(\bs{k}, \bs{p}) c^{\dagger}_{\bs{k} + \bs{q}} c^{\dagger}_{\bs{p} - \bs{q}} c_{\bs{k}} c_{\bs{p}} \,,
\end{equation}
where $\epsilon_c(\bs{k})$ is the electron band dispersion with $C_3$ rotational symmetry, $\mu$ is the chemical potential, and $V_{\bs{q}}(\bs{k}, \bs{p})$ is a band-projected four-fermion interaction. We assume that $\mu$ is chosen so that the band filling of electrons is precisely $\nu = 1/2$. 

With an appropriate choice of $\epsilon_c(\bs{k})$ and $V_{\bs{q}}(\bs{k}, \bs{p})$ (corresponding to a realistic model of twisted MoTe$_2$), numerical simulations have established the existence of a lattice CFL phase~\cite{Goldman2023_ACFL,Dong2023_ACFL}. To describe the low energy dynamics of this phase, we follow Refs.~\cite{Barkeshli2012_latticeCFL_FL, Goldman2023_ACFL,Dong2023_ACFL} and perform a parton decomposition $c(\bs{r}) = f(\bs{r}) \, \Phi(\bs{r})$, where $c$ is the microscopic electron operator, $\Phi$ is a hard-core boson, and $f$ is a fermionic parton that will soon be interpreted as the lattice composite fermion. This parton decomposition introduces a $U(1)$ redundancy $f(\bs{r}) \rightarrow e^{-i\theta(\bs{r})}, \Phi(\bs{r}) \rightarrow e^{i\theta(\bs{r})}$, which can be removed by a $U(1)$ gauge field $a$ under which $f/\Phi$ carries gauge charge $-1/+1$. The physical electric charge carried by $c$ can be assigned to either $f$ or $\Phi$ and we will choose to assign it to $\Phi$ without loss of generality. 

Given the charge assignment above, the low energy parton effective Lagrangian takes the general form   $L_{\rm eff} = L[f,  a] + L[\Phi, A-a] + L_{\rm int}$ where $L[f, a], L[\Phi, A-a]$ describe the minimal coupling of $f, \Phi$ to the gauge fields under which they are charged and $L_{\rm int}$ captures higher order interactions. Since $f, \Phi$ carry opposite gauge charge under $a$, the equation of motion for $a_0$ imposes the constraint $\rho_f = \rho_{\Phi}$, where $\rho_f/\rho_{\Phi}$ are the charge densities associated with $f/\Phi$. As a result, both $f$ and $\Phi$ are at lattice filling $\nu = 1/2$. 

The construction so far is formally exact and merely provides a complicated reformulation of the microscopic electronic problem. To describe a specific electronic phase, we need to choose a mean-field ansatz for $f, \Phi$ which is consistent with their lattice filling and then demonstrate the stability of the mean-field ansatz under the inclusion of gauge fluctuations. For the lattice CFL, the appropriate mean-field is one in which the bosonic parton $\Phi$ forms an interacting bosonic Laughlin state $U(1)_2$ with Hall conductance $\sigma_{xy} =- e^2/2h$ ~\cite{Barkeshli2012_latticeCFL_FL}. This choice is consistent with the lattice filling constraints and leads to a low energy topological quantum field theory in the bosonic sector~\footnote{Throughout the paper, we will use the integral convention $\int_{\tau, \bs{r}} \equiv \int d \tau d^2 \bs{r}$ and $\int_{\omega, \bs{k}} \equiv \int \frac{d \omega d^2 \bs{k}}{(2\pi)^3}$.Our Fourier convention is such that $f(\bm{r},\tau)= \int_{\bm{k},\omega}e^{-i\omega \tau +i\bm{k}\bm{r}}f(\bm{k},\bm{\omega})$ and $\bar{f}(\bm{r},\tau)= \int_{\bm{k},\omega}e^{i\omega \tau -i\bm{k}\bm{r}}\bar{f}(\bm{k},\bm{\omega})$, while all bosonic fields have the same Fourier convention as $f(\bm{r},\tau)$. }
\begin{equation}
    S_{U(1)_2}[\Phi, A-a] = \int_{\tau, \bs{r}} \left[- \frac{2i}{4\pi} \alpha d \alpha + \frac{i}{2\pi} \alpha d (A-a)\right] \,, 
\end{equation}
where $\alpha$ is a dynamical $U(1)$ gauge field and $A d B$ is a short-hand for the Chern-Simons term $\epsilon^{\mu\nu\lambda} A_{\mu} \partial_{\nu} B_{\lambda}$ .
For the fermionic sector, we choose $f$ to be an ordinary Fermi liquid with an interaction-induced dispersion $\xi_f(\bs{k})$ and an effective density-density interaction $V(\bs{r} - \bs{r}')$. The complete Euclidean action then takes the form 
\begin{equation}
    S_{\rm FL}[f,a] = \int_{\tau, \bs{r}} \bar f(\bs{r},\tau) [\partial_{\tau}-i a_0  + \xi_{-i \nabla -\bm{a}}] f(\bs{r},\tau)+  \frac{1}{2} \int_{\tau, \bs{r}, \bs{r}'} \rho_f(\tau, \bs{r}) \, V(\bs{r} - \bs{r}') \, \rho_f(\tau, \bs{r}')
\end{equation} 
After including gauge fluctuations around the mean-field ansatz and shifting the gauge field as $a\rightarrow a+A$, we obtain a complete low energy effective action
\begin{equation}
    S_{\rm eff} = S_{\rm FL}[f, A+a] + \int_{\tau, \bs{r}} \left[-\frac{2i}{4\pi} \alpha d \alpha - \frac{i}{2\pi} \alpha d a \right]\,.
\end{equation}
Since topological order will not be important in this paper, we will work on an infinite plane and integrate out $\alpha$ to reduce the second term to $\int_{\tau, \bs{r}} \frac{i}{8\pi} a da$. Due to the lack of relativistic invariance, it is convenient to choose the Coulomb gauge in which $\nabla \cdot \bs{a} = 0$. If we decompose the spatial part of the gauge field into longitudinal and transverse components
\begin{equation}
    \quad \bm{a}(\bs{q}) = a_L(\bs{q}) \hat{\bm{q}} + a_T(\bs{q}) \hat z \times \hat{\bm{q}}\;,\quad \quad a_{L/T}^*(\bs{q})=  -a_{L/T}(-\bs{q})\;,\label{eq:a_T_T}
\end{equation}
the Coulomb gauge amounts to setting $a_L(\bs{q}) = 0$. The equation of motion for $a_0$ would then impose the constraint $\rho_f = \frac{1}{4\pi} \nabla \times \bs{a} = \frac{i q}{4\pi} a_T(\bs{q})$, which attaches two flux quanta of $\bs{a}$ to each excitation of $f$. The density-density interaction between $\rho_f$ maps to a quadratic term for $a_T(\bs{q})$. Because of this flux attachment constraint, we identify $f$ as the lattice version of a composite fermion (CF)~\cite{Jain1989_CFframework,Lopez1991_CSGL,Halperin1993_HLRtheory}.  Putting all of these ingredients together, we end up with a simplified Euclidean action
\begin{equation}\label{eq:CFL_EFT}
    S_{\rm CFL} = \int_{\tau, \bs{r}} \bar f \left[\partial_{\tau} - i (A_0 + a_0) + \xi_f(-i\nabla - \bs{A} - \bs{a})\right] f + \frac{1}{2} \int_{\bs{q}, \Omega} a^*_{\alpha}(\bs{q}, \Omega) \Pi^{\alpha \beta}_a(\bs{q}, \Omega) \, a_{\beta}(\bs{q}, \Omega) \,, \quad \Pi^{\alpha\beta}_a = \begin{pmatrix}
        0 & -\frac{q}{4\pi} \\  \frac{q}{4\pi} & \frac{q^2 v(\bs{q})}{(4\pi)^2} 
    \end{pmatrix} \,,
\end{equation}
where $\alpha, \beta = 0, T$ and $v(\bs{q})$ is the Fourier transform of the real-space interaction potential $V(\bs{r} - \bs{r}')$. 

As is the case for any effective field theory, the parameters that enter $S_{\rm CFL}$ are difficult to determine analytically from the microscopic model. In fact, this challenge already exists in the conventional continuum CFL, where the CF effective mass and the Landau parameters that control forward-scattering interactions between CFs cannot be straightforwardly deduced from the electron mass and the electron interaction parameters (see however progress in this direction using numerical~\cite{Rezayi1994_CFLnumerics,Morf1995_CFmass} and analytical~\cite{Stern1995_CFmass,Stern1999_CFLdipole,Dong2020_CFL_noncomm} approaches). Nevertheless, exact symmetries of the microscopic model must be inherited by the effective action. In particular, the $C_3$ rotational invariance of the electron band dispersion $\epsilon_c(\bs{k})$ must be inherited by the CF band dispersion $\xi(\bs{k})$. 

\subsection{Random phase approximation for an inversion-asymmetric CFL}\label{subsec:review_RPA}

The effective action defined by Eq.~\eqref{eq:CFL_EFT} belongs to a large class of 2+1D non-Fermi liquid (NFL) metals in which a Fermi surface is strongly coupled to a gapless bosonic mode (in this case, a dynamical $U(1)$ gauge field). A general property of these models is the absence of a small dimensionless parameter. This challenge prompted the development of various deformation schemes that introduce an extra small parameter $\epsilon$, such that the deformed model is solvable at $\epsilon = 0$ and the original model is recovered at $\epsilon = 1$~\cite{Polchinski1993_largeN,Altshuler1994_singular,Nayak1994_epsilon,Lee2009_RPAbad,Metlitski2010_RPAbad,Mross2010_largeN_smalleps,Dalidovich2013_codim,Raghu2015_matrixlargeN,Damia2019_matrix_largeN,Esterlis2019_YukawaSYK0d,Esterlis2021_YukawaSYK2d}. Physical observables could then be computed as an expansion in powers of $\epsilon$. While different expansions involve different types of deformations (e.g. large number of fermion flavors or fractional codimension of the Fermi surface) and make distinct predictions at higher order in $\epsilon$, they all reduce to the standard random phase approximation (RPA) at leading order in $\epsilon$. For this reason, throughout this paper, we will mostly work within the RPA in order to extract the most robust consequences of inversion symmetry breaking that do not depend on the specific choice of deformation scheme. The only exception is Sec.~\ref{subsec:zeroQ_optical}, where a calculation of the singular $\bs{q} = 0$ optical conductivity requires going beyond the RPA. In that context, we will perform the calculation using a specific expansion scheme and then argue that the same result holds in all other available schemes. 

With this background in mind, we will perform a detailed analysis of the inversion-asymmetric lattice CFL within the RPA in the rest of this section. Following Ref.~\cite{Halperin1993_HLRtheory}, we start from Eq.~\eqref{eq:CFL_EFT} and integrate out the CFs, treating them as a free Fermi gas with a $C_3$ symmetric dispersion $\xi(\bs{k})$. The resulting effective Lagrangian is
\begin{gather}
L_{\rm CFL}= \frac{1}{2} \left[A^*_{\alpha}(\bs{q}, \Omega) + a^*_{\alpha}(\bs{q}, \Omega)\right] \Pi_{\rm CF}^{\alpha\beta}(\bs{q}, \Omega) \left[ A_{\beta}(\bs{q}, \Omega) + a_{\beta}(\bs{q}, \Omega)\right] +
    \frac{1}{2} a^*_{\alpha}(\bs{q}, \Omega) \Pi_{a}^{\alpha\beta}(\bs{q}, \Omega)a_{\beta}(\bs{q}, \Omega)\;,\notag\\
  \Pi_{ a}^{\alpha\beta} =\begin{pmatrix}
     0   & -\frac{q}{4\pi}\\
      \frac{q}{4\pi}  & \frac{q^2 v_0(\bm{q})}{(4\pi)^2}
    \end{pmatrix},\quad  \quad \Pi_{\rm CF}^{\alpha\beta} =   \begin{pmatrix}
   \Pi_{\rm CF}^{00}  & \Pi_{\rm CF}^{0T} \\
  \Pi_{\rm CF}^{T0}  & \Pi_{\rm CF}^{TT}
    \end{pmatrix},  \label{eq:CFL_response_function}
\end{gather}
where $\Pi^{\alpha \beta}_{\rm CF}$ is the matrix of density-current correlators for free fermions with a dispersion $\xi(\bs{k})$
\begin{gather}\label{eq:PiCF_def}
     \Pi_{\rm CF}^{00}(\bm{q},\Omega)=\langle \rho(\bm{q},\Omega)\rho(-\bm{q},-\Omega)\rangle,\quad  \Pi_{\rm CF}^{T0}(\bm{q},\Omega)=-i\langle J_T(\bm{q},\Omega)\rho(-\bm{q},-\Omega)\rangle,\\
     \Pi_{\rm CF}^{0T}(\bm{q},\Omega)=-i\langle \rho(\bm{q},\Omega)J_T(-\bm{q},-\Omega)\rangle  ,\quad \Pi_{\rm CF}^{TT}(\bm{q},\Omega)=K_{\rm diam}- \langle J_T(\bm{q},\Omega)J_T(-\bm{q},-\Omega)\rangle  \;.
\end{gather}
Here our sign convention is such that $J_T(\bm{q})= \hat{\bm{q}}_T\cdot J(\bm{q})$ and  $J_T(-\bm{q})= \hat{\bm{q}}_T\cdot J(-\bm{q})$, where $\hat{\bm{q}}_T=(\hat{z}\times \hat{\bm{q}})$ is the transverse vector. We note that this choice is consistent with the conventional definition of the current operator,  $ \bs{J}(\bs{q}) = \int_{\bs{k}} \bs{v}_{\bs{k}} f^{\dagger}(\bs{k} - \bs{q}/2) f(\bs{k} + \bs{q}/2) $, combined with Eq.~\eqref{eq:a_T_T}.

From this RPA effective Lagrangian, we can immediately deduce the physical response functions by integrating out the quadratic theory for $a$
\begin{equation}
    L_{\rm CFL} = \frac{1}{2} A_{\alpha}^* \Pi^{\alpha\beta}_{} A_{\beta} \,, \quad 
    \Pi_{} = \Pi_{\rm a} (\Pi_{\rm CF} + \Pi_{\rm a})^{-1} \Pi_{\rm CF} \,.\label{eq:Pi_CFL_full} 
\end{equation}
Inverting this relation gives 
\begin{equation}
    \Pi^{-1} = \Pi_{\rm a}^{-1} + \Pi_{\rm CF}^{-1} \,,\label{eq:Ioffe-Larkin_2} 
\end{equation}
which is consistent with the more general Ioffe-Larkin rule that holds beyond the RPA~\cite{Ioffe1989_rule}. 

The general structure presented above implies that in order to compute physical electromagnetic response functions, it suffices to calculate the CF response matrix $\Pi_{\rm CF}$. For an arbitrary dispersion relation, we find the following structure of $\Pi_{\rm CF}$ in the quantum critical regime $|\Omega| \ll v_F q$
\begin{equation}
    \begin{aligned}
        \Pi^{0T}_{\rm CF}(\bs{q}, \Omega) = \Pi^{T0}_{\rm CF}(\bs{q}, \Omega) &=-i\varkappa_1(\phi_{\bm{q}}) q^2 - \frac{m \Omega}{2\pi q} \left[\Re f_1(\phi_{\bs{q}}) + i \sgn(\Omega) \Im f_1(\phi_{\bs{q}}) \right]\,, \\
        \Pi^{00}_{\rm CF}(\bs{q}, \Omega) &= \chi_0+\varkappa_0 q^2 - \frac{im \Omega}{2\pi q} \left[ \Re f_0(\phi_{\bs{q}}) + i \sgn(\Omega) \Im f_0(\phi_{\bs{q}})\right]\,, \\
        \Pi^{TT}_{\rm CF}(\bs{q}, \Omega) &= -\varkappa_2(\phi_{\bm{q}}) q^2  +\frac{im\Omega}{2\pi q} \left[\Re f_2(\phi_{\bs{q}}) + i \sgn(\Omega) \Im f_2(\phi_{\bs{q}}) \right] \,,\label{eq:Pi_general_f}
    \end{aligned}
\end{equation}
where $\chi_0$ is the CF compressibility, and the angular functions $f_{\alpha}(\phi_{\bm{q}})$ are defined as 
\begin{equation}
   f_\alpha(\phi_{\bm{q}})= \int\frac{d^2\bm{k} }{2\pi m} \;\frac{  v_T^\alpha(\bm{q}) \delta(\xi_{\bm{k}}) }{i0^+-\bm{v}\cdot \hat{\bm{q}} } \,, \quad \varkappa_\alpha(\phi_{\bm{q}})=\frac{1}{8}\int\frac{d^2\bm{k} }{(2\pi)^2}v_T^\alpha(\bm{q})\left[\delta'(\xi_{\bm{k}})(\hat{\bm{q}}\cdot \nabla_{\bm{k}})^2\xi_{\bm{k}}+\frac{v_L^2(\bm{q})}{3} \delta''(\xi_{\bm{k}})\right] \;,\label{eq:f_def} 
\end{equation}
and $ v_T(\bs{q}) = \bs{v} \cdot \hat{\bs{q}}_T $, $ v_L(\bs{q}) = \bs{v} \cdot \hat{\bs{q}} $. While the explicit evaluation of these integrals is not analytically tractable for a general $C_3$ symmetric dispersion, we illustrate their structure with the specific weakly trigonally warped dispersion 
\begin{equation}
    \xi(\bs{k}) = \frac{k^2}{2m} (1 + \lambda \cos 3 \theta_{\bs{k}}) - \mu \,,
\end{equation}
as introduced in Eq.~\eqref{eq:trigonal_disp}, in which case $f_{\alpha}(\phi_{\bm{q}})$ acquires an especially simple form
\begin{equation}
   f_\alpha(\phi_{\bm{q}})= \int_0^{2\pi} \frac{d\theta_{\bm{k}}}{2\pi }  \frac{ (v_{F,y}\cos\phi_{\bm{q}}-v_{F,x}\sin\phi_{\bm{q}})^\alpha}{(1+\lambda \cos3\theta_{\bm{k}})(i0^+ -  v_{F,x}\cos\phi_{\bm{q}}-  v_{F,y}\sin\phi_{\bm{q}}) } \;,\label{eq:f_def_explicit}
\end{equation}
with $v_{F,x}(\theta_{\bs{k}}), v_{F,y}(\theta_{\bs{k}})$ equal to the Fermi velocity on the Fermi surface at angle $\theta_{\bs{k}}$.  
We then calculate the remaining angular integrals perturbatively in the warping parameter $\lambda$ to investigate the effects of inversion symmetry breaking. \color{black}Leaving details to Appendix~\ref{app:RPA_integrals}, we find the following results up to $\mathcal{O}(\lambda^2)$
% \begin{equation}\label{eq:f_i_definition}
%     f_0(\phi_{\bs{q}}) \approx \sqrt{\frac{m}{2\mu}} \left[-i + 3 \lambda \cos (3 \phi_{\bs{q}})\right] \,, \quad f_1(\phi_{\bs{q}}) = - \frac{7 i \lambda}{2} \sin 3 \phi_{\bs{q}}\,, \quad f_2(\phi_{\bs{q}}) = \sqrt{\frac{2\mu}{m}} \left[-i + 4 \lambda \cos 3 \phi_{\bs{q}}\right] \,.
% \end{equation}
\begin{equation}
    \begin{aligned}
     f_0(\phi_{\bm{q}}) &\approx \sqrt{\frac{m}{2\mu}}\left[-i+ 3\lambda \cos 3\phi_{\bm{q}}+i\lambda^2\left(\frac{21}{2}\cos 6\phi_{\bm{q}}-\frac{3}{8}\right)+\mathcal{O}(\lambda^3)\right]\;,\\
           f_1(\phi_{\bm{q}})&\approx -\frac{7i \lambda}{2} \sin 3\phi_{\bm{q}} +\frac{47\lambda^2}{4 }\sin 6\phi_{\bm{q}}+\mathcal{O}(\lambda^3)\;,\\
              f_2(\phi_{\bm{q}}) &\approx \sqrt{\frac{2\mu}{m}}\left[-i+ 4\lambda \cos 3\phi_{\bm{q}}+i\lambda^2\left(\frac{105}{8}\cos 6\phi_{\bm{q}}-\frac{3}{4}\right)+\mathcal{O}(\lambda^3)\right]\;.
    \end{aligned}\label{eq:f_i_definition}
\end{equation}
% The results at the order $\mathcal{O}(\lambda^2)$ are given in Appendix~\ref{app:RPA_integrals}. 
The compressibility for arbitrary $|\lambda|<1$ is given by $\chi_0= m/(2 \pi \sqrt{1-\lambda^2})$. Similarly, the coefficients $\varkappa_\alpha$ have following expressions: 
\begin{equation}\label{eq:varkappa_exp}
   \varkappa_0=0 ,\quad \varkappa_2=\frac{5-9\sqrt{1-\lambda^2}}{96\pi m},\quad \varkappa_1(\phi_{\bm{q}})= \frac{5\lambda}{256\pi \sqrt{2m\mu}} \sin 3\phi_{\bm{q}}+\mathcal{O}(\lambda^3)\;.
\end{equation}
We note that $\varkappa_0=0$ only holds for the dispersion of the form $\xi_{\bm{k}}\propto k^2 g(\theta_{\bm{k}})$ and is not generic.

After analytically continuing to real frequencies $i\Omega \rightarrow \omega + i 0^+$, the response functions take a simpler form
\begin{equation}\label{eq:PiCF_realfreq}
\begin{aligned}
    \Pi^{0T,R}_{\rm CF}(\bs{q},\omega) &= \Pi^{T0,R}_{\rm CF}(\bs{q}, \omega) = -i\varkappa_1(\phi_{\bs{q}}) q^2+\frac{im\omega}{2\pi q} f_1(\phi_{\bs{q}}) \,, \\
\Pi^{00,R}_{\rm CF}(\bs{q},\omega) &= \frac{m}{2\pi \sqrt{1-\lambda^2}} - \frac{m\omega}{2\pi q} f_0(\phi_{\bs{q}}) \,, \quad \Pi^{TT,R}_{\rm CF}(\bs{q},\omega) =\frac{9\sqrt{1-\lambda^2}-5}{96\pi m} q^2+ \frac{m \omega}{2\pi q} f_2(\phi_{\bs{q}}) \,. 
    \end{aligned}
\end{equation}
To convert to correlators in the longitudinal-transverse basis, we can use the real-frequency continuity equation $J_L(\bs{q}, \omega) =  \frac{\omega}{q} \rho(\bs{q}, \omega)$ and deduce 
\begin{equation}
    \Pi^{LL,R}_{\rm CF}(\bs{q}, \omega)=  -\frac{\omega^2}{q^2} \Pi^{00,R}_{\rm CF}(\bs{q}, \omega) \,,\quad
    \Pi^{LT,R}_{\rm CF}(\bs{q}, \omega)= \Pi^{TL,R}_{\rm CF}(\bs{q}, \omega) =  -\frac{i\omega}{q} \Pi^{0T,R}_{\rm CF} \,.
\end{equation}

The structure of these correlation functions has interesting physical implications. First, we observe that the mixed correlator $\Pi^{0T}_{\rm CF}$ vanishes identically unless $\lambda \neq 0$. Moreover, the form of the mixed correlator is a ``reactive" counterpart to the familiar Landau damping, with a real coefficient multiplying $\omega/q$. This new term modifies the dispersion of the gauge field while preserving the dynamical exponent $z$. A similar reactive effect also appears in $\Pi^{00}_{\rm CF}$ and $\Pi^{TT}_{\rm CF}$ and competes with the usual dissipative Landau damping term which already exists without inversion symmetry breaking. These reactive contributions to $\Pi_{\rm CF}$ will play a crucial role in our discussion of finite-momentum transport in Sec.~\ref{sec:finiteQ_transport}. 

Finally, by combining $\Pi_{\rm CF}$ with $\Pi_a$ and setting $A = 0$, we can deduce the gauge field propagator 
\begin{equation}
     \mathcal{D}^R(\bm{q},\omega) = D^R(\bm{q},\omega) \begin{pmatrix}
       q^2[\frac{ v(\bs{q})}{16\pi^2}-\varkappa_2]  +f_2(\phi_{\bm{q}})\frac{m\omega}{2 \pi q}  & i[\varkappa_1(\phi_{\bm q})q^2-f_1(\phi_{\bm{q}})\frac{m\omega}{2 \pi q}]  +\frac{q}{4\pi}\\
    i[\varkappa_1(\phi_{\bm q})q^2-f_1(\phi_{\bm{q}})\frac{m\omega}{2 \pi q}]   - \frac{q}{4\pi}& \chi_0 -f_0(\phi_{\bm{q}})\frac{m\omega}{2 \pi q}
     \end{pmatrix} \;.
\end{equation}
The overall factor is
\begin{equation}
[D^R(\bm q,\omega)]^{-1}
=
\left(
q^2\left(
\frac{v(\bm q)}{16\pi^2}-\varkappa_2
\right)
+
f_2(\phi_{\bm q})\frac{m\omega}{2\pi q}
\right)
\left(
\chi_0
-f_0(\phi_{\bm q})\frac{m\omega}{2\pi q}
\right)
+
\left(
f_1(\phi_{\bm q})\frac{m\omega}{2\pi q}
-\varkappa_1(\phi_{\bm q})q^2
\right)^2
+\frac{q^2}{16\pi^2}\;.
\end{equation}
Assuming a short-range interaction $v(\bs{q}) = v_0$, $D^R$ reduces in the low energy limit to 
\begin{equation}
      D^R(\bm{q},\omega)  \approx \frac{1}{ q^2 \left(\frac{1+v_0 \chi_0}{16\pi^2} -\chi_0 \varkappa_2\right)+  \frac{\sqrt{2m\mu} \chi_0 }{2\pi }\tilde{f}_2(\phi_{\bm{q}}) \frac{\omega}{q} }\;,
\end{equation}
where we defined a dimensionless integral $\tilde{f}_2(\phi_{\bm{q}})=\sqrt{m/2\mu}f_2(\phi_{\bm{q}})$. Note that this propagator satisfies $z=3$ dynamical scaling (i.e. $\omega\sim q^3$). Crucially, the imaginary part of $\tilde{f}_2(\phi_{\bm{q}})$ is always negative, which guarantees that the pole at $\omega=\omega_{\bm{q}}$ is in the lower half-plane:
\begin{equation}
    \omega_{\bm{q}}=- \frac{q^3(1+v_0 \chi_0 -16\pi^2 \chi_0\varkappa_2 ) }{8\pi \chi_0\sqrt{2m\mu}  } \frac{\tilde{f}_2^*(\phi_{\bm{q}})}{|\tilde{f}_2(\phi_{\bm{q}}) |^2} \;.
\end{equation}
Interestingly, finite trigonal warping $\lambda>0$ implies that the dynamics is not purely overdamped: the pole is shifted away from the imaginary axis resulting in additional oscillations (unless $\operatorname{Re}\tilde{f}_2(\phi_{\bm{q}})$ vanishes at special directions such as $\phi_{\bm{q}}=n \pi/6$, $n\in \mathbb{Z}$). 

From the RPA calculations presented above, we can deduce the following properties of lattice CFLs
\begin{enumerate}
    \item In thermodynamics, the gauge field contribution to the free energy can be computed within the RPA. Since the dynamical exponent is not modified by inversion symmetry breaking, the free energy continues to scale as $F(T) \sim T^{\frac{z+2}{z}}$ and the specific heat scales as $C_V(T) \sim - T \partial^2_T F(T) \sim T^{2/z}$. This scaling of the specific heat is a hallmark feature of CFLs and has been verified both in numerics~\cite{Chen2025_CFLtensor} and recent graphene experiments~\cite{Assouline2026_CFLentropy}. 
    \item The electronic compressibility, which is given by the $q \rightarrow 0$ limit of $\frac{q^2}{(4\pi)^2} \times D^R_{TT}(\bs{q}, \omega = 0)$, is a positive constant. This is an important property that distinguishes CFLs from other candidate quantum Hall states at half-filling. 
    \item The single electron Green's function can be approximated by a product of the CF Green's function and the Green's function of the monopole operator $\mathcal{M}_a^2$, which creates $4\pi$ flux of the gauge field $a$. While the CF Green's function has power-law decay in space and time, the monopole Green's function decays faster than any power law~\cite{Kim1994_monopole} (the precise functional form of the decay is controversial~\cite{He1993_CFL_spectral,Yue2024_CFLspectral}). Therefore, the low energy spectral weight of the electron Green's function is strongly suppressed, consistent with trion sensing experiments~\cite{Anderson2024_trionCFL}. 
\end{enumerate}

Importantly, the properties summarized above are universal to all realizations of lattice CFLs, independent of the presence/absence of inversion symmetry. In the remainder of this paper, we will turn to various transport observables, where the breaking of inversion symmetry leads to qualitatively new effects. 

\section{Transport at zero wave vector}\label{sec:zeroQ_transport}

We now present our first set of results on homogeneous electrical transport of lattice CFLs, both in the optical limit $\omega \gg T$ and in the DC limit $\omega = 0, T \neq 0$. For conceptual clarity, we will work in the ultra-clean limit and ignore the effects of disorder. In what follows, we will first make some general remarks regarding symmetry-enforced non-perturbative constraints on the transport behavior of metallic systems in 2+1D, which are independent of the precise choice of microscopic model. These remarks then set the stage for various interesting transport properties in the rest of the Section enabled by symmetry reduction. 

In the conventional Landau level CFL, electrons with $|\bs{k}|^2/2m$ dispersion are subject to a large magnetic field $B$ such that the Landau level filling of electrons is $\nu = 1/2$. In the presence of interactions that only depend on the position of the electrons, the full Hamiltonian respects magnetic Galilean invariance. One consequence of this symmetry is that the $\bs{q} = 0$ total current operator $J_i$ is proportional to the canonical momentum $\pi_i$
\begin{equation}
    J_i = \frac{1}{m} \pi_i \,. 
\end{equation}
On the other hand, the equations of motion for the canonical momentum $\pi_i$ is independent of the interaction potential $V_{\rm int}$ and completely fixed by the magnetic field $B$ and the probe electric field $\bs{E}$
\begin{equation}
    \dot \pi_i = n E_i + \omega_c \epsilon_{ij} \pi_j \,. 
\end{equation}
Combining these two equations and taking a Fourier transform from $t \rightarrow \omega$, we obtain the exact Heisenberg equations of motion for $J_i(\omega)$ and the exact form of the optical conductivity 
\begin{equation}
    - i \omega J_i(\omega) - \omega_c \epsilon_{ij} J_j(\omega) = \frac{n}{m} E_i \quad \rightarrow \quad \sigma(\omega, T) = \frac{n}{m (\omega^2 - \omega_c^2)} \begin{pmatrix}
        i \omega & - \omega_c \\ \omega_c & i \omega
    \end{pmatrix} \,. 
\end{equation}
Remarkably, the gapless low energy fluctuations of the CF Fermi surface leave no fingerprint in the homogeneous conductivity matrix $\sigma(\omega, T)$, which is an analytic function of $\omega$ independent of $T$. 

In a lattice CFL, Galilean invariance is explicitly broken and homogeneous transport is no longer fixed by UV symmetries. Nevertheless, we can obtain powerful constraints on transport observables from emergent symmetries in the low energy effective description of CFLs. In complete generality, the low energy effective action Eq.~\eqref{eq:CFL_EFT} implies the Ioffe-Larkin composition rule valid at $\bs{q} = 0$~\cite{Ioffe1989_rule}
\begin{equation}
   \rho = \rho_{\rm CF} + \frac{2h}{e^2} \begin{pmatrix}
        0 & 1 \\ -1 & 0 
    \end{pmatrix} 
\end{equation}
where $\rho_{\rm CF}$ is determined by the linear response of CFs to the total gauge field $A + a$ they see. Independent of lattice symmetries, the CF sector forms a Fermi liquid up to corrections from singular gauge fluctuations. Due to the presence of an emergent Fermi surface for CFs, the low energy theory has an emergent infinite-dimensional symmetry associated with the conservation of CF charge density $n(\theta)$ at angle $\theta$ on the Fermi surface~\cite{Else2020_EFL,Shi2022_gifts}. As shown in Ref.~\cite{,Shi2022_loopcurrent} through the memory matrix formalism~\cite{Forster1975_memory,Maebashi1997_memoryI,Maebashi1998_memoryII,Hartnoll2016_holography}, this infinite-dimensional symmetry implies that the CF conductivity $\sigma^{xx}_{\rm CF}(\omega, T = 0)$ must contain a sharp Drude peak $i D/\omega$, where the Drude weight $D$ is fixed by the overlap between the CF current and the conserved charges $n(\theta)$.\footnote{Technically, $n(\theta)$ should be regarded as a distribution, and it is more appropriate to think about the conserved quantities as Fourier modes $n_l = \frac{1}{2\pi} \int d \theta n(\theta) e^{il \theta}$ labeled by the angular momentum.} This Drude peak term is again insensitive to the gapless CF Fermi surface fluctuations. As a result, any residual non-analytic contribution to the conductivity must be subleading relative to $\omega^{-1}$ and take the form 
\begin{equation}
    \sigma^{xx}_{\rm CF}(\omega, T = 0) = \frac{iD}{\omega} + \sigma_{\rm inc}(\omega) \,, \quad \lim_{\omega \rightarrow 0} \omega \sigma_{\rm inc}(\omega) = 0 \,. 
\end{equation}
At nonzero temperature, the infinite-dimensional emergent symmetry is weakly violated and all the emergent conserved charges $n_l$ develop scattering rates $\Gamma_l(T)$ that scale as a power law in $T$. The slowest decaying operator gives the dominant contribution to the conductivity $\sigma^{xx}_{\rm CF}(\omega=0, T) \sim \Gamma_{l_{\rm min}}(T)^{-1}$. In many cases, the slowest decaying operator overlaps with the continuum CF momentum operator, whose relaxation rate is determined by the most efficient Umklapp scattering processes. 

The general argument above implies that there are two potentially interesting sources of non-analytic transport behavior in a lattice CFL:
\begin{enumerate}
    \item The gapless low energy fluctuations in the CFL generate a non-analytic incoherent optical conductivity 
    \begin{equation}
        \sigma^{xx}_{\rm inc}(\omega) = C\,  \omega^{-\alpha} \,, \quad \sigma^{xx}_{\rm CF}(\omega) \sim \frac{iD}{\omega} + C \, \omega^{-\alpha} \approx \frac{iD}{\omega} \left[1 + \frac{C}{iD} \omega^{1-\alpha}\right] \,,
    \end{equation}
    with a scaling exponent $\alpha < 1$. When this is the case, the physical optical resistivity also develops a non-analytic dissipative part
    \begin{equation}
        \Re \, \rho^{xx}(\omega) \approx \Re \, \frac{\omega}{iD} \left[1 - \frac{C}{iD} \omega^{1-\alpha}\right] = \frac{C}{D^2} \omega^{2 - \alpha} \,. 
    \end{equation}
    Does such a term exist in the lattice CFL? It turns out that for an inversion-symmetric convex CF Fermi surface (which is the generic situation considered in the existing literature), the answer is no! In fact, the leading gauge-field-induced contribution to the incoherent conductivity scales as a positive power of $\omega$ (i.e. $\alpha < 0$). This result can be understood from kinematic constraints in 2+1D that forbid certain relaxation channels for inversion-odd operators, of which the current operator is a special case. In lattice CFLs, the general kinematic argument is explicitly verified through perturbative diagrammatic calculations~\cite{Shi2022_loopcurrent,Guo2022_YukawaSYK2d,Shi2023_controlled,Guo2024_fluc_criticalFS,Li2023_NFLoptical,Gindikin2024_optical_NFL,Gindikin2025_NFLstability}.

    Once inversion symmetry is broken, the constraints above are lifted and a non-analytic $\sigma_{\rm inc}(\omega)$ becomes possible. We will show through explicit calculations that this is indeed the case, and that the scaling exponent $\alpha = \frac{2z-4}{z}$. 
    \item Singular Umklapp interactions in the lattice CFL generate a non-analytic DC momentum relaxation rate $\Gamma_P(T) \sim T^{\beta}$ such that $\beta < 2$. If this occurs, the dissipative part of the physical resistivity develops a scaling form
    \begin{equation}
        \Re \, \rho^{xx}(\omega = 0, T) \sim \Gamma_P(T) \sim T^{\beta} \,,
    \end{equation}
    which dominates over the conventional $T^2$ scaling for Fermi liquids. 

    Adapting ideas introduced in Ref.~\cite{Lee2024_powerlaw_resistivity}, we show that this kind of NFL transport is indeed possible when Umklapp processes involve singular $2k_F$ scattering on the CF Fermi surface. The specific value of the exponent $\beta$ depends on the non-universal shape of the Fermi surface and inversion symmetry does not play a prominent role. 
\end{enumerate}

With the conceptual overview above, we are now ready to discuss more detailed results. %We treat optical transport in Sec.~\ref{subsec:zeroQ_optical}, emphasizing the physical intuition behind the singular effect of inversion symmetry breaking on $\sigma_{\rm inc}(\omega)$. This discussion is followed by a more brief treatment of DC transport in Sec.~\ref{subsec:zeroQ_DC}, which is largely based on ideas developed in Ref.~\cite{Lee2024_powerlaw_resistivity}. 

\subsection{Optical resistivity at $T = 0$}\label{subsec:zeroQ_optical}

\begin{figure}
    \centering
    \includegraphics[width=0.6\linewidth]{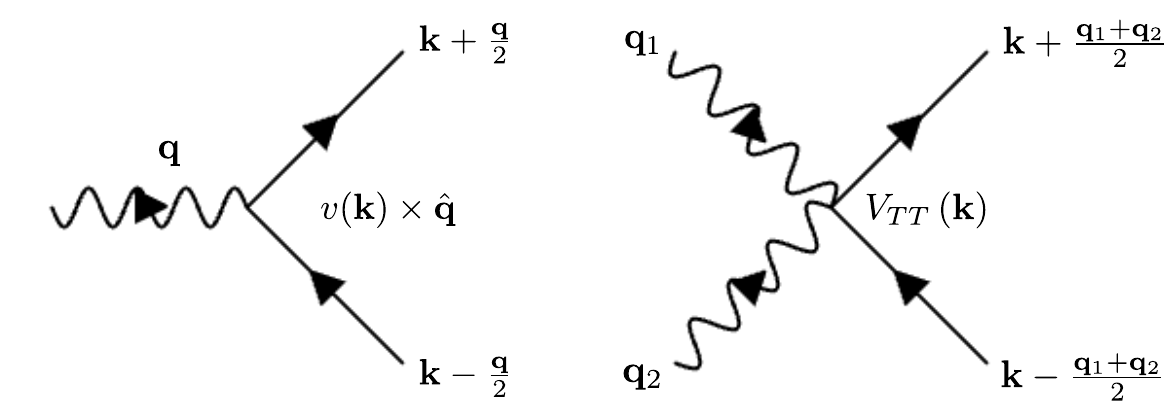}
    \caption{Paramagnetic (left) and diamagnetic (right) vertices for the lattice CFL with arbitrary CF dispersion $\xi(\bs{k})$. $\bs{\hat q}_T$ defines the direction transverse to $\bs{q}$, $\bs{v}_F(\bs{k}) \equiv \nabla_{\bs{k}} \xi(\bs{k})$ is the Fermi velocity, and $V_{TT}$ is the projection of the diamagnetic form factor $V_{ij} \equiv \partial_{k_i} \partial_{k_j} \xi(\bs{k})$ onto the transverse subspace. }
    \label{fig:Feynman_rule}
\end{figure}
Using the Kubo formula, we can express the optical CF conductivity (dropping the arguments $\bs{q} = 0, T = 0$) as
\begin{equation}
    \sigma_{\rm CF}^{ij}(\omega) = \frac{\Pi^{ij,R}_{\rm CF}(\omega)}{-i\omega} \,,
\end{equation}
where $\Pi^{ij,R}_{\rm CF}(\omega)$ is the CF current-current response function as defined in Eq.~\eqref{eq:PiCF_def}. As we have shown in Sec.~\ref{subsec:review_RPA}, the leading RPA contribution to $\Pi^{ij,R}_{\rm CF}(\omega)$ comes entirely from the constant diamagnetic term $K_{\rm diam}$, which contributes a sharp Drude peak. To obtain the singular frequency-dependent correction to the Drude peak, we must include higher-order diagrams that encode the scattering of CFs induced by singular gauge field fluctuations.

Towards that end, we follow the approach of Refs.~\cite{Polchinski1993_largeN,Altshuler1994_singular,Kim1994_RPA_optical} and consider a deformation of the original model with $N$ flavors of CFs with identical dispersion $\xi(\bs{k})$, described by the Euclidean action (with external gauge field $A$ set to zero)
\begin{equation}\label{eq:largeN_action}
    S = S_a + S[f, a] \,,
\end{equation}
\begin{equation}
    S_a =  \frac{N}{2} \int_{\tau, \bs{q}} \frac{V_0}{(4\pi)^2}|\bs{q}|^{z-1} |a_T(\bs{q},\tau)|^2  \,, \quad S[f, a] = \sum_{i=1}^N \int \bar f_i(\bs{k},\tau) \left[\partial_{\tau} - i a_0 + \xi(\bs{k} - \bs{a})\right] f_i(\bs{k},\tau) \,. 
\end{equation}
Here, $S[f, a]$ is the standard minimal coupling between the CF Fermi surface and the $U(1)$ gauge field $a$. As for $S_a$, we have kept only the gapless fluctuations of $a_T$ and dropped contributions from $a_0$, which is screened (i.e. rendered massive) by the Fermi surface fluctuations~\cite{Altshuler1994_singular}. This approximation allows us to extract the leading singular contributions to gauge-invariant correlation functions in the low energy limit. We also assumed that $v(\bm{q})=v_0 |\bm{q}|^{z-3}$, and defined $V_0=v_0+1/\chi_0$ for $z=3$ and $V_0=v_0$ for $2\leq z<3$. %The dynamical exponent $z$ controls the range of density-density interactions $V(r)$ in the system, with $z=2$ corresponding to $1/r$ Coulomb interactions and $z = 3$ corresponding to short-range interactions with $V(r)$ decaying faster than $1/r^2$. 

Expanding the action $S[f, a]$ up to quadratic order in $a$, we find a paramagnetic coupling $S_{\rm para}[f, a]$ as well as a diamagnetic coupling $S_{\rm diam}[f, a]$. Higher-order terms in $a$ are irrelevant in the low energy limit and can be neglected for the purpose of extracting the dominant singular frequency dependence of $\Pi^{ij}_{\rm CF}(\bs{q}=0,\omega)$. The total action, therefore, takes the approximate form
\begin{equation}\label{eq:Seff_transport}
    S_{\rm eff} = S_a + \sum_{i=1}^N \int_{\bs{k}, \tau} \bar f_i(\bs{k}, \tau) \left[\partial_{\tau} + \xi(\bs{k})\right] f_i(\bs{k}, \tau) + S_{\rm para}[f,a] + S_{\rm diam}[f, a] \,,
\end{equation}
where 
\begin{equation}
    S_{\rm para}[f, a] = \sum_{i=1}^N \int_{\bs{k}, \bs{q}, \tau} \bar f_i(\bs{k}+\bs{q}/2, \tau) \left[- i a_0(\bs{q},\tau) -\nabla_{\bs{k}}\xi(\bs{k}) \cdot \bs{a}(\bs{q}, \tau)\right] f_i(\bs{k} - \bs{q}/2, \tau) \,,
\end{equation}
\begin{equation}\label{eq:S_diam_a}
    S_{\rm diam}[f, a] = \frac{1}{2} \sum_{i=1}^N \int_{\bs{k}, \bs{q},\tau} \bar f_i(\bs{k}+  \frac{\bs{q}_1 + \bs{q}_2}{2}, \tau) f_i(\bs{k} -  \frac{\bs{q}_1 + \bs{q}_2}{2}, \tau) a_i(\bs{q}_1, \tau) a_j(\bs{q}_2, \tau) V_{ij}\left(\bs{k}\right) + \mathcal{O}(q_i^2) \,.
\end{equation}
In the last line, we defined a diamagnetic form factor $V_{ij} = \partial_{k_i} \partial_{k_j} \xi(\bs{k})$. Note that corrections of order $q_i^2$ in the last line vanish identically for a Galilean invariant dispersion $\xi(\bs{k}) = k^2/2m$ and the form factor $V$ reduces to $V_{ij} = m^{-1} \delta_{ij}$. For a general dispersion $\xi(\bs{k})$, we will drop the $\mathcal{O}(q_i^2)$ corrections as they only give rise to subleading corrections in the low energy limit. The resulting Feynman rules then take the simple form in Fig.~\ref{fig:Feynman_rule}.

Using the Feynman rules in Fig.~\ref{fig:Feynman_rule}, we can compute every physical observable as a power series in $1/N$, where each series coefficient selects a finite number of Feynman diagrams. Although this expansion scheme ceases to be controlled in the ultimate low energy limit $\omega \ll E_F N^{-3}$~\cite{Lee2009_RPAbad,Metlitski2010_RPAbad}, the frequency-scaling of $\sigma_{\rm CF}^{ij}$ extracted from this expansion agrees with more sophisticated controlled expansions as shown in Refs.~\cite{Mross2010_largeN_smalleps,Shi2023_controlled,Guo2022_YukawaSYK2d,Guo2024_fluc_criticalFS}. Thus we will work within the ordinary large $N$ deformation defined by Eq.~\ref{eq:Seff_transport} and compute $\sigma^{ij}_{\rm CF}$ to leading two orders in the $1/N$ expansion.

\begin{figure*}
    \centering
    \includegraphics[width = 0.8\textwidth]{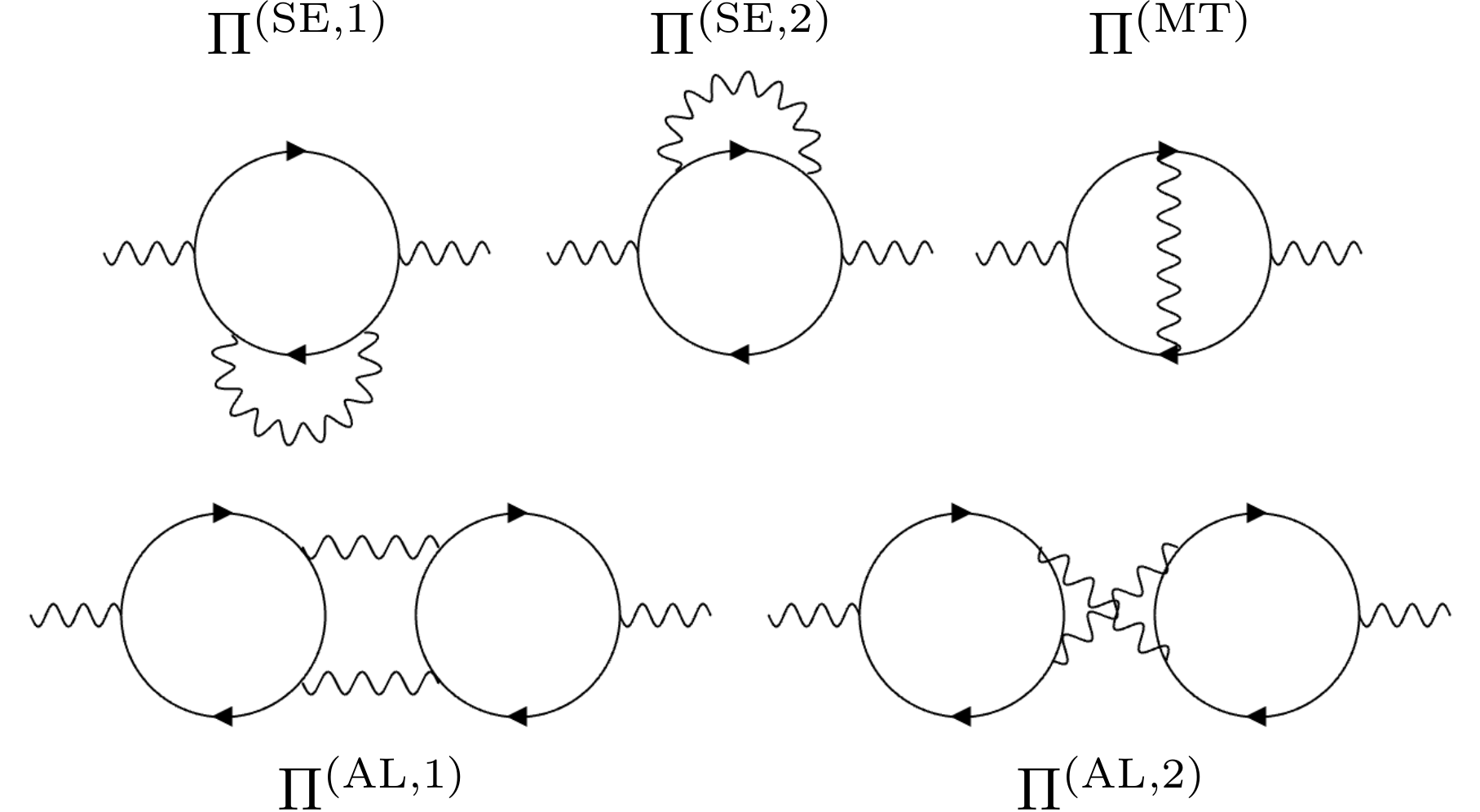}
    \caption{The set of paramagnetic diagrams contributing to $\sigma^{ij}_{\rm CF}$ at $\mathcal{O}(N^{-1})$. All vertex factors are taken to be $f(\bs{k}, \hat{\bs{q}}) = \bs{v}_F(\bs{k}) \times \hat{\bs{q}}$ in accordance with Fig.~\ref{fig:Feynman_rule}. Diagrams that involve diamagnetic vertices are subleading in the low energy limit, as we show in Appendix~\ref{app:detail_optical}.}
    \label{fig:Feynman_O(1:N)_para}
\end{figure*}

At leading order in the large $N$ expansion, the relevant diagrams are the RPA diagrams computed in Sec.~\ref{subsec:review_RPA}. In the homogeneous $\bs{q} = 0$ limit, these diagrams evaluate to a sharp Drude peak 
\begin{equation}
    \sigma^{xx, (0)}_{\rm CF}(\bs{q}=0, \omega) = \frac{i D_0}{\omega} \,, \quad D_0 = \Pi^{TT,R}_{\rm CF}(\bs{q}=0, \omega) \,. 
\end{equation}
The subleading $\mathcal{O}(N^{-1})$ contribution $\sigma^{xx, (1)}_{\rm CF}$ includes a large number of Feynman diagrams. In Appendix~\ref{app:detail_optical}, we show that all diagrams involving at least one insertion of the diamagnetic vertex in Fig.~\ref{fig:Feynman_rule} contribute a term in the conductivity that is at most $\mathcal{O}(\omega^{\frac{3-z}{z}})$ in the $\omega \rightarrow 0$ limit. The remaining diagrams that do not involve any diamagnetic vertices are shown in Fig.~\ref{fig:Feynman_O(1:N)_para}. 
In Appendix~\ref{app:detail_optical}, we show that the sum of these diagrams always dominates over the diagrams that involve diamagnetic vertices. As a result, we can safely neglect the diamagnetic vertices at this order in the $1/N$ expansion. 

The IR scaling form of the $\mathcal{O}(N^{-1})$ optical conductivity contains two qualitatively different terms
\begin{equation}
    \sigma_{\rm CF}^{xx,(1)}(\omega) = \frac{1}{-i \omega N} \left[\Pi_{\rm CF, on-shell}^{xx,R}(\omega)+ \Pi_{\rm CF, virtual}^{xx,R}(\omega)\right] \,,\label{eq:sigma_CF_1_full}
\end{equation}
where the ``on-shell" and ``virtual" contributions are given on the Matsubara axis as
\begin{equation}\label{eq:zeroQ_BnB_decomposition}
    \begin{aligned}
    &\Pi_{\rm CF, on-shell}^{xx}(\Omega) \\
    &= \frac{1}{2\Omega^2}\int_{\bs{q}',\Omega'} \mathcal{D}_{TT}(\bs{q}', \Omega') \mathcal{D}_{TT}(\bs{q}', \Omega'-\Omega) \bigg\{\left[\Re \, \pi^{\scaleto{(\rm AL)}{6pt}}(\bs{q'},\Omega' - \Omega) - \Re \, \pi^{\scaleto{(\rm AL)}{6pt}}(\bs{q'},\Omega') \right]^2 \\
    &-\left[\Re \, \pi_0(\bs{q}', \Omega' - \Omega) - \Re \, \pi_0(\bs{q}', \Omega')\right] \cdot \left[\Re \, \pi^{\scaleto{(\rm MT)}{6pt}}(\bs{q'},\Omega' - \Omega) - \Re \, \pi^{\scaleto{(\rm MT)}{6pt}}(\bs{q'},\Omega') \right]\bigg\} \,, \\
    &\Pi_{\rm CF, virtual}^{xx}(\Omega) \\ &= -\frac{1}{2\Omega^2}\int_{\bs{q}',\Omega'} \mathcal{D}_{TT}(\bs{q}', \Omega') \mathcal{D}_{TT}(\bs{q}', \Omega'-\Omega) \bigg\{\left[\Im \, \pi^{\scaleto{(\rm AL)}{6pt}}(\bs{q'},\Omega' - \Omega) - \Im \, \pi^{\scaleto{(\rm AL)}{6pt}}(\bs{q'},\Omega') \right]^2 \\
    &-\left[\Im \, \pi_0(\bs{q}', \Omega' - \Omega) - \Im \, \pi_0(\bs{q}', \Omega')\right] \cdot \left[\Im \, \pi^{\scaleto{(\rm MT)}{6pt}}(\bs{q'},\Omega' - \Omega) - \Im \, \pi^{\scaleto{(\rm MT)}{6pt}}(\bs{q'},\Omega') \right]\bigg\} \\
    &+ \frac{i}{2\Omega^2} \int_{\bs{q}', \Omega'} \mathcal{D}_{TT}(\bs{q}', \Omega') \mathcal{D}_{TT}(\bs{q}', \Omega'-\Omega) \\
    &\bigg\{2\left[\Re \, \pi^{\scaleto{(\rm AL)}{6pt}}(\bs{q'},\Omega' - \Omega) - \Re \, \pi^{\scaleto{(\rm AL)}{6pt}}(\bs{q'},\Omega') \right]\left[\Im \, \pi^{\scaleto{(\rm AL)}{6pt}}(\bs{q'},\Omega' - \Omega) - \Im \, \pi^{\scaleto{(\rm AL)}{6pt}}(\bs{q'},\Omega') \right] \\
    &-\left[\Re \, \pi_0(\bs{q}', \Omega' - \Omega) - \Re \, \pi_0(\bs{q}', \Omega')\right] \cdot \left[\Im \, \pi^{\scaleto{(\rm MT)}{6pt}}(\bs{q'},\Omega' - \Omega) - \Im \, \pi^{\scaleto{(\rm MT)}{6pt}}(\bs{q'},\Omega') \right] \\
    &- \left[\Im \, \pi_0(\bs{q}', \Omega' - \Omega) - \Im \, \pi_0(\bs{q}', \Omega')\right] \cdot \left[\Re \, \pi^{\scaleto{(\rm MT)}{6pt}}(\bs{q'},\Omega' - \Omega) - \Re \, \pi^{\scaleto{(\rm MT)}{6pt}}(\bs{q'},\Omega')\right]\bigg\} \,.
    \end{aligned}
\end{equation}
We note that here ``real'' and ``imaginary'' parts refer to the decomposition of the Matsubara bubbles before analytic continuation. Both of these expressions involve two factors of the gauge field propagator $\mathcal{D}$ and a specific combination of the generalized one-loop fermion bubble diagrams 
\begin{equation*}
    \begin{aligned}
    \pi_0(\bs{q}', \Omega') &\equiv \Pi^{TT}_{\rm CF}(\bs{q}', \Omega') = K_{\rm diam} + \int f(\bs{k}, \hat{\bs{q}})^2 G(\bs{k} + \frac{\bs{q}}{2}, \omega + \frac{\Omega}{2}) G(\bs{k} - \frac{\bs{q}}{2}, \omega - \frac{\Omega}{2})\,, \\
    \pi^{\scaleto{(\rm MT)}{6pt}}(\bs{q}', \Omega') &= \int_{\bs{k}, \omega} [f(\bs{k} + \frac{\bs{q'}}{2}, \hat{\bs{q}}) - f(\bs{k} - \frac{\bs{q'}}{2}, \hat{\bs{q}})]^2 f(\bs{k}, \hat{\bs{q}}')^2 \, G(\bs{k} + \frac{\bs{q'}}{2}, \omega+\Omega') G(\bs{k} - \frac{\bs{q'}}{2}, \omega) \,, \\
    \pi^{\scaleto{(\rm AL)}{6pt}}(\bs{q'},\Omega') &= \int_{\bs{k}, \omega} \left[f(\bs{k} + \frac{\bs{q'}}{2}, \hat{\bs{q}}) - f(\bs{k} - \frac{\bs{q'}}{2}, \hat{\bs{q}})\right] f(\bs{k},\hat{\bs{q}}')^2 G(\bs{k} + \frac{\bs{q'}}{2}, \omega+\Omega') G(\bs{k} - \frac{\bs{q'}}{2}, \omega) \,,
    \end{aligned}
\end{equation*}
with $f(\bs{k}, \hat{\bs{q}}) = \bs{v}_F(\bs{k}) \times \hat{\bs{q}}$. As we show in Appendix~\ref{app:detail_optical}, singular contributions to the conductivity come from the kinetic regime $|\Omega'| \ll v_F |\bs{q}'|$, where the real and imaginary parts of the general one-loop fermion bubble
\begin{equation}
    \pi_F(\bs{q}', \Omega') = \int_{\bs{k}, \omega} F(\bs{k}, \bs{q}') G(\bs{k} + \frac{\bs{q'}}{2}, \omega+\Omega') G(\bs{k} - \frac{\bs{q'}}{2}, \omega)= \Re \,\pi_F(\bs{q}', \Omega') + i \Im \,\pi_F(\bs{q}', \Omega') 
\end{equation}
can be expressed as angular integrals
\begin{equation}
    \begin{aligned}
    \Re \,\pi_F(\bs{q}',\Omega') &= I_F(\bs{q}') + \frac{|\Omega'|}{4\pi |\bs{q}'|} \int d \theta J(\theta) F(\theta, \bs{q}') \delta[\hat{\bs{q}}' \cdot \bs{v}_F(\theta)]  \,, \\
    \Im \,\pi_F(\bs{q}',\Omega') &= - \frac{\Omega'}{4 \pi^2 |\bs{q}'|} \int d \theta J(\theta) F(\theta,\bs{q}') \mathcal{P} \frac{1}{\hat{\bs{q}}' \cdot \bs{v}_F(\theta)} \,.
    \end{aligned}
\end{equation}
Here, $\bs{v}_F(\theta)$ is the Fermi velocity at angle $\theta$ on the CF Fermi surface, $F(\theta, \bs{q}')$ is the general form factor $F(\bs{k}, \bs{q}')$ evaluated at momentum $\bs{k}_F(\theta)$ on the CF Fermi surface, and $J(\theta)$ is the Jacobian associated with the change of variables from $\{k_x, k_y\}$ to $\{\xi(k_x, k_y), \theta(k_x, k_y)\}$. 

The real and imaginary parts of $\pi_F$ have different physical interpretations. In $\Re \,\pi_F$, the angular integral multiplying $|\Omega'|/|\bs{q}'|$ localizes to discrete angular patches on the CF Fermi surface where $\hat{\bs{q}}'$ is orthogonal to the Fermi velocity $\bs{v}_F(\theta)$ and hence tangential to the CF Fermi surface. This localization implies that $\Re \,\pi_F$ encodes on-shell small-angle scattering of CFs induced by the long-wavelength gauge fluctuations. In contrast, the angular integral multiplying $\Omega'/|\bs{q}'|$ in $\Im \,\pi_F$ is a principal-value integral that receives contributions from all $\theta$. This principal-value integral encodes virtual off-shell scattering processes that take CFs away from the Fermi surface.

\begin{figure}[t!]
    \centering
    \includegraphics[width=0.9\linewidth]{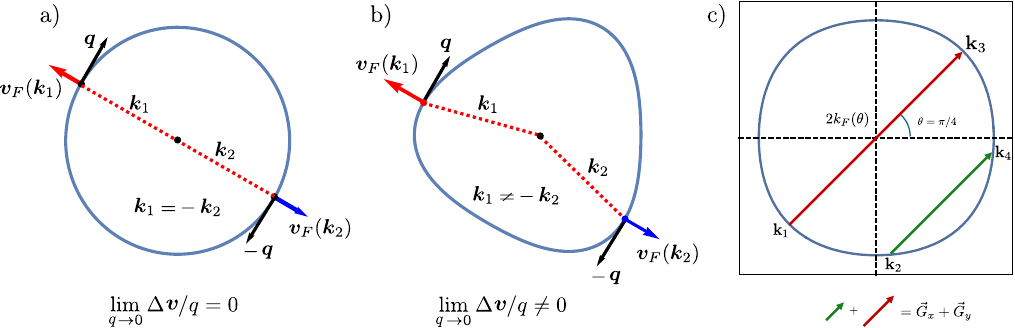}
    \caption{On-shell scattering kinematics controlling optical and DC transport.
(a)
For a convex inversion-symmetric Fermi surface, the on-shell scattering with small momentum transfer $\bm q$ selects two antipodal
tangential patches, $\bm k_1=-\bm k_2$, and the leading total velocity change
vanishes,
$\lim_{q\to0}\Delta\bm v/q=0$.
(b) For a generic convex $C_3$-symmetric Fermi surface without inversion
symmetry, the tangential on-shell patches are no
longer antipodal and the cancellation fails,
$\lim_{q\to0}\Delta\bm v/q\neq0$, allowing the collisions to relax
the current.
c) A typical four-fermion Umklapp process in the lattice CFL that contributes to the DC resistivity. Singular temperature dependence can arise when the $2k_F$ vertex function of CFs is sufficiently enhanced.}
\label{fig:FS_scattering}
\end{figure}

The decomposition of $\pi_F$ gives a clean physical interpretation of the two terms appearing in $\sigma^{xx, (1)}_{\rm CF}$. The first term $\Pi^{xx}_{\rm CF, on-shell}$ in Eq.~\eqref{eq:sigma_CF_1_full} depends only on $\Re\, \pi_F$ for several different choices of $F$. Physically, this term captures real on-shell gauge-field-mediated collisions that cause current relaxation at $\omega \neq 0$. These processes have a direct kinetic interpretation in terms of a Fermi-golden-rule scattering rate, or equivalently as the collision integral of a Boltzmann equation. On a convex Fermi surface, for every $\bs{q}'$, there are generically two angles $\theta_{\bs{q}'}^{(1)}, \theta_{\bs{q}'}^{(2)}$ on the Fermi surface that satisfy $\bs{q}' \cdot \bs{v}_F(\theta) = 0$. In the absence of inversion symmetry, these two points are not related by $\theta_{\bs{q}'}^{(1)} = \theta_{\bs{q}'}^{(2)} + \pi$. Therefore, the most general form of the angular integral is 
\begin{equation}
    \Re \, \pi_F(\bs{q}',\Omega') \approx I_F(\bs{q}') + \frac{|\Omega'|}{4\pi |\bs{q}'|} \sum_{i=1}^2 \frac{J(\theta_{\bs{q}'}^{(i)}) F(\theta_{\bs{q}'}^{(i)}, \bs{q}')}{\left|\hat{\bs{q}}' \cdot \partial_{\theta} \bs{v}_F(\theta_{\bs{q}'}^{(i)})\right|} \,. 
\end{equation}
Plugging this answer back into $\Pi^{xx}_{\rm CF, on-shell}$, we find
\begin{equation}
    \begin{aligned}
    \Pi^{xx}_{\rm CF, on-shell} &= -\frac{1}{2\Omega^2} \int_{\bs{q}',\Omega'} \mathcal{D}_{TT}(\bs{q}', \Omega') \mathcal{D}_{TT}(\bs{q}', \Omega'-\Omega) \left(\frac{|\Omega'-\Omega| - |\Omega'|}{4\pi |\bs{q}'|}\right)^2 \\
    &\bigg[\left(\sum_{i=1}^2 \frac{J(\theta_{\bs{q}'}^{(i)}) f(\theta_{\bs{q}'}^{(i)}, \hat{\bs{q}}')^2}{\left|\hat{\bs{q}}' \cdot \partial_{\theta} \bs{v}_F(\theta_{\bs{q}'}^{(i)})\right|} \right) \cdot \left(\sum_{j=1}^2 \frac{J(\theta_{\bs{q}'}^{(j)}) \left[\bs{q}' \cdot \nabla_{\bs{k}} f(\theta_{\bs{q}'}^{(j)}, \hat{\bs{q}})\right]^2 f(\theta_{\bs{q}'}^{(j)}, \hat{\bs{q}}')^2}{\left|\hat{\bs{q}}' \cdot \partial_{\theta} \bs{v}_F(\theta_{\bs{q}'}^{(j)})\right|}\right) \\
    &- \left(\sum_{i=1}^2 \frac{J(\theta_{\bs{q}'}^{(i)}) \left[\bs{q}' \cdot \nabla_{\bs{k}} f(\theta_{\bs{q}'}^{(i)}, \hat{\bs{q}})\right] f(\theta_{\bs{q}'}^{(i)}, \hat{\bs{q}}')^2}{\left|\hat{\bs{q}}' \cdot \partial_{\theta} \bs{v}_F(\theta_{\bs{q}'}^{(i)})\right|} \right) \cdot \left(\sum_{j=1}^2 \frac{J(\theta_{\bs{q}'}^{(j)}) \left[\bs{q}' \cdot \nabla_{\bs{k}} f(\theta_{\bs{q}'}^{(j)}, \hat{\bs{q}})\right] f(\theta_{\bs{q}'}^{(j)}, \hat{\bs{q}}')^2}{\left|\hat{\bs{q}}' \cdot \partial_{\theta} \bs{v}_F(\theta_{\bs{q}'}^{(j)})\right|} \right) \bigg] \,. 
    \end{aligned}
\end{equation}
Now observe that the $i = j$ terms in the second and third lines cancel each other exactly. Therefore, we are left with the $i \neq j$ terms 
\begin{equation}
    \begin{aligned}
        \Pi^{xx}_{\rm CF, on-shell} &= -\frac{1}{2\Omega^2}\int_{\bs{q}',\Omega'} \mathcal{D}_{TT}(\bs{q}', \Omega') \mathcal{D}_{TT}(\bs{q}', \Omega'-\Omega) \left(\frac{|\Omega'-\Omega| - |\Omega'|}{4\pi |\bs{q}'|}\right)^2 \\
        &\cdot \frac{J(\theta_{\bs{q}'}^{(1)})  J(\theta_{\bs{q}'}^{(2)})  f(\theta_{\bs{q}'}^{(1)}, \hat{\bs{q}}')^2 f(\theta_{\bs{q}'}^{(2)}, \hat{\bs{q}}')^2}{\left|\hat{\bs{q}}' \cdot \partial_{\theta} \bs{v}_F(\theta_{\bs{q}'}^{(1)})\right| \cdot \left|\hat{\bs{q}}' \cdot \partial_{\theta} \bs{v}_F(\theta_{\bs{q}'}^{(2)})\right|} \left[\bs{q}' \cdot \nabla_{\bs{k}} f(\theta_{\bs{q}'}^{(1)}, \hat{\bs{q}}) - \bs{q}' \cdot \nabla_{\bs{k}} f(\theta_{\bs{q}'}^{(2)}, \hat{\bs{q}})\right]^2 \,. 
    \end{aligned}
\end{equation}

The final expression admits a very transparent physical interpretation. The term inside the square bracket is the total change in velocity induced by a small-$|\bs{q}'|$ scattering event for a pair of CFs at $\theta^{(1)}_{\bs{q}'}$ and $\theta^{(2)}_{\bs{q}'}$. The structure of $\Pi^{xx}_{\rm CF, on-shell}$ is therefore reminiscent of a semiclassical transport calculation based on the quantum Boltzmann equation (QBE) for CFs~\cite{Kim1995_QBE}. 

The connection between the Kubo response and the QBE anticipates the crucial role that inversion symmetry plays in optical transport. According to the QBE, the most singular low energy scattering processes involve CFs on the CF Fermi surface, with incoming momenta $\bs{k}_1 = - \bs{k}_2$ and outgoing momenta $\bs{k}_1 + \bs{q}, \bs{k}_2 - \bs{q}$ satisfying
\begin{equation}\label{eq:FS_constraint}
    \xi(\bs{k}_1 + \bs{q}) = \xi(\bs{k}_1) = \xi(\bs{k}_2) = \xi(\bs{k}_2 - \bs{q}) = 0 \,,
\end{equation}
with $|\bs{q}| \ll k_F$, as illustrated in Fig.~\ref{fig:FS_scattering}(a). In a convex inversion-symmetric system, the change in the total CF velocity in this channel vanishes
\begin{equation}\label{eq:velocity_change}
    \Delta \bs{v} = \bs{v}_F(\bs{k}_1 + \bs{q}) + \bs{v}_F(\bs{k}_2 - \bs{q}) - \bs{v}_F(\bs{k}_1) - \bs{v}_F(\bs{k}_2) = 0 \,.
\end{equation}
Since the QBE conductivity is proportional to a weighted integral of $|\Delta \bs{v}|^2$, we immediately see that inversion symmetry forces a vanishing QBE contribution to the conductivity. Once inversion symmetry is broken, the two patches tangential to $\bs{q}$ and selected by Eq.~\ref{eq:FS_constraint} are generically no longer antipodal and the cancellation in Eq.~\eqref{eq:velocity_change} is lifted, as shown in Fig.~\ref{fig:FS_scattering}(b). Consequently, low-energy gauge-field-mediated small-$\bm{q}$ scattering can produce a singular contribution to $\Pi^{xx}_{\rm CF, on-shell}(\omega)$.

The second term $\Pi^{xx}_{\rm CF, virtual}$ in Eq.~\eqref{eq:sigma_CF_1_full} is beyond any Boltzmann calculation, as every term in $\Pi^{xx}_{\rm CF, virtual}$ contains at least one factor of $\Im \,\pi_F$ which encodes virtual off-shell processes. % depends on virtual off-shell processes encoded in the principal value integral in $\Im \,\pi_F$. 
Remarkably, we find that in the IR limit, the contribution from $\Pi^{xx}_{\rm CF, virtual}$ has the same scaling form as $\Pi^{xx}_{\rm CF, on-shell}$, but with a different prefactor. In other words, the on-shell collision and virtual polarization processes are equally important in the optical transport of lattice CFLs with broken inversion. 

In Appendix~\ref{app:detail_optical}, we evaluate the integrals over $\bs{q}', \Omega'$ explicitly and extract the final scaling form of both terms:
\begin{equation}
    \begin{aligned}
    \Pi_{\rm CF, on-shell}^{xx,R}(\omega)-\Pi_{\rm CF, on-shell}^{xx,R}(0) &= \begin{cases} 0 & \textrm{Inversion preserved, convex FS} \\ C (-i\omega)^{(4-z)/z} & \textrm{Inversion preserved, concave FS} \\ C_{\rm on-shell} (-i\omega)^{(4-z)/z} & \textrm{Inversion broken, convex or concave FS} \end{cases} \,,\\
    \Pi_{\rm CF, virtual}^{xx,R}(\omega)-\Pi_{\rm CF, virtual}^{xx,R}(0) &= \begin{cases} 0 & \textrm{Inversion preserved, convex or concave FS} \\ C_{\rm virtual} (-i\omega)^{(4-z)/z} & \textrm{Inversion broken, convex or concave FS} \end{cases} \,, 
    \end{aligned}
\end{equation}
where $C, C_{\rm on-shell}, C_{\rm virtual}$ are positive constants. For the weakly trigonally warped CF dispersion Eq.~\eqref{eq:trigonal_disp}, the Fermi surface is convex and the coefficients $C_{\rm on-shell}, C_{\rm virtual}$ can be explicitly evaluated at the leading order in $\lambda^2$ for the dynamical exponent $z = 3$. After absorbing the frequency-independent part of $\Pi_{\rm CF,on-shell}^{xx}+\Pi_{\rm CF,virtual}^{xx}$ into a renormalization of the Drude weight, we find
\begin{equation}\label{eq:sigmaCF_subleadingN}
    \sigma^{xx, (1)}_{\rm CF}(\omega) = \frac{1}{N} (C_{\rm on-shell} + C_{\rm virtual}) (-i\omega)^{-2/3} = \frac{0.963\lambda^2\mu^{2/3}}{N \left(1+\chi_0 v_0-16\pi^2 \chi_0\varkappa_2\right)^{4/3}} (-i\omega)^{-2/3} \,, \quad \chi_0 = \frac{m}{2\pi \sqrt{1 - \lambda^2}} \,,
\end{equation}
where the exact integral expression for the numerical coefficient is given in Appendix~\ref{app:detail_optical}, $m$ is the CF effective mass, $\lambda$ is the warping parameter, $\mu$ is the chemical potential, and $v_0$ is the microscopic short-range density-density interaction strength. We note that only the first term in the expression for $\Pi^{xx}_{\rm CF, virtual}$ in Eq.~\eqref{eq:zeroQ_BnB_decomposition} contributes at this order.

Putting together $\sigma^{xx, (0)}_{\rm CF}$ and $\sigma^{xx,(1)}_{\rm CF}$, extrapolating back to $N = 1$, and taking into account that $\chi_0\approx m/2\pi$ and $\varkappa_2\approx -1/(24\pi m)$ up to corrections of order $\lambda^2$, we find
\begin{equation}\label{eq:sigmaCF_maintext}
    \sigma^{xx}_{\rm CF}(\omega) = \frac{iD}{\omega} + \frac{0.963\lambda^2\mu^{2/3}}{\left(4/3+m v_0/2\pi\right)^{4/3}} (-i\omega)^{-2/3} \,, 
\end{equation}
where $D$ is the renormalized CF Drude weight. Since the Hall conductance of CFs is at most $\mathcal{O}(1)$ in the $\omega \rightarrow 0$ limit~\cite{Kivelson1997_HallCF,Wang2017_PH_DiracCFL,Kumar2019_HallCF}, we can extract the longitudinal CF resistivity by directly inverting the longitudinal CF conductivity. The final result is
\begin{equation}\label{eq:main_transport_optical_detail}
    \rho^{xx}(\omega) = \rho^{xx}_{\rm CF}(\omega) \approx \frac{-i \omega}{D} - \frac{0.963\lambda^2\mu^{2/3}}{D^2 \left(4/3+m v_0/2\pi\right)^{4/3}} (-i\omega)^{4/3} \quad \rightarrow \quad \Re \,\rho^{xx}(\omega) \approx \frac{1.926\lambda^2 \pi^2}{\mu^{4/3} \left(4/3+m v_0/2\pi\right)^{4/3}} \omega^{4/3} \,,
\end{equation}
where we also used the fact that $D=D_0 + \mathcal{O}(\lambda^2)=m/2\pi+\mathcal{O}(\lambda^2)$. For more general values of $z$, the integrals can no longer be done explicitly, but the frequency-dependence can be extracted by a scaling argument. The more general result is
\begin{equation}
    \Re \, \rho^{xx}(\omega) = C(z) \lambda^2 \omega^{4/z} \,, \quad C(z) > 0 \,, 
\end{equation}
as advertised in Eq.~\eqref{eq:main_transport_optical}. 

Before moving forward, we make a few conceptual remarks about this key result:
\begin{enumerate}
    \item The coefficient of the $\omega^{4/z}$ term is always proportional to $\lambda^2$ and vanishes when inversion symmetry is restored. The absence of a $\omega^{4/z}$ term in inversion-symmetric CFLs is consistent with existing calculations in Refs.~\cite{Shi2022_loopcurrent,Guo2022_YukawaSYK2d,Li2023_NFLoptical,Shi2023_controlled,Guo2024_fluc_criticalFS,Gindikin2024_optical_NFL,Gindikin2025_NFLstability}. 
    \item Although the optical resistivity scales as $\omega^{4/z}$, it is incorrect to infer that the DC resistivity scales as $T^{4/z}$, as such a scaling would eliminate the Drude peak in $\sigma^{xx}_{\rm CF}(\omega)$ and violate the non-perturbative constraints from continuum momentum conservation. As we will soon explain in Sec.~\ref{subsec:zeroQ_DC}, a nonzero DC resistivity in the clean limit can only arise from CF Umklapp scattering processes that relax the continuum momentum. The temperature scaling of the Umklapp-induced DC resistivity is not related to the $4/z$ exponent that we find in $\Re \,\rho^{xx}(\omega)$. 
    \item Finally, we note that the essential ingredients underlying the singular optical conductivity Eq.~\eqref{eq:main_transport_optical_detail} are the explicit breaking of inversion symmetry and the interaction between a Fermi surface with a gapless bosonic mode. These ingredients are universally present in a large class of non-Fermi liquids (NFL)s in 2+1 dimensions associated with the onset of a $\bs{q} = 0$ order parameter in an itinerant electronic system~\cite{Moriya1973,Hertz1976,Millis1993}. In these metallic symmetry-breaking phase transitions, there is no flux attachment, and the physical charge conductivity mimics the behavior of the CF conductivity in a lattice CFL. We therefore conclude that 
    \begin{equation}
        \sigma^{xx}_{\rm NFL}(\omega) = \frac{iD_{\rm NFL}}{\omega} + C_{\rm NFL}(z) \lambda^2 \, (-i\omega)^{\frac{4-2z}{z}} \,, \quad \textrm{with broken inversion symmetry.}
    \end{equation}
    In this different context, the constants $D_{\rm NFL}$ and $C_{\rm NFL}$ remain positive and are determined by the interaction between electrons and the gapless boson. 
\end{enumerate} 

\subsection{DC resistivity at nonzero temperature}\label{subsec:zeroQ_DC}

In this section, we move away from the optical limit $\omega \neq 0, T = 0$ and turn to DC transport in a lattice CFL at $T \neq 0$. As explained in the beginning of Sec.~\ref{sec:zeroQ_transport}, the dominant mechanism for current relaxation in the absence of disorder is Umklapp scattering. For any lattice potential, the lowest order Umklapp process involves two incoming CFs with momenta $\bs{k}_1, \bs{k}_2$ and two outgoing CFs with momenta $\bs{k}_3, \bs{k}_4$ such that 
\begin{equation}
    \Delta \bs{k} = \bs{k}_3 + \bs{k}_4 - (\bs{k}_1 + \bs{k}_2) = \bs{G} \neq 0 \,, 
\end{equation}
where $\bs{G}$ is a reciprocal lattice vector. When $\bs{G} \neq 0$, these scattering events violate the conservation of continuum momentum and lead to the relaxation of CF current. If we postulate a Umklapp-induced current relaxation rate $\Gamma_U(T)$, then the CF conductivity must scale as
\begin{equation}
    \sigma^{xx}_{\rm CF}(\omega, T) = \frac{D}{-i \omega + \Gamma_U(T)} \quad \rightarrow \quad \sigma^{xx}_{\rm CF}(\omega = 0, T) = \frac{D}{\Gamma_U(T)} \,, 
\end{equation}
where $D$ is the Drude weight in the absence of a lattice potential. Assuming that $\Gamma_U(T)$ vanishes as $T \rightarrow 0$, the temperature scaling of the physical DC resistivity is thus completely fixed by $\Gamma_U(T)$
\begin{equation}
    \rho^{xx}(T) \equiv \rho^{xx}_{\rm CF}(\omega=0, T) = \frac{\Gamma_U(T)}{D} \,. 
\end{equation}

If the CFs formed a Fermi liquid, conventional arguments would imply that $\Gamma_U(T) \sim T^2$~\cite{Lawrence1973_FLUmklapp,MacDonald1981_FLUmklapp}. However, we know that gauge fluctuations give singular corrections to many physical observables in a lattice CFL. It is natural to ask: do any of these singular effects modify the Umklapp scattering rate?

A key insight along this direction was provided by P.~Lee in a closely related doped spin liquid model where a spinon Fermi surface coexists with an ordinary Fermi liquid of doped holes~\cite{Lee2024_powerlaw_resistivity}. Here, we adapt his argument to the lattice CFL, which only has a single CF Fermi pocket. Working in the regime of weak trigonal warping, where the Fermi surface is convex, we know that for each $\theta$, there is a unique $P(\theta)$ such that $\bs{v}_F(\theta)$ is parallel to $\bs{v}_F(P(\theta))$. We define the generalized ``$2k_F$" vector at angle $\theta$ to be $\bs{K}_{2k_F}(\theta) = \bs{k}_F(\theta) - \bs{k}_F(P(\theta))$ and the collection of all $\bs{K}_{2k_F}(\theta)$ to be the ``$2k_F$" surface. The special kinematic structure of this ``$2k_F$" surface gives rise to a singular density susceptibility $\Pi_f$ of the lattice CFL~\cite{Altshuler1994_singular}
\begin{equation}\label{eq:2kF_Pi}
    \Pi_f(\bs{K}_{2k_F}(\theta) + \bs{q}, \omega) = \omega^{2/z - 2 \sigma(\theta)} \pi_{2k_F}\left(\theta, \hat{\bs{q}}, \frac{\omega}{q^{z/2}}\right) \,.
\end{equation}
Here, $\pi_{2k_F}\left(\theta, \hat{\bs{q}}, \frac{\omega}{q^{z/2}}\right)$ is a scaling function that depends in general on the geometry of the Fermi surface and $\sigma(\theta)$ is an exponent that captures the enhancement of the $2k_F$ vertex at angle $\theta$ due to gauge fluctuations. For a rotationally invariant dispersion, the dependence of $\sigma$ and $\pi_{2k_F}$ on $\theta$ disappears.

The singular ``$2k_F$" vertex function in lattice CFLs enhances the scattering rate of CFs induced by Umklapp processes. To model the dominant Umklapp interaction, we introduce a four-fermion term in the CFL Hamiltonian
\begin{equation}
    \delta H_{\rm int}[\bs{G}] = V_{\bs{G}} \, \sum_{\bs{k}_1, \bs{Q}, \bs{k}_2} f^{\dagger}_{\bs{k}_1 + \bs{Q}} \, f_{\bs{k}_1} \, f^{\dagger}_{\bs{k}_2 - \bs{Q} + \bs{G}} f_{\bs{k}_2} \,, 
\end{equation}
where $\bs{G}$ is a reciprocal lattice vector and $\bs{k}_1, \bs{k}_2, \bs{Q}$ are summed over the first Brillouin zone. Physically, $\delta H_{\rm int}[\bs{G}]$ describes processes in which a pair of CFs get scattered by $\bs{Q}$ and $\bs{G} - \bs{Q}$ respectively, such that the total momentum is conserved mod $\bs{G}$. When the CF Fermi surface is sufficiently large (which is the case at half-filling), all values of $\bs{k}_1, \bs{k}_2$ on the CF Fermi surface can participate in this kind of scattering for some appropriate choice of $\bs{G}$.

For a weakly interacting Fermi liquid, the transport scattering rate can be calculated perturbatively in $V_{\bs{G}}$. At leading nontrivial order, it is given by an integral over the product of density susceptibilities at $\bs{Q}$ and $\bs{G} - \bs{Q}$
\begin{equation}\label{eq:GammaU_FL}
    \Gamma_U(T) \propto \frac{|V_{\bs{G}}|^2}{T} \int_{\bs{Q}, \omega} \frac{\Im \Pi_f(\bs{Q},\omega)\,  \Im \Pi_f(\bs{G}-\bs{Q}, \omega)}{\sinh^2 (\omega/2T)} \,. 
\end{equation}
In a lattice CFL, we expect a variety of singular self-energy and vertex corrections to Eq.~\eqref{eq:GammaU_FL} due to low energy gauge fluctuations. However, it turns out that the CF density susceptibility $\Pi_f(\bs{q}, \omega)$ retains its Fermi liquid form for generic values of $\bs{q}$ except when $\bs{q}$ is close to the ``$2k_F$" surface~\cite{Altshuler1994_singular}. As a result, for generic $\bs{Q}$ in the integration range of Eq.~\eqref{eq:GammaU_FL}, both $\Pi_f(\bs{Q}, \omega)$ and $\Pi_f(\bs{G} - \bs{Q}, \omega)$ are Fermi-liquid-like and the induced scattering rate is proportional to $T^2$. Singular contributions to the scattering rate only arise when one of $\bs{Q}$ and $\bs{G} - \bs{Q}$ lies on the ``$2k_F$" surface, so that $\Pi_f(\bs{Q})$ or $\Pi_f(\bs{G}-\bs{Q})$ is singularly enhanced. At weak trigonal warping, we can approximate $\sigma(\theta)$ by its value in a CFL with rotational symmetry. Following Ref.~\cite{Lee2024_powerlaw_resistivity} and using the form of $\Pi_f$ in Eq.~\eqref{eq:2kF_Pi}, we find a CF transport scattering rate that scales as
\begin{equation}
    \Gamma_U(T) = \begin{cases}
        T^{\frac{z+4}{z} - 2 \sigma} & \sigma > \frac{2}{z} - \frac{1}{2} \\ 
        T^2 & \sigma \leq \frac{2}{z} - \frac{1}{2} 
    \end{cases} \,.
\end{equation}
Within the standard large $N$ expansion, $\sigma$ can be computed up to $\mathcal{O}(1/N^2)$~\cite{Altshuler1994_singular}
\begin{equation}
    2 \sigma = \frac{1}{N} + \frac{1}{\pi^2 N^2} \ln^3 N + \mathcal{O}\left(\frac{1}{N^3}\right) \,. 
\end{equation}
Extrapolating to $N = 1$, we find $2 \sigma = 1$, which implies a transport scattering rate 
\begin{equation}
    \Gamma_U(T) \sim T^{\frac{z+4}{z} - 2 \sigma} =  T^{\frac{4}{z}} \,. 
\end{equation}
This exponent happens to be the same as the exponent appearing in the zero-temperature optical resistivity $\Re \,\rho^{xx}(\omega) \sim \omega^{4/z}$. However, we caution that this equivalence should be critically examined. While the general scaling structure $\Gamma_U(T) \sim T^{\frac{z+4}{z} - 2 \sigma}$ is robust, the exponent $\sigma = 1/2$ extracted from the large $N$ expansion may not be reliable. In fact, even for $N = \infty$, the standard large $N$ expansion fails to be controlled in the low energy limit~\cite{Lee2009_RPAbad}. Thus, it would be important to obtain a more accurate estimate of $\sigma$ in the future by systematically analyzing gauge fluctuations beyond the RG-improved scattering rate calculation performed here.

% \ZS{It is kind of amusing that this exponent is different from $4/z$ but coincides with it for $z = 3$.}
% \ZS{Comment on caveats in this calculation. Provide more diagrammatic details. }

\section{Transport at finite wave vector}\label{sec:finiteQ_transport}

In this section, we switch gears and study the transport behavior of an inversion-asymmetric lattice CFL at nonzero momentum, which requires a proper treatment of charge screening induced by an inhomogeneous source field. Generally, the conductivity tensor with respect to the external field, $\sigma_{\rm ext}^{\mu\nu}$ (here $\mu,\nu=L,T$), is defined as
\begin{equation}\label{eq:J_Kubo_sigma}
    J^\mu(\bs{q},\omega) = \sigma_{\rm ext}^{\mu\nu}(\bs{q},\omega) E_{{\rm ext},\nu}(\bs{q}, \omega)  \,, \quad \sigma_{\rm ext}^{\mu\nu}(\bs{q}, \omega) = \frac{\Pi^{\mu\nu}(\bs{q}, \omega)}{-i\omega} \,,
\end{equation}
where $\Pi(\bs{q}, \omega)$ is directly related to the physical retarded current-current correlators (throughout this section, we work exclusively at real frequencies and therefore suppress the superscript $R$). However, the conductivity $\sigma$ measured in transport experiments is the response of the current $J^{\mu}$ to the total electric field, which should include the screening field $E_{{\rm screen}, \nu}$ generated by the current $J^{\mu}$
\begin{equation}\label{eq:J_phy_sigma}
    J^\mu(\bs{q},\omega) = \sigma^{\mu\nu}(\bs{q},\omega) E_{{\rm tot},\nu}(\bs{q}, \omega)  \,, \quad E_{{\rm tot},\nu}(\bs{q}, \omega) = E_{{\rm ext},\nu}(\bs{q}, \omega) + E_{{\rm screen},\nu}\,.
\end{equation}
Since the EM gauge field $A_{\mu}$ lives in 3D, the screening vector potential is related to the current by $A_{{\rm screen}, \mu} = \Pi_{{\rm Maxwell}, \mu\nu}^{-1} J^{\nu}$, where $\Pi_{\rm Maxwell}$ is the polarization function of the EM gauge field projected onto the 2D plane hosting the lattice CFL. In condensed matter experiments, the accessible $\omega$ and $q$ always satisfy $\omega \ll c q$ where $c$ is the speed of light. Thus, it is reasonable to neglect the transverse screening fields, which are suppressed relative to the longitudinal fields by factors of $\frac{\omega}{c q}$. Under this approximation, we have 
\begin{equation}
    A_{\rm screen, 0}(\bs{q}, \omega) = v(\bs{q}) J^0(\bs{q}, \omega) = \frac{q v(\bs{q})}{\omega} \, J^L(\bs{q}, \omega) \,, \quad A_{{\rm screen},T}(\bs{q}, \omega) \approx 0 \,. 
\end{equation}
Comparing the formula for $\sigma_{\rm ext}$ and $\sigma$, we conclude that
\begin{equation}\label{eq:screened_rho}
    \rho(\bs{q},\omega) \equiv \sigma^{-1}(\bs{q},\omega) = \rho_{\rm ext}(\bs{q},\omega) - \begin{pmatrix}
        \frac{i q^2 v(\bs{q})}{\omega} & 0 \\ 0 & 0 
    \end{pmatrix} \,. 
\end{equation}
We stress that our considerations above apply to finite-$q$ transport in any electronic system. 

With the general discussion of screening in mind, we specialize to the lattice CFL. Surprisingly, unlike in the zero-momentum setting, where interesting effects of inversion symmetry breaking only appear in higher loop Feynman diagrams for $\sigma^{ij}$, the finite-momentum conductivity already exhibits striking features attributed to broken inversion symmetry within the RPA. In what follows, we will fix a momentum direction $\bs{q} = q (\cos \phi_{\bs{q}}, \sin \phi_{\bs{q}})$ which defines a longitudinal (L) - transverse (T) basis. Within the RPA, $\rho_{\rm ext}$ is related to the CF response functions $\rho_{\rm CF}$ via
\begin{equation}
    \rho_{\rm ext}(\bs{q}, \omega) = \rho_{\rm CF}(\bs{q},\omega) + \begin{pmatrix}
        \frac{iq^2 v(\bs{q})}{\omega} & 4\pi \\ - 4\pi & 0 
    \end{pmatrix} \,, 
\end{equation}
in units where $\hbar=e^2=1$. Using \eqref{eq:screened_rho}, we further deduce the screened resistivity
\begin{equation}
    \rho(\bs{q},\omega) = \rho_{\rm CF}(\bs{q}, \omega) + 4\pi \begin{pmatrix}
        0 & 1 \\ -1 & 0
    \end{pmatrix} \,, \quad \rho_{\rm CF}(\bs{q}, \omega) = \sigma_{\rm CF}(\bs{q},\omega)^{-1} \,. 
\end{equation}
Using the RPA results from Sec.~\ref{subsec:review_RPA}, we can immediately infer the CF conductivity tensor $\sigma^{ij}_{\rm CF}(\bs{q}, \omega)$ in the L-T basis
\begin{equation}\label{eq:sigmaCF_finiteQ}
    \begin{aligned}
    \sigma^{LL}_{\rm CF}(\bs{q}, \omega) &= \frac{\Pi^{LL}_{\rm CF}(\bs{q}, \omega)}{-i\omega} = -\frac{i\omega}{q^2} \Pi^{00}_{\rm CF}(\bs{q}, \omega) \approx - \frac{i\omega}{q^2} \left[\chi_0 - \frac{m \omega}{2\pi q} f_0(\phi_{\bs{q}})\right] \,, \\
\sigma^{LT}_{\rm CF}(\bs{q}, \omega) = \sigma^{TL}_{\rm CF}(\bs{q}, \omega)&= \frac{\Pi^{LT}_{\rm CF}(\bs{q}, \omega)}{-i\omega} =  \frac{\Pi^{0T}_{\rm CF}(\bs{q}, \omega)}{q} \approx  -i\varkappa_1(\phi_{\bs q})q+\frac{i m \omega}{2\pi q^2} f_1(\phi_{\bs{q}}) \;,
    \\
    \sigma^{TT}_{\rm CF}(\bs{q}, \omega) &= \frac{\Pi^{TT}_{\rm CF}(\bs{q}, \omega)}{-i\omega} \approx -\frac{i\varkappa_2q^2}{\omega}+ \frac{i m}{2\pi q} f_2(\phi_{\bs{q}}) \,, 
    \end{aligned}
\end{equation}
where the angular functions $f_0, f_1, f_2$ are defined as in Eq.~\eqref{eq:f_i_definition}. 

Plugging the form of $\sigma_{\rm CF}$ into $\rho_{\rm ext}$ and performing matrix inversion, we immediately extract the Kubo conductivity $\sigma_{\rm ext}$ in units $\hbar=e^2=1$
\begin{equation}\label{eq:sigmaext_finiteQ}
\begin{aligned}
\sigma_{\rm ext}^{LL}(\bs q,\omega)
&=
\frac{i\omega q}{
2\pi\Upsilon(\bs q,\omega)
}
\left[
m\omega f_0(\phi_{\bs q})-2\pi q\chi_0
\right],
\\[4pt]
\sigma_{\rm ext}^{LT/TL}(\bs q,\omega)
&=
\frac{q^2}{\Upsilon(\bs q,\omega)}
\bigg\{
i\left[
\frac{m\omega}{2\pi}f_1(\phi_{\bs q})
-\varkappa_1(\phi_{\bs q})q^3
\right]
\mp 4\pi
\bigg[
\left(
\chi_0-\frac{m\omega}{2\pi q}f_0(\phi_{\bs q})
\right)
\left(
\frac{m\omega}{2\pi q}f_2(\phi_{\bs q})
-\varkappa_2q^2
\right)
\\
&\hspace{47mm}
+
\left(
\frac{m\omega}{2\pi q}f_1(\phi_{\bs q})
-\varkappa_1(\phi_{\bs q})q^2
\right)^2
\bigg]
\bigg\},
\\[4pt]
\sigma_{\rm ext}^{TT}(\bs q,\omega)
&=
\frac{iq^4}{
\omega\Upsilon(\bs q,\omega)
}
\bigg\{
\left[
1+v(\bs q)
\left(
\chi_0-\frac{m\omega}{2\pi q}f_0(\phi_{\bs q})
\right)
\right]
\left(
\frac{m\omega}{2\pi q}f_2(\phi_{\bs q})
-\varkappa_2q^2
\right)
+
v(\bs q)
\left(
\frac{m\omega}{2\pi q}f_1(\phi_{\bs q})
-\varkappa_1(\phi_{\bs q})q^2
\right)^2
\bigg\},
\end{aligned}
\end{equation}
where 
\begin{equation}\label{eq:Upsilon_finiteQ}
\begin{aligned}
\Upsilon(\bs q,\omega)
={}&
q^4
\left[
1+v(\bs q)
\left(
\chi_0-\frac{m\omega}{2\pi q}f_0(\phi_{\bs q})
\right)
\right]
+
16\pi^2q^2
\bigg[
\left(
\chi_0-\frac{m\omega}{2\pi q}f_0(\phi_{\bs q})
\right)
\left(
\frac{m\omega}{2\pi q}f_2(\phi_{\bs q})
-\varkappa_2q^2
\right)
\\
&\hspace{29mm}
+
\bigg(
\frac{m\omega}{2\pi q}f_1(\phi_{\bs q})
-\varkappa_1(\phi_{\bs q})q^2
\bigg)^2
\bigg].
\end{aligned}
\end{equation}
In the $\omega\rightarrow 0$ limit, we find the following asymptotic behavior to leading order in the small $\bs{q}$ expansion
\begin{equation}\label{eq:sigmaext_static}
\begin{aligned}
\sigma_{\rm ext}(\bs q,\omega\to0)
\approx{}&
\frac{1}{1-\frac{16\pi^2\chi_0\varkappa_2}{1+v(\bs{q})\chi_0} }\left[
-i\varkappa_2
\frac{q^2}{\omega}
+
\frac{im}{2\pi(1-\frac{16\pi^2\chi_0\varkappa_2}{1+v(\bs{q})\chi_0} )}
\frac{f_2(\phi_{\bs q})}{q}
\right]
\begin{pmatrix}
0&0\\
0&1
\end{pmatrix}
\\[4pt]
&+
\frac{4\pi\chi_0\varkappa_2}{1+v(\bs{q})\chi_0-16\pi^2\chi_0\varkappa_2}
\begin{pmatrix}
0&1\\
-1&0
\end{pmatrix}
-
\frac{iq\varkappa_1(\phi_{\bs q})}{1+v(\bs{q})\chi_0-16\pi^2\chi_0\varkappa_2}\,
\begin{pmatrix}
0&1\\
1&0
\end{pmatrix}\;.
\end{aligned}
\end{equation}
Using the relation between $\sigma$ and $\sigma_{\rm ext}$, the scaling of $\sigma$ in the same regime can be obtained by setting $v(\bs{q}) \rightarrow 0$ in Eq.~\eqref{eq:sigmaext_static}. Since different components of the conductivity matrix encode different physical effects, we will discuss them separately in the next few sections.

\subsection{Hall channel}\label{subsec:finiteQ_Hall}

The most direct manifestation of inversion symmetry breaking occurs in the mixed longitudinal-transverse channel. It is important, however, to distinguish the raw Kubo conductivity $\sigma_{\rm ext}$ from the transport conductivity $\sigma$. The latter includes screening effects, whereas the former does not.

Using the full physical response obtained from the Ioffe--Larkin composition rule, the mixed components of the external Kubo conductivity in the static limit are
\begin{equation}\label{eq:sigmaext_mixed_static}
\sigma_{\rm ext}^{LT/TL}(\bs q,\omega\to0)
=
\frac{
\pm 4\pi\chi_0\varkappa_2
-iq\varkappa_1(\phi_{\bs q})
}{
1+\chi_0v(\bs q)-16\pi^2\chi_0\varkappa_2
}+\mathcal{O}(q^2)\;.
\end{equation}
Note that the $\varkappa_2$ contribution is already present in the inversion-symmetric theory, whereas the symmetric term proportional to $\varkappa_1$ requires broken inversion symmetry.

After accounting for screening, the static transport conductivity $\sigma^{LT/TL}$ takes the form
\begin{equation}
    \sigma^{LT/TL}(\bs{q}, \omega \rightarrow 0) = \frac{\pm 4 \pi \chi_0 \kappa_2 - i q \kappa_1(\phi_{\bs{q}})}{1 - 16 \pi^2 \chi_0 \kappa_2} + \mathcal{O}(q^2) \,. 
\end{equation}
Relative to the optical conductivity at $\bs{q} = 0$, this correction occurs already at the order $\mathcal{O}(\lambda)$, and thus it is much stronger in magnitude at weak trigonal warping. At this order, the $q$-linear mixed correction is purely
imaginary and satisfies
$\sigma^{LT}=-(\sigma^{TL})^*$, which is non-dissipative (does not contribute to Joule heating). The real symmetric term appearing at
order $\lambda^2$ gives the leading dissipative contribution in the
mixed channel. 

It is also useful to consider the kinematic regime
\begin{equation}
v_Fq(q/k_F)^{z-1}\ll\omega\ll v_Fq,
\end{equation}
which is relevant for surface acoustic wave probes because they involve frequencies of the order $\omega\approx v_sq \ll v_F q$, where $v_s$ is the speed of sound. In this regime, the static-gradient terms proportional to
$\varkappa_1$ and $\varkappa_2$ are parametrically subleading, and the
effects of longitudinal screening are suppressed to the order displayed. Consequently, $\sigma^{LT}(\bs q,\omega)\approx \sigma_{\rm ext}^{LT}(\bs q,\omega)$ and $\sigma^{TL}(\bs q,\omega)\approx \sigma_{\rm ext}^{TL}(\bs q,\omega)$ in this regime
\begin{equation}\label{eq:sigma_mixed_kinematic}
\begin{aligned}
\sigma^{LT}(\bs q,\omega)
&\approx
\sigma_{\rm ext}^{LT}(\bs q,\omega)
\approx
-\frac{1}{4\pi}
+\frac{iqf_1(\phi_{\bs q})}
{16\pi^2\chi_0f_2(\phi_{\bs q})}
+\cdots,
\\
\sigma^{TL}(\bs q,\omega)
&\approx
\sigma_{\rm ext}^{TL}(\bs q,\omega)
\approx
\frac{1}{4\pi}
+\frac{iqf_1(\phi_{\bs q})}
{16\pi^2\chi_0f_2(\phi_{\bs q})}
+\cdots.
\end{aligned}
\end{equation}
Here the omitted terms are suppressed by higher powers of $q$,
$\omega/(v_Fq)$, or
$v_Fq(q/k_F)^{z-1}/\omega$. 

\subsection{Longitudinal channel}\label{subsec:finiteQ_longitudinal}

We next consider the longitudinal channel. In the static limit, the external Kubo conductivity vanishes linearly with $\omega$
\begin{equation}
\sigma_{\rm ext}^{LL}(\bs q,\omega\rightarrow 0)
\approx 
-\frac{i\omega\chi_0}{q^2[1+\chi_0v(\bs q)-16\pi^2\chi_0\varkappa_2
]}
\,, \quad \sigma^{LL}(\bs{q}, \omega \rightarrow 0) \approx - \frac{i\omega \chi_0}{q^2 \left[1 - 16 \pi^2 (\chi_0 \varkappa_2 - \varkappa_1 q^2)\right]} \,. 
\end{equation}
The screened conductivity $\sigma^{LL}$ can obtain be obtained by replacing $v(\bs{q}) \rightarrow 0$ in $\sigma^{LL}_{\rm ext}$.

In the kinematic regime
$v_Fq(q/k_F)^{z-1}\ll\omega\ll v_Fq$, the screening correction is
parametrically suppressed, and $\sigma^{LL}(\bs q,\omega), \sigma^{LL}_{\rm ext}(\bs{q},\omega)$ share the same leading-order expansion 
\begin{equation}
    \begin{aligned}
    \sigma^{LL}(\bs q,\omega) \approx \sigma^{LL}_{\rm ext}(\bs q,\omega) 
    &=
    -\frac{iq}{8\pi m f_2(\phi_{\bs q})}
    +\mathcal O(q^3)
    \\
    &\approx
    \frac{q}{8\pi\sqrt{2m\mu}}
    \left[
    1-4i\lambda\cos3\phi_{\bs q}
    +\lambda^2
    \left(
    \frac{41}{8}\cos6\phi_{\bs q}
    -\frac{35}{4}
    \right)
    \right]
    +\mathcal O(q\lambda^3,q^3).
\end{aligned}
\end{equation} 
\color{black}
Clearly, the leading contribution to $\sigma^{LL}$ is already non-vanishing even with inversion symmetry. This is the famous $q$-linear conductivity that was invoked by Ref.~\cite{Halperin1993_HLRtheory} to explain the surface acoustic wave anomaly of the half-filled Landau level. The main role of inversion symmetry breaking is to generate corrections to $f_2(\phi_{\bs{q}})$ that appear at the same order in $q$. This large correction can be directly probed in the modulation of surface acoustic waves near a lattice CFL, as we explain later in Sec.~\ref{sec:discussion}. 

%Another interesting feature of these formulae is that the leading frequency-dependent correction proportional to $\omega$ is suppressed by $\lambda^2$, since $f_1^2 = \mathcal{O}(\lambda^2)$ while $f_2 = \mathcal{O}(\lambda^0)$. As a result, the strongest effects of inversion symmetry breaking in $\sigma^{LL}$ are largely frequency-independent in the entire dynamical range $0 < \omega \ll v_F q$. 

\subsection{Transverse channel}\label{subsec:finiteQ_transverse}

Finally, the transverse channel displays the most pronounced
difference between the raw Kubo and transport responses. At fixed
nonzero $q$, the raw external conductivity contains an equilibrium
magnetization contribution proportional to $q^2/\omega$:
\begin{equation}\label{eq:sigmaext_TT_static}
\begin{aligned}
\sigma_{\rm ext}^{TT}(\bs q,\omega\to0)
={}&
-\frac{i q^2}{\omega}
\frac{
(1+\chi_0v(\bs q))\varkappa_2
}{
(1+\chi_0v(\bs q)-16\pi^2\chi_0\varkappa_2
)}
+
\frac{imf_2(\phi_{\bs q})}{2\pi q}
\left[
\frac{
1+\chi_0v(\bs q)
}{
1+\chi_0v(\bs q)-16\pi^2\chi_0\varkappa_2
}
\right]^2
+\mathcal O(q^4/\omega, q,\omega).
\end{aligned}
\end{equation}
Thus $\sigma_{\rm ext}^{TT}$ does not possess a finite static limit at fixed $q$. The apparent $1/\omega$ divergence is the equilibrium reactive magnetization response to a static transverse vector potential. $\sigma^{TT}(\bs{q}, \omega \rightarrow 0)$ can again be obtained by replacing $v(\bs{q}) \rightarrow 0$. At linear order in $\lambda$, the
inversion-breaking correction is imaginary and reactive, whereas
terms of order $\lambda^2$ also modify the real, dissipative part.

In the kinematic regime
$v_Fq(q/k_F)^{z-1}\ll\omega\ll v_Fq$, the external Kubo conductivity takes the form
\begin{equation}\label{eq:sigmaext_TT_kinematic}
\sigma_{\rm ext}^{TT}(\bs q,\omega)
\approx
\frac{iq^2}{\omega}
\frac{1+\chi_0v(\bs q)}
{16\pi^2\chi_0}+
\frac{imq}{32\pi^3\chi_0^2}
\frac{
f_0(\phi_{\bs q})f_2(\phi_{\bs q})
-f_1^2(\phi_{\bs q})
}{
f_2(\phi_{\bs q})
}
\;.
\end{equation}
$\sigma^{TT}(\bs{q}, \omega)$ can again be obtained by setting $v(\bs{q}) \rightarrow 0$. Remarkably, corrections produced by the static $\varkappa_1$ and $\varkappa_2$
terms are suppressed in this regime by
$v_Fq(q/k_F)^{z-1}/\omega$. %Thus screening and equilibrium-current effects remain important in the transverse channel even when $v_Fq(q/k_F)^{z-1}\ll\omega\ll v_Fq$. Compared with the transport conductivity, the external response contains an additional $q^2/\omega$ contribution determined by the compressibility and interaction potential. 
Apart from the different coefficients multiplying $iq^2/\omega$, the two responses share the same leading $q$-linear contribution. The leading inversion-breaking correction is imaginary at order $\lambda$ and hence non-dissipative, while order-$\lambda^2$ terms also modify the dissipative part.

\section{Discussion}\label{sec:discussion}

\subsection{Experimental probes of zero wave vector transport}

The most experimentally accessible prediction of our paper is a non-analytic scaling of optical and DC resistivities in the low-temperature and low-frequency regime. For inversion-asymmetric CFLs, we have shown that the optical resistivity scales as a fractional power of the drive frequency $\Re \, \rho(\omega) \sim \omega^{4/z}$, with a prefactor that is proportional to the strength of inversion symmetry breaking. Although this calculation is done at $T = 0$, the same scaling form should hold whenever $T \ll \omega \ll E_F$, where $E_F$ is the Fermi energy of the CFs. In experimental platforms where signatures of lattice CFLs have been observed, measurements of the neighboring FCI charge gaps give a crude estimate of the CF Fermi energy $E_F \sim 30K$~\cite{Park2025_FCIgap}. On the other hand, the lowest electron temperature that can be achieved in dilution refrigerators is $T \sim 50 mK$. Given that the lowest temperature and the Fermi energy are separated by roughly three orders of magnitudes, it is in principle possible to extract a scaling form of $\Re\, \rho(\omega)$ over one decade of $\omega$ (roughly between 10 and 100 GHz) which satisfies the hierarchy of scales $T \ll \omega \ll E_F$. It would be interesting to extend existing microwave and THz spectroscopy techniques to this intermediate frequency regime (see e.g. Refs.~\cite{Yoshioka2020_coherentTHz,Potts2023_onchipTHz,Seo2024_onchipTHz,Chen2025_onchipTHz,Vukelich2026_broadbandTHz}). 

Independent of inversion symmetry, we have also argued that Umklapp effects can lead to a power-law DC resistivity $\Re \, \rho(T) \sim T^{\frac{z+4}{z} - 2 \sigma}$ in any lattice CFL, where $\sigma$ is the scaling exponent associated with the $2k_F$ vertex function. If $\frac{z+4}{z} - 2 \sigma < 2$, this contribution dominates over the conventional $T^2$ Fermi liquid resistivity and should persist down to $T = 0$. For an inversion-asymmetric CFL, one can combine a measurement of $\Re \, \rho(\omega)$ with a measurement of $\Re \, \rho(T)$ to extract both the dynamical critical exponent $z$ and the $2k_F$ exponent $\sigma$. Since our theoretical calculation of $\Re \, \rho(\omega)$ and $\Re \, \rho(T)$ involve higher-loop singular gauge fluctuations in the CFL state, this pair of experiments would serve as the first definitive test of the HLR framework for composite Fermi liquids beyond the random phase approximation.

\subsection{Experimental probes of finite wave vector transport}\label{sec:discussion_Q}

Signatures of inversion symmetry breaking in finite-$\bm{q}$ transport are more subtle. The most promising experiments that we can imagine is a direct measurement of surface acoustic wave modulation. Assuming that the incoming surface acoustic wave has a velocity $v_s = \omega/q$ which is much less than the Fermi velocity, the velocity shift $\Delta v_s$ and attenuation rate $\kappa$ of the surface acoustic wave are directly related to the longitudinal conductivity $\sigma^{LL}(\bs{q}, \omega = v_s q)$ via the compact formula~\cite{Simon1996_SAWderivation}
\begin{equation}
    \frac{\Delta v_s}{v_s} - \frac{i \kappa}{q} = \frac{\alpha(qd)^2}{2} \frac{1}{1 + i \sigma^{LL}(\bs{q},\omega)/\sigma_m(\bs{q})} \,, \quad \sigma_m(\bs{q}) = \frac{\epsilon_{\rm eff}(\bs{q}) v_s}{2\pi} \,,
\end{equation}
where $\alpha(qd)$ is the piezoelectric coupling constant that depends on $qd$, $d$ is the distance from
the surface, and $\epsilon_{\rm eff}(\bs{q})$ is an effective dielectric function related to the interaction potential as $V(\bs{q}) = \frac{2\pi}{\epsilon_{\rm eff}(\bs{q}) q}$. As shown in Sec.~\ref{sec:finiteQ_transport}, inversion symmetry breaking indeed leads to a strong anisotropic correction to $\sigma^{LL}(\bs{q}, \omega)$ in the kinematic regime $v_Fq(q/k_F)^{z-1}\ll \omega \ll v_F q$
\begin{equation}
    \sigma^{LL}(\bs{q}, \omega = v_s q) \approx -i q \left[\Gamma_I + i \Gamma_R\right] \,, \quad \Gamma_I = \frac{\lambda \cos 3 \phi_{\bs{q}}}{2\pi\sqrt{2\mu m} }  \,, \quad \Gamma_R \approx \frac{1}{8\pi\sqrt{2m\mu}}
\left[
1
+\lambda^2
\left(
\frac{41}{8}\cos6\phi_{\bs q}
-\frac{35}{4}
\right)
\right] \,. 
\end{equation}
Plugging $\sigma^{LL}$ into the surface acoustic wave propagation formula gives 
\begin{equation}
    \frac{\Delta v_s}{v_s} - \frac{i \kappa}{q} = \frac{\alpha(qd)^2}{2} \frac{\sigma_m}{\sigma_m + q \Gamma_I + i q \Gamma_R} = \frac{\alpha(qd)^2}{2} \left[\frac{\sigma_m (\sigma_m + q \Gamma_I)}{(\sigma_m + q \Gamma_I)^2 + \Gamma_R^2 q^2} - i \frac{q \sigma_m \Gamma_R}{(\sigma_m + q \Gamma_I)^2 + \Gamma_R^2 q^2}\right] \,. 
\end{equation}
Clearly, $\alpha(qd), \sigma_m, v_s$ are determined by the experimental setup and independent of the intrinsic properties of the lattice CFL. Therefore, for every $\bs{q}$, we have two real unknowns $\Gamma_I, \Gamma_R$ and two equations. By scanning over $\bs{q}$, we can completely map out the functions $\Gamma_I(\phi_{\bs{q}}), \Gamma_R(\phi_{\bs{q}})$. 

When the warping parameter $\lambda$ vanishes, inversion symmetry is restored and the formula above simplifies to
\begin{equation}
    \frac{\Delta v_s}{v_s} - \frac{i \kappa}{q}\bigg|_{\textrm{inversion-symmetric}} = \frac{\alpha(qd)^2}{2}\left[\frac{\sigma_m^2}{\sigma_m^2 + \Gamma_R^2 q^2} - i \frac{q \sigma_m \Gamma_R}{\sigma_m^2 + \Gamma_R^2 q^2}\right] \,,
\end{equation}
in agreement with the original HLR theory~\cite{Halperin1993_HLRtheory}. At weak inversion breaking and small $q/k_F$, we can expand the velocity shift and the attenuation rate to linear order in $\Gamma_I$ and to leading order in $q/k_F$. The results simplify dramatically
\begin{equation}
    \begin{aligned}
    \frac{\Delta v_s(\bs{q})}{v_s}\bigg|_{\textrm{inversion-broken}} &\approx \frac{\alpha(qd)^2}{2} \left[\frac{\sigma_m^2}{\sigma_m^2 + \Gamma_R^2 q^2} - \frac{q \sigma_m (\sigma_m^2 - \Gamma_R^2q^2 )}{(\sigma_m^2 + \Gamma_R^2 q^2)^2} \Gamma_I \right] \approx \frac{\alpha(qd)^2}{2} \left[1 - \frac{q}{\sigma_m} \Gamma_I\right] \,,  \\
    \kappa(\bs{q})\bigg|_{\textrm{inversion-broken}} &\approx \frac{\alpha(qd)^2}{2} \left[\frac{q^2 \sigma_m \Gamma_R}{\sigma_m^2 + \Gamma_R^2 q^2} - \frac{2 q^3 \sigma_m^2 \Gamma_R}{(\sigma_m^2 + \Gamma_R^2 q^2)^2} \Gamma_I \right] \approx \frac{\alpha(qd)^2 q^2 \Gamma_R}{2 \sigma_m} \left[1 - \frac{2q}{\sigma_m} \Gamma_I\right] \,.
    \end{aligned}
\end{equation}
Since $\Gamma_R$ is even under inversion and $\Gamma_I$ is odd under inversion, a simple way to extract the inversion-breaking component is to antisymmetrize $\Delta v_s(\bs{q})/v_s$ and $\kappa(\bs{q})$ with respect to $q$
\begin{equation}
    \frac{\Delta v_s(\bs{q}) - \Delta v_s(-\bs{q})}{v_s} \approx - \frac{\alpha(qd)^2 q}{\sigma_m} \Gamma_I(\bs{q}) \,, \quad \kappa(\bs{q}) - \kappa(-\bs{q}) \approx - \frac{2 \alpha(qd)^2 q^3 \Gamma_R}{\sigma_m^2} \Gamma_I(\bs{q}) \,. 
\end{equation}
These anisotropic shifts to $\Delta v_s(\bs{q})$ and $\kappa(\bs{q})$ are promising targets for near-term experiments in moire realizations of lattice CFLs. 

Other components of the finite-$Q$ conductivity tensor $\sigma^{LT}(\bs{q},\omega)$ and $\sigma^{TT}(\bs{q}, \omega)$ are more challenging to access. Measurements of both quantities require a momentum-resolved detection of transverse current patterns in the lattice CFL. While recent advances in NV center magnetometry allow direct measurements of the dissipative transverse current response (e.g. $\Re \,\sigma^{TT}(\bs{q}, \omega)$~\cite{Rondin2014_NVreview,Agarwal2017_magnetic_noise}), leading-order signatures of inversion-breaking derived in Sec.~\ref{sec:finiteQ_transport} occur only in $\Im \,\sigma^{LT}(\bs{q}, \omega)$ and $\Im\, \sigma^{TT}(\bs{q}, \omega)$, which are non-dissipative. Dissipative anisotropic contributions are parametrically weaker and contribute at the order $\lambda^2$, which might be more difficult to detect. Even if phase-sensitive probes can be engineered to detect non-dissipative transverse response, existing measurement geometries can select a narrow window of $|\bs{q}|$ but usually average over all orientations $\hat{\bs{q}}$. Such a directional average washes out inversion-odd terms proportional to $\cos 3 \phi_{\bs{q}}$ and $\sin 3 \phi_{\bs{q}}$ in $\Im\, \sigma^{LT}(\bs{q}, \omega)$ and $\Im\, \sigma^{TT}(\bs{q},\omega)$. Given these challenges, an important future target is to search for more sophisticated experimental designs sensitive to both phase and orientation information contained in the transverse current. 

\subsection{Further theoretical developments}

In addition to direct experimental predictions, our work opens up many new theoretical directions in the study of lattice CFLs. In the immediate future, there are several natural extensions of our results that are worth pursuing. In DC transport, we showed through a scaling argument that Umklapp scattering could induce a power-law DC resistivity $T^{\alpha}$ in the lattice CFL. It would be satisfying to directly verify this scaling argument by an explicit calculation of all the relevant transport diagrams within some controlled diagrammatic expansion scheme. Towards that end, it may be fruitful to combine the expansion in the number $N$ of CF flavors with an expansion in the deviation of the dynamical exponent $z$ from $2$, holding $(z-2) N$ fixed. Using the rigorous classification of transport diagrams in Ref.~\cite{Shi2023_controlled}, it is conceivable that the DC resistivity can be calculated to leading two orders in the $1/N$ expansion. In optical transport, it would be interesting to search for mechanisms beyond inversion symmetry breaking that would generate the $\Re \, \rho(\omega) \sim |\omega|^{4/z}$ scaling law. A promising idea is to consider CF Fermi surfaces with nontrivial topology, including annular Fermi surfaces, or other Fermi surfaces that bound disconnected regions of the Brillouin zone. This kind of nontrivial Fermi surface topology has already been observed in metallic phases of moire materials such as rhombohedral multilayer graphene~\cite{Zhou2021_flavormetal}. It is then natural to ask whether lattice CFLs within the same material platform could realize CF Fermi surfaces with nontrivial topology. From the perspective of optical transport, singular gauge-field-mediated scattering processes involving CFs on different sheets of the Fermi surface could relax current efficiently while conserving energy and momentum~\cite{Gindikin2024_twovalley}. Therefore, we expect the optical conductivity to show a similar scaling form $\Re \, \rho(\omega) \sim |\omega|^{4/z}$, which is a concrete target for a direct diagrammatic calculation.

As far as finite wave vector transport is concerned, an important extension would be to include a weak nonzero effective magnetic field experienced by the CFs, generated by a perpendicular magnetic field or doping. This would introduce a CF cyclotron scale and allow one to study associated commensurability resonances \cite{Mirlin1997_SAW0}. Static periodic modulations (e.g. produced by a patterned gate or grating) have provided powerful probes of the CF Fermi surface in a conventional Landau-level setting \cite{Mirlin1998_SAW1}. Similar measurements could directly probe the size, shape, and anisotropy of the CF Fermi surface in Chern bands.

Going beyond linear-response electrical transport, it would be interesting to search for qualitatively new effects of inversion symmetry breaking in other physical observables. In the past few years, an emerging body of work has uncovered exciting nonreciprocal effects in Coulomb drag, nonlinear hydrodynamic transport, collective modes of superconductors etc. in systems with inversion-asymmetric electron dispersion~\cite{Kirkinis2025_nonreciprocal_hydro,Zverevich2025_nonreciprocal_Coulombdrag,Kirkinis2025_photogalvanic_nonreciprocal,Levitan2025_trigonal_SC}. We expect many of these effects to have more subtle analogues in the CF context, where an inversion-asymmetric CF Fermi surface remains strongly coupled to a dynamical gauge field.

Another important direction is to understand the role of composite fermion Berry curvature, which we have neglected so far. Berry curvature can modify both the current operator and the coupling of CFs to the emergent gauge field through anomalous-velocity and orbital-moment terms. It would be particularly interesting to determine whether these additional effects can generate an incoherent optical conductivity even for a convex inversion-symmetric CF Fermi surface, as well as how they modify the finite-$Q$ response.

We also note a more speculative possibility suggested by our calculation. The gauge-flux stiffness, defined as the coefficient of $q^2$ in the inverse gauge-field propagator, $\left[\mathcal D^{TT}(\bs q,\omega)\right]^{-1} \sim \left[(1+\chi_0v_0)/(16\pi^2\chi_0)-\varkappa_2\right]q^2 +c_4q^4+\gamma|\omega|/q$, can change sign at sufficiently strong trigonal warping if the paramagnetic CF contribution exceeds the positive stiffness generated by the Chern--Simons and interaction terms. Physically, this possibility arises because sufficiently strong trigonal warping renders the orbital response of the CF Fermi sea paramagnetic. The underlying mechanism is more general than inversion breaking and could also occur for sufficiently nonconvex inversion-symmetric dispersions~\cite{Vignale1991_paramagnet_FL}. If the total $q^2$ coefficient vanishes while $c_4>0$, the system reaches a gauge-flux Lifshitz point with the unusual dynamical exponent $z=5$. Beyond this point, the uniform CFL becomes unstable toward a state with spontaneous non-uniform emergent magnetic flux. Because the Chern--Simons constraint ties this flux to the physical charge density, while its spatial average is fixed by the filling, the resulting state is likely a spatially modulated flux-density state (different mechanisms for charge/flux-ordering instabilities of spinon Fermi surfaces/CFLs have been considered in Refs.~\cite{Wang1990_fluxDW_tJ,Morse1990_fluxDW,Hlubina1992_fluxDW_spinonFS,Jian2020_2kFDW_CFL}). The $C_3$ symmetry makes both a single-$Q$ stripe and a three-$Q$ triangular flux-density pattern natural candidates. Determining which state is selected (and whether the transition remains continuous) requires an analysis of the higher-gradient and nonlinear terms.

Finally, the breaking of inversion symmetry opens up new possibilities for paired phases of CFs and phase transitions between such paired phases and the CFL phase. In a previous work~\cite{Shi2026_CBFL}, we have demonstrated that the onset of a generic pair condensate in an inversion-asymmetric CFL does not fully gap the CF Fermi surface and instead leads to gapless Bogoliubov Fermi pockets of the CFs. The resulting electronic phase is termed the Composite Bogoliubov Fermi liquid and exhibits a number of striking properties that deviate from both the lattice CFL and the fully gapped topological phases at the same filling. However, the dynamical mechanism that would lead to pairing of CFs was not explored in Ref.~\cite{Shi2026_CBFL}. Given the more complete theory of lattice CFLs with broken inversion symmetry developed in this work, it would be interesting to revisit the CF pairing problem in detail and clarify the interplay between gauge-field-mediated CF interactions and inversion-asymmetric Fermi surface fluctuations.

\begin{acknowledgments}
    We would like to thank Andrey Chubukov, Bert Halperin, Eslam Khalaf, Alex Levchenko, Ady Stern, Dmitrii Maslov for helpful discussions. PAN is supported in part by a Harvard Quantum Initiative postdoctoral fellowship at Harvard University. ZDS is supported by a Leinweber Institute for Theoretical Physics postdoctoral fellowship at Stanford University and in part by the Gordon and Betty Moore Foundation EPiQS initiative, Grant GBMF8686.01. 
\end{acknowledgments}

\appendix

\section{RPA calculation of composite fermion response functions}\label{app:RPA_integrals}

In this appendix, we perform a calculation of the free CF response function $\Pi_{\rm CF}$ defined in Eq.~\eqref{eq:CFL_response_function}, filling in some of the steps omitted in the main text. 

We first analyze the mixed transverse current-density correlator, which is most strongly affected by the presence of inversion symmetry breaking. We find
\begin{equation}
    \langle J_T(\bm{q},\Omega)\rho(-\bm{q},-\Omega)\rangle =-\int\frac{d\omega}{2\pi}\int\frac{d^2\bm{k}}{(2\pi)^2} v_T(\bm{q}) G(\bm{k}+\bm{q}/2,\omega+\Omega/2)G(\bm{k}-\bm{q}/2,\omega-\Omega/2)\;,
\end{equation}
where we defined the transverse component of the Fermi velocity
\begin{equation}
    v_T(\bm{q})= \bm{v}\cdot \hat{\bm{q}}_T,\quad  v_L(\bm{q})= \bm{v}\cdot \hat{\bm{q}},\quad \bm{v}= \bm{\nabla}_{\bm{k}}\xi_{\bm{k}},
\end{equation}
and the (mean-field) Euclidean CF Green's function is $G(\bm{k},\omega)=1/(i\omega-\xi_{\bm{k}})$. After performing the frequency integral and approximating $\xi_{\bm{k}+\bm{q}}\approx \xi_{\bm{k}} + \bm{v}\cdot \bm{q}+  (\bm{q}\cdot \nabla_{\bm{k}})^2\xi_{\bm{k}}/2+(\bm{q}\cdot \nabla_{\bm{k}})^3\xi_{\bm{k}}/6$ to the cubic order in $q$, we find 
\begin{equation}
\begin{aligned}
   \langle J_T(\bm{q},\Omega)\rho(-\bm{q},-\Omega)\rangle&\approx\int\frac{d^2\bm{k}}{(2\pi)^2} \frac{v_T(\bm{q}) \sum_{\eta=\pm}\eta\theta[-\xi_{\bm k} -\eta \bm{v}\cdot \bm{q}/2 - (\bm{q}\cdot \nabla_{\bm{k}})^2\xi_{\bm{k}}/8 -\eta (\bm{q}\cdot \nabla_{\bm{k}})^3\xi_{\bm{k}}/48]}{i\Omega-\bm{v}\cdot \bm{q} -(\bm{q}\cdot \nabla_{\bm{k}})^3\xi_{\bm{k}}/24} \\
   &\approx \int\frac{d^2\bm{k} }{(2\pi)^2}v_T(\bm{q}) \delta(\xi_{\bm{k}})+\frac{q^2}{8}\int\frac{d^2\bm{k} }{(2\pi)^2}v_T(\bm{q})\left[\delta'(\xi_{\bm{k}})(\hat{\bm{q}}\cdot \nabla_{\bm{k}})^2\xi_{\bm{k}}+\frac{v_L^2(\bm{q})}{3} \delta''(\xi_{\bm{k}})\right] \\
   &- \frac{i\Omega}{q} \int\frac{d^2\bm{k} }{(2\pi)^2} \frac{ v_T(\bm{q}) \delta(\xi_{\bm{k}}) }{i\Omega/q-\bm{v}\cdot \hat{\bm{q}} } +\mathcal{O}(\Omega q, \;q^4)\;.
    \end{aligned}
\end{equation}
Here we used the following expansion up to the cubic order in $q$
\begin{equation}
\begin{aligned}
    \theta[-\xi_{\bm k} -\bm{v}\cdot \bm{q}/2 &- (\bm{q}\cdot \nabla_{\bm{k}})^2\xi_{\bm{k}}/8- (\bm{q}\cdot \nabla_{\bm{k}})^3\xi_{\bm{k}}/48] \approx \theta(-\xi_{\bm k} ) -\frac{(\bm{v}\cdot \bm{q})}{2}\delta(\xi_{\bm{k}})- \frac{(\bm{q}\cdot \nabla_{\bm{k}})^2\xi_{\bm{k}}}{8} \delta(\xi_{\bm{k}}) \\
    &-\frac{(\bm{v}\cdot \bm{q})^2}{8}\delta'(\xi_{\bm{k}}) -\frac{(\bm{v}\cdot \bm{q})^3}{48} \delta''(\xi_{\bm{k}})-\frac{(\bm{v}\cdot \bm{q})(\bm{q}\cdot \nabla_{\bm{k}})^2\xi_{\bm{k}}}{16}\delta'(\xi_{\bm{k}})- \frac{(\bm{q}\cdot \nabla_{\bm{k}})^3\xi_{\bm{k}}}{48} \delta(\xi_{\bm{k}}).
    \end{aligned}
\end{equation}
We note that the terms proportional to $(\bm{q}\cdot \nabla_{\bm{k}})^3\xi_{\bm{k}}$ cancel out in the correlator.

In the presence of the inversion symmetry, we can always decompose the integration over $\bm{k}$ in terms of its transverse $k_T$ and longitudinal $k_L$ components. Then we can see that the integrand is odd under the change of variables $k_T\rightarrow -k_T$, so the correlator must be identically zero. As a trivial example, the correlator is zero for a circular Fermi surface. 

We now specialize to the weakly trigonally warped dispersion  
\begin{equation}\label{eq:xi_C3}
     \xi_{\bm{k}} = \frac{k^2}{2m}\left(1+\lambda \cos 3\theta_{\bm{k}}\right)-\mu\;,
\end{equation}
as defined in Eq.~\eqref{eq:trigonal_disp}. We remind the reader that $\mu$ is the chemical potential, $\cos3\theta_{\bm{k}}= (k_x^3-3k_y^2 k_x)/k^3$, and we assume $|\lambda|<1$ for stability. In this case, $k$ on the Fermi surface as a function of $\theta_{\bm{k}}$ is
\begin{equation}
    k_F(\theta_{\bm{k}})= \sqrt{\frac{2m \mu}{1+\lambda \cos 3\theta_{\bm{k}}}}\;,
\end{equation}
and the velocity components on the Fermi surface are
\begin{equation}
\begin{aligned}
    v_{F,x}&= \frac{ k_F(\theta_{\bm{k}})}{m}\left(\cos\theta_{\bm{k}}+\frac{5\lambda}{4}\cos 2\theta_{\bm{k}} -\frac{\lambda}{4}\cos 4\theta_{\bm{k}}\right),\\
     v_{F,y}&= \frac{ k_F(\theta_{\bm{k}})}{m}\left(\sin\theta_{\bm{k}}-\frac{3\lambda}{2}\sin 2\theta_{\bm{k}} -\frac{\lambda}{2}\cos 3\theta_{\bm{k}} \sin \theta_{\bm{k}} \right)\;.
    \end{aligned}
\end{equation}
In addition, on the Fermi surface, we find
\begin{equation}
    v_T(\bm{q})= -v_{F,x}\sin\phi_{\bm{q}}+v_{F,y}\cos\phi_{\bm{q}}\;,\quad \bm{v}\cdot \bm{q}=q( v_{F,x}\cos\phi_{\bm{q}}+  v_{F,y}\sin\phi_{\bm{q}}),
\end{equation}
where we used $\phi_{\bm{q}}$ to parameterize the angle of $\bm{q}$. Similarly,
\begin{equation}
  (\hat{\bm{q}}\cdot \nabla_{\bm{k}})^2\xi_{\bm{k}}   = \frac{1}{m}+\frac{ \lambda }{8 m}\Big[15 \cos (\theta_{\bm{k}} +2 \phi_{\bm{q}} )+3\cos (5 \theta_{\bm{k}} -2 \phi_{\bm{q}} )-10 \cos (3 \theta_{\bm{k}} )\Big]\;.
\end{equation}

Plugging in all of these formulae, we find 
\begin{equation}
\begin{aligned}
  &\langle J_T(\bm{q},\Omega)\rho(-\bm{q},-\Omega)\rangle =\varkappa_1(\phi_{\bm{q}}) q^2-\frac{1}{2\pi}\int_0^{2\pi} \frac{d\theta_{\bm{k}}}{2\pi} \int_0^{\infty} dk k \; \delta\left( \frac{k^2}{2m}\left(1+\lambda\cos3\theta_{\bm{k}}\right)-\mu\right)\frac{ v_T(\bm{q}) (\bm{v}\cdot \bm{q})}{i\Omega-\bm{v}\cdot \bm{q} }\\
   &= \frac{ m}{2\pi }\int_0^{2\pi} \frac{d\theta_{\bm{k}}}{2\pi} \frac{(v_{F,y}\cos\phi_{\bm{q}}-v_{F,x}\sin\phi_{\bm{q}})}{1+\lambda \cos3\theta_{\bm{k}}}+\varkappa_1(\phi_{\bm{q}}) q^2-\frac{i\Omega m}{2\pi q} \int_0^{2\pi} \frac{d\theta_{\bm{k}}}{2\pi }  \frac{ (v_{F,y}\cos\phi_{\bm{q}}-v_{F,x}\sin\phi_{\bm{q}})}{(1+\lambda \cos3\theta_{\bm{k}})(i\Omega/q-  v_{F,x}\cos\phi_{\bm{q}}-  v_{F,y}\sin\phi_{\bm{q}}) }
    \end{aligned}
\end{equation}
The first term here vanishes identically for any $|\lambda|<1$. The last term corresponds to Landau damping. Let us consider the limit $|\Omega|/q \ll 1$. In this case, we can replace $\Omega/q$ in the denominator by $0^{\pm}$ (depending on the sign of $\Omega$). The coefficient $\varkappa_1(\phi_{\bm{q}})$ in front of $q^2$ can be calculated similarly as
\begin{equation}
    \begin{aligned}
\varkappa_\alpha(\phi_{\bm{q}})&=\frac{1}{8}\int\frac{d^2\bm{k} }{(2\pi)^2}v_T^\alpha(\bm{q})\left[\delta'(\xi_{\bm{k}})(\hat{\bm{q}}\cdot \nabla_{\bm{k}})^2\xi_{\bm{k}}+\frac{v_L^2(\bm{q})}{3} \delta''(\xi_{\bm{k}})\right]\\
&=\frac{\alpha\mu^{\alpha/2-1}}{32\pi m^{\alpha/2}}\int_0^{2\pi}\frac{d\theta_{\bm{k}}}{2\pi (1+\lambda\cos 3\theta_{\bm{k}})}\left[ \frac{1+\alpha/2}{3}v_L^2(\bm{q})v_T^\alpha(\bm{q})    -v_T^\alpha(\bm{q}) (\hat{\bm{q}}\cdot \nabla_{\bm{k}})^2\xi_{\bm{k}}\right]_{\substack{k=k_F(\theta_k)\\ m=\mu=1}}\;.
    \end{aligned}
    \label{eq:varkappa_simplified}
\end{equation}
Here we eliminated the $k-$integral using an identity
\begin{equation}
    \int_0^{\infty} dk g(k)\delta^{(n)}(f(k)-\mu) = (-1)^n\partial^n_\mu     \int_0^{\infty} dk g(k)\delta(f(k)-\mu), 
\end{equation}
where $g(k)$ has no explicit $\mu$-dependence before imposing the Fermi-surface constraint, and then we used the fact that $ (\hat{\bm{q}}\cdot \nabla_{\bm{k}})^2\xi_{\bm{k}} $ is independent of the magnitude of $k$ and $\mu$, while $v_T$ and $v_L$ (after evaluated on the Fermi surface) both depend on $\mu$ via a simple overall prefactor $\sqrt{\mu/m}$.

\begin{figure}[t!]
    \centering
\includegraphics[width=0.8\linewidth]{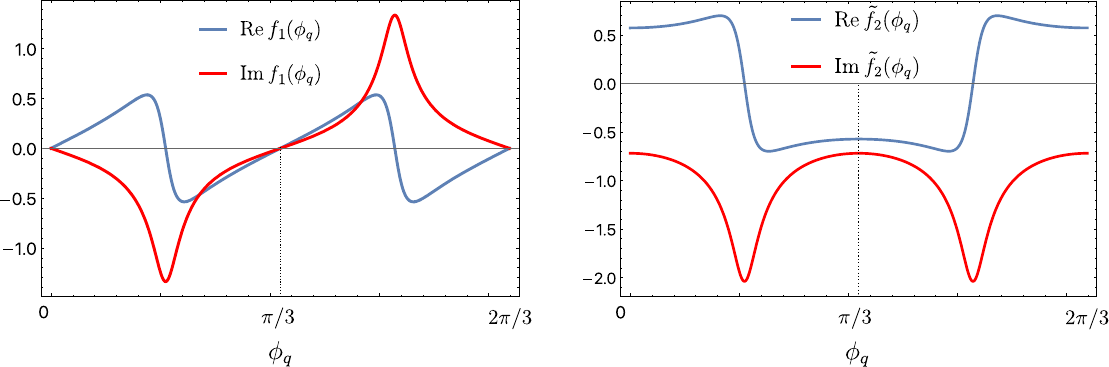}
    \caption{Angular dependence of $f_1(\phi_{\bm{q}})$ (left panel) and $\tilde{f}_2(\phi_{\bm{q}})=\sqrt{m/2\mu}f_2(\phi_{\bm{q}})$ (right panel) defined in Eq.~\eqref{eq:f_def}, for $\lambda=0.2$. }
    \label{fig:fig_f}
\end{figure}

Finally, we obtain 
\begin{equation}
   \langle J_T(\bm{q},\Omega)\rho(-\bm{q},-\Omega)\rangle \approx \varkappa_1(\phi_{\bm{q}}) q^2 -\frac{i m\Omega}{2 \pi q} \left(\operatorname{Re}f_1(\phi_{\bm{q}}) +i \operatorname{sgn}(\Omega) \operatorname{Im}f_1(\phi_{\bm{q}}) \right)\;,\label{eq:J_rho_correlator}
\end{equation}
where the angle-dependent functions $  f_\alpha(\phi_{\bm{q}})$ and $ \varkappa_\alpha(\phi_{\bm{q}})$  are defined in Eq.~\eqref{eq:f_def_explicit} of the main text, and simplified for our dispersion in Eq.~\eqref{eq:f_def_explicit} and Eq.~\eqref{eq:varkappa_simplified}.  The imaginary part of $ f_1(\phi_{\bm{q}})$ has period $2\pi/3$ and vanishes for $\phi_{\bm{q}}=n \pi/3$ where $n\in \mathbb{Z}$. In contrast, its real part vanishes at $\phi_{\bm{q}}=n \pi/6$, and its period is $\pi/3$. The real part can also develop additional zeros for sufficiently large $\lambda$, but they are absent at small $\lambda$. We also note that $\operatorname{Re}f_1(\phi_{\bm{q}}+\pi)=\operatorname{Re}f_1(\phi_{\bm{q}})$ and $\operatorname{Im}f_1(\phi_{\bm{q}}+\pi)=-\operatorname{Im}f_1(\phi_{\bm{q}})$. This implies that $\langle J_T(\bm{q},\Omega)\rho(-\bm{q},-\Omega)\rangle$ is odd under the simultaneous change $\bm{q}\rightarrow -\bm{q}$ and $\Omega\rightarrow -\Omega$. The behavior of $f_1$ is illustrated in Fig.~\ref{fig:fig_f} for $\lambda=0.2$.

By direct inspection, we also find $ \langle J_T(\bm{q},\Omega)\rho(-\bm{q},-\Omega)\rangle = \langle \rho(\bm{q},\Omega)J_T(-\bm{q},-\Omega)\rangle $ (although we stress that this relation does not have to hold in general, with the usual Hall response being a counter-example). We note that the fact that $\langle J_T(\bm{q},\Omega)\rho(-\bm{q},-\Omega)\rangle$ has both real and imaginary parts is not in contradiction with general properties of mixed correlators. Indeed, using spectral representation for any two operators $A$ and $B$, we find $[\langle A(\Omega)B(-\Omega)\rangle]^* = \langle B^\dagger(-\Omega)A^\dagger(\Omega)\rangle$. Thus, from $J_T(\bm{q})^\dagger=J_T(-\bm{q})$ and  $\rho(\bm{q})^\dagger=\rho(-\bm{q})$ it follows that $\langle \rho(\bm{q},-\Omega)J_T(-\bm{q},\Omega)\rangle^* = \langle J_T(\bm{q},\Omega)\rho(-\bm{q},-\Omega)\rangle$. In combination with the relation $ \langle J_T(\bm{q},\Omega)\rho(-\bm{q},-\Omega)\rangle = \langle \rho(\bm{q},\Omega)J_T(-\bm{q},-\Omega)\rangle $, this implies $ \langle J_T(\bm{q},-\Omega)\rho(-\bm{q},\Omega)\rangle^* = \langle J_T(\bm{q},\Omega)\rho(-\bm{q},-\Omega)\rangle$, which our result in Eq.~\eqref{eq:J_rho_correlator} clearly obeys.

After expanding $ f_1(\phi_{\bm{q}})$ at small $\lambda$, we find
\begin{equation}
    f_1(\phi_{\bm{q}})\approx -\frac{7i \lambda}{2} \sin 3\phi_{\bm{q}} +\frac{47\lambda^2}{4 }\sin 6\phi_{\bm{q}}+\mathcal{O}(\lambda^3)\;.
\end{equation}
Such perturbative expansions of angular integrals can be evaluated systematically using the following identity
\begin{equation}
    \int_0^{2\pi} \frac{d\theta}{2\pi} \frac{e^{im\theta}}{i\epsilon -\cos\theta} = i^{|m|-1} \frac{(\epsilon -\sqrt{\epsilon^2+1})^{|m|}}{\sqrt{\epsilon^2+1}},\quad 1>\epsilon>0,\quad m\in \mathrm{Z}\;,
\end{equation}
and similarly
\begin{equation}
    \int_0^{2\pi} \frac{d\theta}{2\pi} \frac{e^{im\theta}}{(i\epsilon -\cos\theta)^n} =  \frac{i^{n-1}}{(n-1)!}\frac{\partial^{n-1}}{\partial \epsilon^{n-1}}\int_0^{2\pi} \frac{d\theta}{2\pi} \frac{e^{im\theta}}{i\epsilon -\cos\theta} = \frac{i^{|m|+n-2}}{(n-1)!} \frac{\partial^{n-1}}{\partial \epsilon^{n-1}}\frac{(\epsilon -\sqrt{\epsilon^2+1})^{|m|}}{\sqrt{\epsilon^2+1}} \;.\label{eq:identity_small_lambda}
\end{equation}
After taking the limit $\epsilon \rightarrow 0^+$, we find
\begin{equation}
    \begin{aligned}
\int_0^{2\pi}\frac{d\theta}{2\pi}
\frac{e^{im\theta}}{i0^+-\cos\theta}
&=
(-i)^{|m|+1},\\
\int_0^{2\pi}\frac{d\theta}{2\pi}
\frac{e^{im\theta}}{(i0^+-\cos\theta)^2}
&=
-|m|\,(-i)^{|m|},\\
\int_0^{2\pi}\frac{d\theta}{2\pi}
\frac{e^{im\theta}}{(i0^+-\cos\theta)^3}
&=
\frac{\left(1-m^2\right)}{2}(-i)^{|m|+1}\,.
\end{aligned}
\end{equation}

The typical angular integral in Eq.~\eqref{eq:f_def}
can then be expanded in powers of $\lambda$, which will lead to powers of $1/(i0^+-\cos(\theta-\phi_{\bm{q }}))$ multiplied by angular factors. Finally, after shifting  $\theta\rightarrow \theta+\phi_{\bm{q }}$, collecting all factors, and using Eq.~\eqref{eq:identity_small_lambda} we can evaluate any order in $\lambda$. The expansion of $\varkappa_\alpha(\phi_{\bm{q}})$ can be performed similarly, and the result is given in the main text, Eq.~\eqref{eq:varkappa_exp}.

Therefore, this mixed correlator is only finite due to the presence of non-zero trigonal warping.  After performing analytic continuation to the real frequency axis from the upper half plane $\Omega>0$, i.e.  $i\Omega\rightarrow \omega+i0^+$, we obtain
\begin{equation}
    \Pi^{T0,R}_{\rm CF}(\bm{q},\omega)=-i\varkappa_1(\phi_{\bm{q}}) q^2+if_1(\phi_{\bm{q}})\frac{m\omega}{2 \pi q} \;.
\end{equation}
as advertised in the first equation of Eq.~\eqref{eq:PiCF_realfreq}. 

\begin{comment}

By direct inspection, we see that $\langle J_T(\bm{q},\Omega)\rho(-\bm{q},-\Omega)\rangle=\langle \rho(\bm{q},\Omega)J_T(-\bm{q},-\Omega)\rangle$. We note that the fact that $\langle J_T(\bm{q},\Omega)\rho(-\bm{q},-\Omega)\rangle$ has both real and imaginary parts is not in contradiction with general properties of mixed correlators. Indeed, using spectral representation for any two operators $A$ and $B$, we find $[\langle A(\Omega)B(-\Omega)\rangle]^* = \langle B^\dagger(-\Omega)A^\dagger(\Omega)\rangle$. Thus it follows that $\langle \rho(\bm{q},-\Omega)J_T(-\bm{q},\Omega)\rangle^* = \langle J_T(\bm{q},\Omega)\rho(-\bm{q},-\Omega)\rangle$. In total, this implies $ \langle J_T(\bm{q},-\Omega)\rho(-\bm{q},\Omega)\rangle^* = \langle J_T(\bm{q},\Omega)\rho(-\bm{q},-\Omega)\rangle$, which our result in Eq.~\eqref{eq:J_rho_correlator} clearly obeys. 
\end{comment}

A similar calculation for the density-density correlator yields 
\begin{equation}
\begin{aligned}
     \langle \rho(\bm{q},\Omega)\rho(-\bm{q},-\Omega)\rangle
   = \frac{ m}{2\pi }\int_0^{2\pi} \frac{d\theta_{\bm{k}}}{2\pi} \frac{1}{1+\lambda \cos3\theta_{\bm{k}}}+\varkappa_0 q^2-\frac{i \Omega m}{2\pi q} \int_0^{2\pi} \frac{d\theta_{\bm{k}}}{2\pi }  \frac{ 1}{(1+\lambda \cos3\theta_{\bm{k}})(i\Omega/q{-}  v_{F,x}\cos\phi_{\bm{q}}{-}  v_{F,y}\sin\phi_{\bm{q}}) }.
    \end{aligned}
\end{equation}
The first term here equals $\chi_0=(m/2\pi)/\sqrt{1-\lambda^2}$, which is just the density of states at the Fermi level. Thus, for $|\Omega|\ll q$ we find
\begin{equation}
    \Pi^{00}_{\rm CF}(\bs{q}, \Omega) \equiv \langle \rho(\bm{q},\Omega)\rho(-\bm{q},-\Omega)\rangle \approx \chi_0 +\varkappa_0 q^2- \frac{i\Omega m}{2\pi q} \left[\operatorname{Re}f_0(\phi_{\bm{q}}) +i \operatorname{sgn}(\Omega) \operatorname{Im}f_0(\phi_{\bm{q}}) \right]\;.
\end{equation}
We note that for our dispersion, $\varkappa_0=0$. Expanding $f_0(\phi_{\bm{q}})$ at small $\lambda$, we find
\begin{equation}
    f_0(\phi_{\bm{q}}) \approx \sqrt{\frac{m}{2\mu}}\left[-i+ 3\lambda \cos 3\phi_{\bm{q}}+i\lambda^2\left(\frac{21}{2}\cos 6\phi_{\bm{q}}-\frac{3}{8}\right)+\mathcal{O}(\lambda^3)\right]\;.
\end{equation}
Analytically continuing $i\Omega \rightarrow \omega + i 0^+$ reproduces the second equation of Eq.~\eqref{eq:PiCF_realfreq} in the main text
\begin{equation}
    \Pi^{00,R}_{\rm CF}(\bs{q}, \omega) = \chi_0 - \frac{m\omega}{2\pi q} f_0(\phi_{\bs{q}}) \,. 
\end{equation}
Finally, the transverse current-current correlator becomes 
\begin{equation}
\begin{aligned}
     \langle J_T(\bm{q},\Omega)J_T(-\bm{q},-\Omega)\rangle
  & = \frac{ m}{2\pi }\int_0^{2\pi} \frac{d\theta_{\bm{k}}}{2\pi} \frac{(v_{F,y}\cos\phi_{\bm{q}}-v_{F,x}\sin\phi_{\bm{q}})^2}{1+\lambda \cos3\theta_{\bm{k}}}+\varkappa_2 q^2\\
   &-\frac{i\Omega m}{2\pi q} \int_0^{2\pi} \frac{d\theta_{\bm{k}}}{2\pi }  \frac{(v_{F,y}\cos\phi_{\bm{q}}-v_{F,x}\sin\phi_{\bm{q}})^2}{(1+\lambda \cos3\theta_{\bm{k}})(i\Omega/q-  v_{F,x}\cos\phi_{\bm{q}}-  v_{F,y}\sin\phi_{\bm{q}}) }.\label{eq:JTJT}
    \end{aligned}
\end{equation}
The first term here is independent of $\phi_{\bm{q}}$ and will be canceled by the diamagnetic contribution to the current. Indeed, the latter is separately given by
\begin{equation}
   K_{\rm diam} = \int\frac{d^2\bm{k}}{(2\pi)^2} \theta(-\xi_{\bm{k}}) \;\hat{\bm{q}}_T^T\left(\frac{\partial^2 \xi_{\bm{k}}}{\partial {k_i} \partial {k_j}}\right)\hat{\bm{q}}_T = \frac{\mu}{8\pi}\left(\frac{9}{\sqrt{1-\lambda^2}}-5\right)\;,
\end{equation}
which exactly cancels the first term in Eq.~\eqref{eq:JTJT}. Thus, we obtain 
\begin{equation}
  \langle J_T(\bm{q},\Omega)J_T(-\bm{q},-\Omega)\rangle
   \approx K_{\rm diam} +\varkappa_2(\phi_{\bm{q}}) q^2- \frac{i\Omega m}{2\pi q} \left[\operatorname{Re}f_2(\phi_{\bm{q}}) +i \operatorname{sgn}(\Omega) \operatorname{Im}f_2(\phi_{\bm{q}}) \right]\;.
\end{equation}
and 
\begin{equation}
    f_2(\phi_{\bm{q}}) \approx \sqrt{\frac{2\mu}{m}}\left[-i+ 4\lambda \cos 3\phi_{\bm{q}}+i\lambda^2\left(\frac{105}{8}\cos 6\phi_{\bm{q}}-\frac{3}{4}\right)+\mathcal{O}(\lambda^3)\right]\;.\label{eq:f_2_expansion}
\end{equation}
The full behavior of $ f_2(\phi_{\bm{q}})$ is shown in Fig.~\ref{fig:fig_f} for $\lambda=0.2$. After analytic continuation, the full transverse-transverse response function then becomes
\begin{equation}
    \Pi^{TT,R}_{\rm CF}(\bs{q}, \omega) =-\varkappa_2(\phi_{\bm{q}}) q^2+\frac{m\omega}{2\pi q} f_2(\phi_{\bs{q}}) \,,
\end{equation}
as advertised in the third equation of Eq.~\eqref{eq:PiCF_realfreq}.

\section{Detailed evaluation of Feynman diagrams for the homogeneous optical conductivity}\label{app:detail_optical}

In this appendix, we provide all the technical details for the derivation of optical conductivity in Sec.~\ref{subsec:zeroQ_optical}. As explained in Sec.~\ref{subsec:zeroQ_optical}, the fundamental action that we will work with is a large $N$ generalization of the CFL action, Eqs.~\eqref{eq:Seff_transport}-\eqref{eq:S_diam_a}, in which $N$ species of CFs with a generic dispersion $\xi(\bs{k})$ are coupled to a single $U(1)$ gauge field $a$.
% \begin{equation}\label{eq:Seff_app_optical}
%     S_{\rm eff} = S_a + \sum_{i=1}^N \int_{\bs{k}, \tau} \bar f_i(\bs{k}, \tau) \left[\partial_{\tau} + \xi(\bs{k})\right] f_i(\bs{k}, \tau) + S_{\rm para}[f,a] + S_{\rm diam}[f, a] \,,
% \end{equation}
% where 
% \begin{equation}
%     S_{\rm para}[f, a] = \sum_{i=1}^N \int_{\bs{k}, \bs{q}, \tau} \bar f_i(\bs{k}+\bs{q}/2, \tau) \left[- a_0(\bs{q},\tau) + \nabla_{\bs{k}}\xi(\bs{k}) \cdot \bs{a}(\bs{q}, \tau)\right] f_i(\bs{k} - \bs{q}/2, \tau) \,,
% \end{equation}
% \begin{equation}
%     S_{\rm diam}[f, a] = \frac{1}{2} \sum_{i=1}^N \int_{\bs{k}, \bs{q},\tau} \bar f_i(\bs{k}+\frac{\bs{q}_1 + \bs{q}_2}{2}, \tau) f_i(\bs{k} - \frac{\bs{q}_1 + \bs{q}_2}{2}, \tau) a_i(\bs{q}_1, \tau) a_j(\bs{q}_2, \tau) V_{ij}\left(\bs{k} + \frac{\bs{q}_1 + \bs{q}_2}{2}\right) + \mathcal{O}(q_i^2) \,, 
% \end{equation}
% where the diamagnetic form factor is given by
% \begin{equation}
%     V_{ij} = \partial_{k_i} \partial_{k_j} \xi(\bs{k}) \,. 
% \end{equation}

From the effective action Eqs.~\eqref{eq:Seff_transport}-\eqref{eq:S_diam_a}, we can read off the Feynman rules and construct all Feynman diagrams that contribute to the Euclidean gauge field self-energy $\Pi(\bs{q}=0, \omega)$, which is equivalent to the irreducible CF response function. The upshot of our calculations is the following scaling structure of $\Pi^{xx}_{\rm CF}(\bs{q}=0, \omega)$ to leading two orders in the $1/N$ expansion: 
\begin{equation}\label{eq:new_RPA_Pi}
    \Pi^{xx}_{\rm CF}(\bs{q}=0,\Omega) = D + \frac{1}{N} \, C(z) \, |\Omega|^{\frac{4-z}{z}} \,. 
\end{equation}
where $D$ is a positive constant and $C_I(z)$ is a constant that depends on the choice of dynamical exponent $z$. Upon analytically continuing $i\Omega \rightarrow \omega + i 0^+$, we find the retarded gauge field self energy 
\begin{equation}
    \Pi^{xx}_{\rm CF}(\bs{q}=0,\omega) = D + \frac{1}{N} C(z) (-i \omega)^{\frac{4-z}{z}} \,,
\end{equation}
which leads to a CF conductivity as stated in Eq.~\eqref{eq:sigmaCF_maintext} of the main text 
\begin{equation}
    \sigma^{xx}_{\rm CF}(\bs{q}=0,\omega) \equiv \frac{\Pi^{xx}_{\rm CF}(\bs{q}=0,\omega)}{-i\omega} = \frac{i D}{\omega} + \frac{1}{N} C(z) (-i\omega)^{\frac{4-2z}{z}} \,. 
\end{equation}
In what follows, we will derive this result by working through all the relevant Feynman diagrams. 

\subsection{Organization of diagrams for $\Pi_{\rm CF}$}

Within the large-$N$ expansion, the fermion propagator remains free at leading order in $1/N$
\begin{equation}\label{eq:RPA_G_transport}
    G(\bs{k}, \omega) \approx \frac{1}{i\omega - \xi(\bs{k})} \,,
\end{equation}
while the effective propagator for the gauge field can be worked out using the vertex factor $f(\bs{k}, \hat{\bs{q}}) \equiv \bs{v}_F(\bs{k}) \times \hat{\bs{q}}$ and the bare density-density interaction $v(\bs{q}) = v_0 |\bs{q}|^{z-3}$
\begin{equation}\label{eq:RPA_D_transport}
    \begin{aligned}
    \mathcal{D}_{TT}(\bs{q}, \Omega) &= \left[\frac{V_0}{(4\pi)^2} |\bs{q}|^{z-1} + \pi_0(\bs{q}, \Omega)\right]^{-1} \,, \\
    \pi_0(\bs{q}, \Omega) &\equiv \Pi^{TT}_{\rm CF}(\bs{q}, \Omega) = K_{\rm diam} + \int f(\bs{k}, \hat{\bs{q}})^2 G(\bs{k} + \frac{\bs{q}}{2}, \omega + \frac{\Omega}{2}) G(\bs{k} - \frac{\bs{q}}{2}, \omega - \frac{\Omega}{2}) \\
    &= -\varkappa_2|\bm{q}|^2-\frac{m \operatorname{Im}f_2(\phi_{\bm{q}})}{2\pi}\frac{|\Omega|}{|\bs{q}|} + \frac{m \operatorname{Re}f_2(\phi_{\bm{q}})}{2\pi} \frac{i\Omega}{|\bs{q}|} \,,
    \end{aligned}
\end{equation}
where we defined $V_0=v_0+1/\chi_0$ for $z=3$ and $V_0=v_0$ for $2\leq z<3$ (in this case, the $\varkappa_2|\bm{q}|^2$ term in  $\Pi^{TT}_{\rm CF}(\bs{q}, \Omega)$ is irrelevant and can be omitted).
Crucially, in contrast to the inversion-symmetric case where $\pi_0$ is real, the inversion-asymmetric case generically contains an imaginary term proportional to $\tilde \Gamma_{\hat{\bs{q}}}$. Before proceeding further, let us derive three fundamental identities that follow from the structure of $G$ and $D$ 
\begin{equation}\label{eq:transport_fund_identities}
    \begin{aligned}
    G(\bs{k}, \omega) - G(\bs{k}, \omega + \Omega) &= i\Omega \, G(\bs{k}, \omega) G(\bs{k}, \omega + \Omega) \,, \\
    \mathcal{D}_{TT}(\bs{q}, \Omega) - \mathcal{D}_{TT}(\bs{q}, \Omega + \Omega') &= \left[\pi_0(\bs{q},\Omega+\Omega') - \pi_0(\bs{q}, \Omega)\right] \mathcal{D}_{TT}(\bs{q}, \Omega) \mathcal{D}_{TT}(\bs{q}, \Omega + \Omega') \\
    \mathcal{D}_{TT}(\bs{q}, \Omega) &= \mathcal{D}_{TT}(-\bs{q}, - \Omega) \,. 
    \end{aligned}
\end{equation}
These identities will be used repeatedly in subsequent calculations. 

We now proceed to organize the diagrams that we need to compute to extract \eqref{eq:new_RPA_Pi}. At leading order in the $1/N$ expansion, $\Pi_{\rm CF}$ receives two contributions from the diagrams in Fig.~\ref{fig:Feynman_O(1)} with a single fermion loop and no internal gauge field propagator. At $\mathcal{O}(N^{-1})$, the number of relevant diagrams becomes large and can be classified into three categories. The class of diagrams that provide the dominant low frequency scaling are diagrams that only involve the paramagnetic current vertex, as enumerated in Fig.~\ref{fig:Feynman_O(1:N)_para}. We will compute these diagrams first. The remaining diagrams enumerated in Fig.~\ref{fig:Feynman_O(1:N)_diam} involve at least one insertion of the diamagnetic vertex. We will show that the contribution from every diagram in Fig.~\ref{fig:Feynman_O(1:N)_diam} is subleading relative to the sum of diagrams in Fig.~\ref{fig:Feynman_O(1:N)_para}. Combining these arguments gives the answer in \eqref{eq:new_RPA_Pi}. 

\subsection{Evaluation of $\Pi_{\rm CF}$ at $\mathcal{O}(1)$: reproducing the RPA}

\begin{figure*}
    \centering
    \includegraphics[width = 0.6\textwidth]{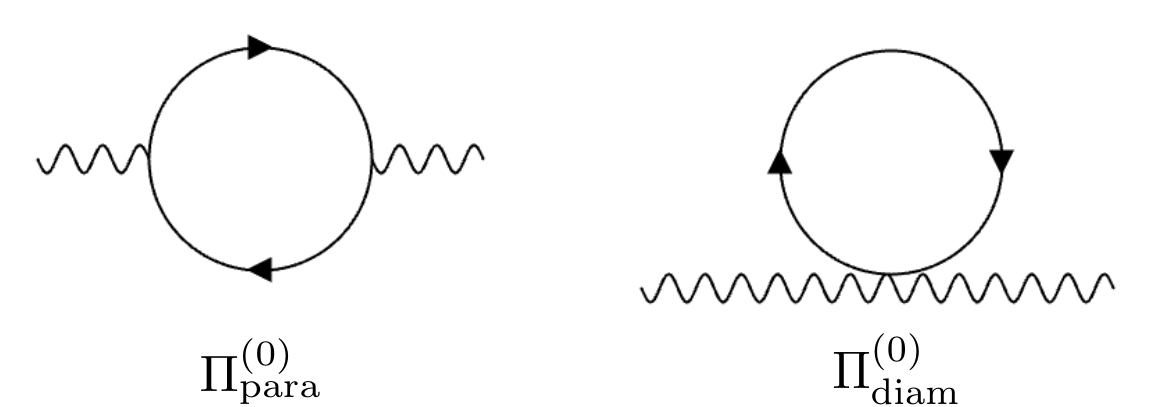}
    \caption{At leading order in the $1/N$ expansion, there are two diagrams that contribute to the gauge field self energy. The first $\Pi^{(0)}_{\rm para}$ involves two insertions of the paramagnetic current vertex, while the second $\Pi^{(0)}_{\rm diam}$ involves a single insertion of the diamagnetic current vertex.}
    \label{fig:Feynman_O(1)}
\end{figure*}
At leading order in the $1/N$ expansion, the diagrams contributing to the gauge field self energy are given in Fig.~\ref{fig:Feynman_O(1)}. The first diagram involves only the paramagnetic current vertex and can be written as 
\begin{equation}
    \Pi^{(0)}_{\rm para}(\bs{q} = 0, \Omega) = \int_{\bs{k},\omega} f(\bs{k}, \hat{\bs{q}})^2 G(\bs{k}, \omega+\Omega) G(\bs{k},\omega) = \frac{1}{i\Omega} \int_{\bs{k},\omega} f(\bs{k}, \hat{\bs{q}})^2  \left[G(\bs{k},\omega) - G(\bs{k}, \omega+\Omega)\right] \,.
\end{equation}
A simple change of variables $\omega \rightarrow \omega - \Omega$ in the second term shows that this integral vanishes identically. The second diagram involves only the diamagnetic current vertex and gives the Drude weight corresponding to fermions with $\xi(\bs{k})$ dispersion. We will denote the Drude weight by $D_0$. Hence, the total gauge field self energy at leading order in the $1/N$ expansion is 
\begin{equation}
    \Pi^{(0)}(\bs{q}=0,\Omega) = \Pi^{(0)}_{\rm para}(\bs{q}=0,\Omega) + \Pi^{(0)}_{\rm diam}(\bs{q}=0,\Omega) = D_0 \,. 
\end{equation}

\subsection{Evaluation of $\Pi_{\rm CF}$ at $\mathcal{O}(1/N)$: paramagnetic contributions}

The corrections to $\Pi_{\rm CF}(\bs{q}=0,\Omega)$ at $\mathcal{O}(1/N)$ are organized into paramagnetic and diamagnetic parts. We first treat diagrams that only involve paramagnetic current vertices, as shown in Fig.~\ref{fig:Feynman_O(1:N)_para}. Following conventions in the literature, we will refer to the two fermion self-energy corrections as $\Pi^{(\rm SE,1)},\Pi^{(\rm SE, 2)}$, the Maki-Thompson correction as $\Pi^{(\rm MT)}$, and the Aslamazov-Larkin corrections as $\Pi^{(\rm AL,1)},\Pi^{(\rm AL,2)}$ (the arguments of these functions are always understood to be $\bs{q}=0$ and $\Omega$). 

The calculation of these diagrams largely parallels the calculation in Appendix.~F of Ref.~\cite{Shi2022_loopcurrent}. It turns out that although each of $\Pi^{(\rm SE,1)},\Pi^{(\rm SE, 2)}, \Pi^{(\rm MT)}, \Pi^{(\rm AL,1)},\Pi^{(\rm AL,2)}$ is a highly singular function of $\omega$ that diverges faster than $\omega^{-1}$ as $\omega \rightarrow 0$, there are interesting cancellations between the singularities enforced by $U(1)$ gauge invariance. The final result is less singular than $\omega^{-1}$ and can be expressed as
\begin{equation}\label{eq:para_simplified}
    \begin{aligned}
    \Pi^{(\rm SE,1)} + \Pi^{(\rm SE, 2)} + \Pi^{(\rm MT)} 
    &= - \frac{1}{2\Omega^2} \int_{\bs{q'},\Omega'} \mathcal{D}_{TT}(\bs{q'},\Omega') \mathcal{D}_{TT}(\bs{q'},\Omega'-\Omega)\left[\pi_0(\bs{q}', \Omega' - \Omega) - \pi_0(\bs{q}', \Omega')\right] \\
    &\hspace{2cm} \cdot \left[\pi^{\scaleto{(\rm MT)}{6pt}}(\bs{q'},\Omega' - \Omega) - \pi^{\scaleto{(\rm MT)}{6pt}}(\bs{q'},\Omega') \right] \,,\\
    \Pi^{(\rm AL,1)} + \Pi^{(\rm AL,2)} &= \frac{1}{2\Omega^2} \int_{\bs{q'}, \Omega'} \mathcal{D}_{TT}(\bs{q'}, \Omega') \mathcal{D}_{TT}(\bs{q'},\Omega'-\Omega) \left[\pi^{\scaleto{(\rm AL)}{6pt}}(\bs{q'},\Omega' - \Omega) - \pi^{\scaleto{(\rm AL)}{6pt}}(\bs{q'},\Omega')\right]^2 \,.
    \end{aligned}
\end{equation}
Here $\mathcal{D}_{TT}$ is the transverse-transverse component of the Euclidean gauge field propagator and $\pi_0, \pi^{\scaleto{(\rm MT)}{6pt}}, \pi^{\scaleto{(\rm AL)}{6pt}}$ are one-loop fermion bubble diagrams  
\begin{equation}
    \begin{aligned}
    \pi^{\scaleto{(\rm MT)}{6pt}}(\bs{q'},\Omega') &= \int_{\bs{k}, \omega} [f(\bs{k} + \frac{\bs{q'}}{2}, \hat{\bs{q}}) - f(\bs{k} - \frac{\bs{q'}}{2}, \hat{\bs{q}})]^2f(\bs{k}, \hat{\bs{q}}')^2 G(\bs{k} - \frac{\bs{q'}}{2}, \omega) G(\bs{k} + \frac{\bs{q'}}{2}, \omega+\Omega') \,, \\
    \pi^{\scaleto{(\rm AL)}{6pt}}(\bs{q'},\Omega') &= \int_{\bs{k}, \omega} [f(\bs{k} + \frac{\bs{q'}}{2}, \hat{\bs{q}}) - f(\bs{k} - \frac{\bs{q'}}{2}, \hat{\bs{q}})] f(\bs{k}, \hat{\bs{q}}')^2 G(\bs{k} - \frac{\bs{q'}}{2}, \omega) G(\bs{k} + \frac{\bs{q'}}{2}, \omega+\Omega') \,.
    \end{aligned}
\end{equation}
To complete the calculation, we will evaluate these one-loop diagrams up to quadratic order in $\lambda$ and then plug them into $\Pi^{(\rm SE,1)} + \Pi^{(\rm SE, 2)} + \Pi^{(\rm MT)}$ and $\Pi^{(\rm AL,1)} + \Pi^{(\rm AL, 2)}$. 

\subsubsection{Evaluation of a general fermion one-loop bubble}

The first step is to evaluate the most general one-loop integral
\begin{equation}
    \pi_F(\bs{q'}, \Omega', \hat{\bm{q}}) = \int_{\bs{k}, \omega} F(\bs{k},\bs{q'},\hat{\bm{q}})G(\bs{k} + \frac{\bs{q'}}{2}, \omega+\Omega') G(\bs{k} - \frac{\bs{q'}}{2}, \omega) \,,
\end{equation}
for two choices of the vertex $F_{MT}(\bs{k}, \bs{q'},\hat{\bm{q}}) = [f(\bs{k} + \frac{\bs{q'}}{2}, \hat{\bs{q}}) - f(\bs{k} - \frac{\bs{q'}}{2}, \hat{\bs{q}})]^2 f(\bs{k}, \hat{\bs{q}}')^2$ and $F_{AL}(\bs{k}, \bs{q'},\hat{\bm{q}}) = [f(\bs{k} + \frac{\bs{q'}}{2}, \hat{\bs{q}}) - f(\bs{k} - \frac{\bs{q'}}{2}, \hat{\bs{q}})]f(\bs{k}, \hat{\bs{q}}')^2$, where  $f(\bs{k}, \hat{\bs{q}}) \equiv \bs{v}_F(\bs{k}) \times \hat{\bs{q}}$. We note that these vertices depend on the full vector $\bm{q}'$ but only on the direction of $\bm{q}$. In the $q'\ll k_F$ and $|\Omega'|/q' \ll 1$ regime, we can approximate
\begin{equation}
     \pi_F(\bs{q'}, \Omega', \hat{\bm{q}})\approx -\int\frac{d^2\bm{k} }{(2\pi)^2}F(\bs{k}, \bs{q'},\hat{\bm{q}})  \delta(\xi_{\bm{k}}) + \frac{i\Omega'}{q'} \int\frac{d^2\bm{k} }{(2\pi)^2} \frac{F(\bs{k}, \bs{q'},\hat{\bm{q}}) \delta(\xi_{\bm{k}}) }{i0^+ \operatorname{sgn}(\Omega')-\bm{v}\cdot \hat{\bm{q}}' }  \,.
\end{equation}
Now let us make a change of variables from $\left[k_x, k_y\right]$ to $\left[\xi_{\bs{k}}, \theta_{\bs{k}}\right]$. The Jacobian associated with this change of variables is denoted by $J(\xi, \theta)$. Since the dominant contributions to the integral come from the vicinity of the Fermi surface, we can approximate $\xi = 0$ inside the Jacobian and the form factor $F(\bs{k}, \bs{q}', \hat{\bs{q}})$. This approximation allows us to integrate over $\xi$ and obtain 
\begin{equation}
    \pi_F(\bs{q}', \Omega', \hat{\bs{q}}) = \Re \, \pi_F(\bs{q}',\Omega',\hat{\bs{q}}) + i \Im\, \pi_F(\bs{q}', \Omega', \hat{\bs{q}}) \,, 
\end{equation}
where the real and imaginary parts simplifies in the $\Omega' \ll q'$ regime
\begin{equation}
    \begin{aligned}
    \Re \, \pi_F(\bs{q}',\Omega') &\approx I_F(\bs{q}', \hat{\bs{q}}) + \frac{|\Omega'|}{4\pi |\bs{q}'|} \int d \theta J(\theta) F(\theta, \bs{q}') \delta[\hat{\bs{q}}' \cdot \bs{v}_F(\theta)]  \,, \\
    \Im\, \pi_F(\bs{q}',\Omega',\hat{\bs{q}}) &= - \frac{\Omega'}{4 \pi^2 |\bs{q}'|} \int d \theta J(\theta) F(\theta,\bs{q}') \mathcal{P} \frac{1}{\hat{\bs{q}}' \cdot \bs{v}_F(\theta)} \,. 
    \end{aligned}
\end{equation}
Substituting these expressions into Eq.~\eqref{eq:para_simplified} gives Eq.~\eqref{eq:zeroQ_BnB_decomposition} in Sec.~\ref{subsec:zeroQ_optical}. 

So far, our calculation has assumed a completely general form factor $J(\theta), F(\theta, \bs{q}', \hat{\bs{q}})$. Now let us specialize to the weakly trigonally warped dispersion $\xi_{\bs{k}} = \frac{k^2}{2m} (1 + \lambda \cos 3 \theta_{\bs{k}}) - \mu$ and the form factors entering $\pi^{\scaleto{(\rm AL)}{6pt}}$ and $\pi^{\scaleto{(\rm MT)}{6pt}}$. The frequency-independent term evaluates to
\begin{equation}
    I_F(\bs{q}', \hat{\bs{q}}) = - \int\frac{d^2\bm{k} }{(2\pi)^2}F(\bs{k}, \bs{q'},\hat{\bm{q}})  \delta(\xi_{\bm{k}}) \approx  \frac{m|\bm{q}'|^n}{2\pi}\int_0^{2\pi} \frac{d\theta}{2\pi} \frac{[(\hat{q}' \cdot \nabla_{\bm{k}}) (\bm{v}\times \hat{q})]^n_{k=k_F(\theta)}}{1+\lambda \cos 3\theta}   [v_F^2-(\bm{v}_F\cdot \hat{\bm{q}}')^2]
\end{equation}
where $n=1$ for AL and $n=2$ for MT. Since this term does not enter Eq.~\eqref{eq:zeroQ_BnB_decomposition}, we will not evaluate the angular integrals explicitly. 

The more interesting frequency-dependent terms can be evaluated order by order in $\lambda$, where each order requires repeated uses of the integral identities in Eq.~\eqref{eq:identity_small_lambda}. The final results for MT and AL are
\begin{equation}\label{eq:one_loop_bubble_lambda_exp}
    \begin{aligned}
    \Re \, \pi^{\scaleto{(\rm MT)}{6pt}}(\bs{q}', \Omega', \hat{\bs{q}}) - I_{\rm MT}(\bs{q}', \hat{\bs{q}}) &= \frac{|\Omega'||\bs{q}'|}{2\pi m} \sqrt{\frac{2\mu}{m}} \bigg\{\sin^2 (\phi_{\bs{q}} - \phi_{\bs{q}'}) - \frac{\lambda^2}{8} \big[\frac{45}{4}\cos(2\phi_{\bs{q}}+4\phi_{\bs{q}'}) \\
    &+\frac{33}{4}\cos(2\phi_{\bs{q}}-8\phi_{\bs{q}'}) -\frac{57}{2}\cos(6\phi_{\bs{q}'}) +12\cos\left(2\phi_{\bs{q}}-2\phi_{\bs{q}'}\right) -21 \big] \bigg\} \,, \\
    \Im \, \pi^{\scaleto{(\rm MT)}{6pt}}(\bs{q}', \Omega', \hat{\bs{q}}) &= - \frac{\Omega' |\bs{q}'|}{2\pi m} \sqrt{\frac{2\mu}{m}} \frac{9 \lambda}{8} \sin (\phi_{\bs{q}} - \phi_{\bs{q}'}) \sin(\phi_{\bs{q}} + 2 \phi_{\bs{q}'}) \,, \\ 
     \Re \, \pi^{\scaleto{(\rm AL)}{6pt}}(\bs{q}', \Omega', \hat{\bs{q}}) - I_{\rm AL}(\bs{q}', \hat{\bs{q}}) &= \frac{|\Omega'|}{2\pi} \sqrt{\frac{2\mu}{m}} \bigg\{\sin (\phi_{\bs{q}} - \phi_{\bs{q}'}) \\
     &\hspace{2cm} - \frac{\lambda^2}{4} \big[-5\sin(\phi_{\bs{q}}+5\phi_{\bs{q}'})
     +7\sin(\phi_{\bs{q}}-7\phi_{\bs{q}'})
     +\frac{5}{2}\sin(\phi_{\bs{q}}-\phi_{\bs{q}'})\big]\bigg\} \,, \\
     \Im \,\pi^{\scaleto{(\rm AL)}{6pt}}(\bs{q}', \Omega', \hat{\bs{q}}) &= \frac{\Omega'}{2\pi} \sqrt{\frac{2\mu}{m}} \lambda \left[\frac{7}{16}\sin(\phi_{\bs{q}}+2\phi_{\bs{q}'}) +\sin(\phi_{\bs{q}}-4\phi_{\bs{q}'}) \right] \,.
    \end{aligned}
\end{equation}
We thus have at our disposal all the relevant one-loop integrals that go into the optical conductivity. 

\subsubsection{Using the general fermion one-loop bubble to deduce $\Pi_{\rm CF}$}

Using the one-loop integrals in Eq.~\eqref{eq:one_loop_bubble_lambda_exp} in combination with Eq.~\eqref{eq:f_2_expansion}, we obtain
\begin{equation}
\begin{aligned}
 &\left[\pi_0(\bs{q}', \Omega'-\Omega) - \pi_0(\bs{q}',\Omega')\right] \left[\pi^{\scaleto{(\rm MT)}{6pt}}(\bs{q}', \Omega'-\Omega) - \pi^{\scaleto{(\rm MT)}{6pt}}(\bs{q}',\Omega')\right] - \left[\pi^{\scaleto{(\rm AL)}{6pt}}(\bs{q}', \Omega'-\Omega) - \pi^{\scaleto{(\rm AL)}{6pt}}(\bs{q}',\Omega')\right]^2
 =\\
 &\hspace{10em}=\frac{\lambda^2 \mu }{2\pi^2 m}
\left[
(|\Omega'-\Omega|-|\Omega'|)^2\,
 A(\phi_{\bs{q}},\phi_{\bs{q}'})
+
\Omega^2\,
 B(\phi_{\bs{q}},\phi_{\bs{q}'})
\right]
+\mathcal O(\lambda^3) .
\end{aligned}
\end{equation}
where $\pi_0\equiv \Pi_{\rm CF}^{TT}$, and
\begin{equation}
\begin{aligned}
A(\phi_{\bs{q}},\phi_{\bs{q}'})
&=
\left[
\frac{5}{2}\cos(\phi_{\bs{q}}+2\phi_{\bs{q}'})
-
\cos(\phi_{\bs{q}}-4\phi_{\bs{q}'})
\right]^2 ,
\\
B(\phi_{\bs{q}},\phi_{\bs{q}'})
&=
\left[
\frac{25}{16}\sin(\phi_{\bs{q}}+2\phi_{\bs{q}'})
+
\sin(\phi_{\bs{q}}-4\phi_{\bs{q}'})
\right]^2 .
\end{aligned}
\end{equation}
Feeding this term into Eq.~\eqref{eq:para_simplified}, we arrive at 
\begin{equation}
     \Pi^{xx, (1)}_{\rm CF}(\Omega) \approx - \frac{\lambda^2 \mu }{4\pi^2 m \Omega^2} \int_{\bs{q'},\Omega'}  \mathcal{D}_{TT}(\bs{q'},\Omega') \mathcal{D}_{TT}(\bs{q'},\Omega'-\Omega)\left[
(|\Omega'-\Omega|-|\Omega'|)^2\,
 A(\phi_{\bs{q}},\phi_{\bs{q}'})
+
\Omega^2\,
 B(\phi_{\bs{q}},\phi_{\bs{q}'})
\right]\label{eq:Pi_total}
\end{equation}
where $\mathcal{D}_{TT}(q,\Omega)$ is the $\lambda=0$ limit of the transverse-transverse gauge field propagator 
\begin{equation}
   \mathcal{D}_{TT}(q,\Omega)= \frac{1}{a q^2 +\frac{b |\Omega|}{q}},\quad\quad  a= \frac{1+\chi_0 v_0}{16\pi^2 \chi_0}-\varkappa_2, \quad\quad  b=\frac{\sqrt{2m\mu}}{2\pi}\;.
\end{equation}
The angular integrals over $\phi_{\bs{q}'}$ are independent of $\phi_{\bs{q}}$ and produce the following numerical factors
\begin{equation}
    \int_0^{2\pi}\frac{d\phi_{\bs{q}'}}{2\pi} A(\phi_{\bs{q}},\phi_{\bs{q}'}) = \frac{29}{8},\quad \quad     \int_0^{2\pi}\frac{d\phi_{\bs{q}'}}{2\pi} B(\phi_{\bs{q}},\phi_{\bs{q}'}) = \frac{881}{512}\;.
\end{equation}
The remaining integral contains UV divergences. To isolate the IR contribution, let us consider $\Pi^{xx, (1)}_{\rm CF}(\Omega>0) - \Pi^{xx, (1)}_{\rm CF}(0^+) $ and rescale $q\rightarrow \Omega^{1/3}x$, $\Omega'\rightarrow \Omega y$. Then we find
\begin{equation}
\begin{aligned}
     \Pi^{xx, (1)}_{\rm CF}(\Omega) - \Pi^{xx, (1)}_{\rm CF}(0) &\approx - \frac{29\lambda^2 \mu \Omega^{1/3}}{32\pi^2 m a^{4/3} b^{2/3}} \underbrace{\int_{-\infty}^{+\infty}\frac{dy}{2\pi}\int_0^{+\infty} \frac{dx}{2\pi} x^3\left\{ \frac{ (|y-1|-|y|)^2}{(x^3 + |y|)(x^3 + |y-1|)}- \frac{1 }{(x^3 + |y|)^2} \right\}}_{=-\sqrt{3}/(14\pi )}\,\\
     &- \frac{881\lambda^2 \mu \Omega^{1/3}}{2048\pi^2 m a^{4/3} b^{2/3}}  \underbrace{\int_{-\infty}^{+\infty}\frac{dy}{2\pi}\int_0^{+\infty} \frac{dx}{2\pi} x^3\left\{ \frac{1}{(x^3 + |y|)(x^3 + |y-1|)}- \frac{1 }{(x^3 + |y|)^2} \right\}}_{\equiv \mathcal{C}_2}\;.
     \end{aligned}
\end{equation}
After evaluating the remaining numerical coefficient
\begin{equation}
    \mathcal C_2
=
\frac{1}{12\pi^2}
\int_0^\infty du\,u^{1/3}
\left[
\frac{4(u+1)}{2u+1}
\log\left(1+\frac{1}{u}\right)
-\frac{2}{u}
\right]
\approx -0.028\;,
\end{equation}
we obtain
\begin{equation}
\begin{aligned}
      \Pi^{xx, (1)}_{\rm CF}(\Omega) - \Pi^{xx, (1)}_{\rm CF}(0) &= \left[\frac{29\sqrt{3}}{112\pi} -\frac{881}{512}\mathcal{C}_2\right]  \frac{\lambda^2 \mu \Omega^{1/3}}{4\pi^2 m a^{4/3} b^{2/3}} =\left[\frac{29\sqrt{3}}{112\pi} - \frac{881}{512}\mathcal{C}_2\right]  \frac{2^{7/3}
\lambda^2\mu^{2/3}\Omega^{1/3}
}
{
\left(1+\chi_0 v_0-16\pi^2 \chi_0\varkappa_2\right)^{4/3}
}\\
&\approx 0.963\times 
\frac{
\lambda^2\mu^{2/3}\Omega^{1/3}
}
{
\left(1+\chi_0 v_0-16\pi^2 \chi_0\varkappa_2\right)^{4/3}
}\;,
\end{aligned}
\label{eq:PimPi}
\end{equation}
where we used $\chi_0 =  m/(2\pi) + \mathcal{O}(\lambda^2)$. Moreover, $\varkappa_2\approx -1/(24\pi m)+\mathcal{O}(\lambda^2)$ and thus $1-16\pi^2 \chi_0\varkappa_2\approx 4/3+\mathcal{O}(\lambda^2)$.

Generalizing this calculation to an arbitrary dynamical exponent $2 < z \leq 3$ gives a different scaling form $\Omega^{\frac{4-z}{z}}$, with a prefactor that can be evaluated for every $z$. This general form is the result quoted in Eq.~\eqref{eq:sigmaCF_subleadingN} of the main text.

\subsection{Evaluation of $\Pi_{\rm CF}$ at $\mathcal{O}(1/N)$: diamagnetic contributions}

\begin{figure*}
    \centering
    \includegraphics[width = 0.8\textwidth]{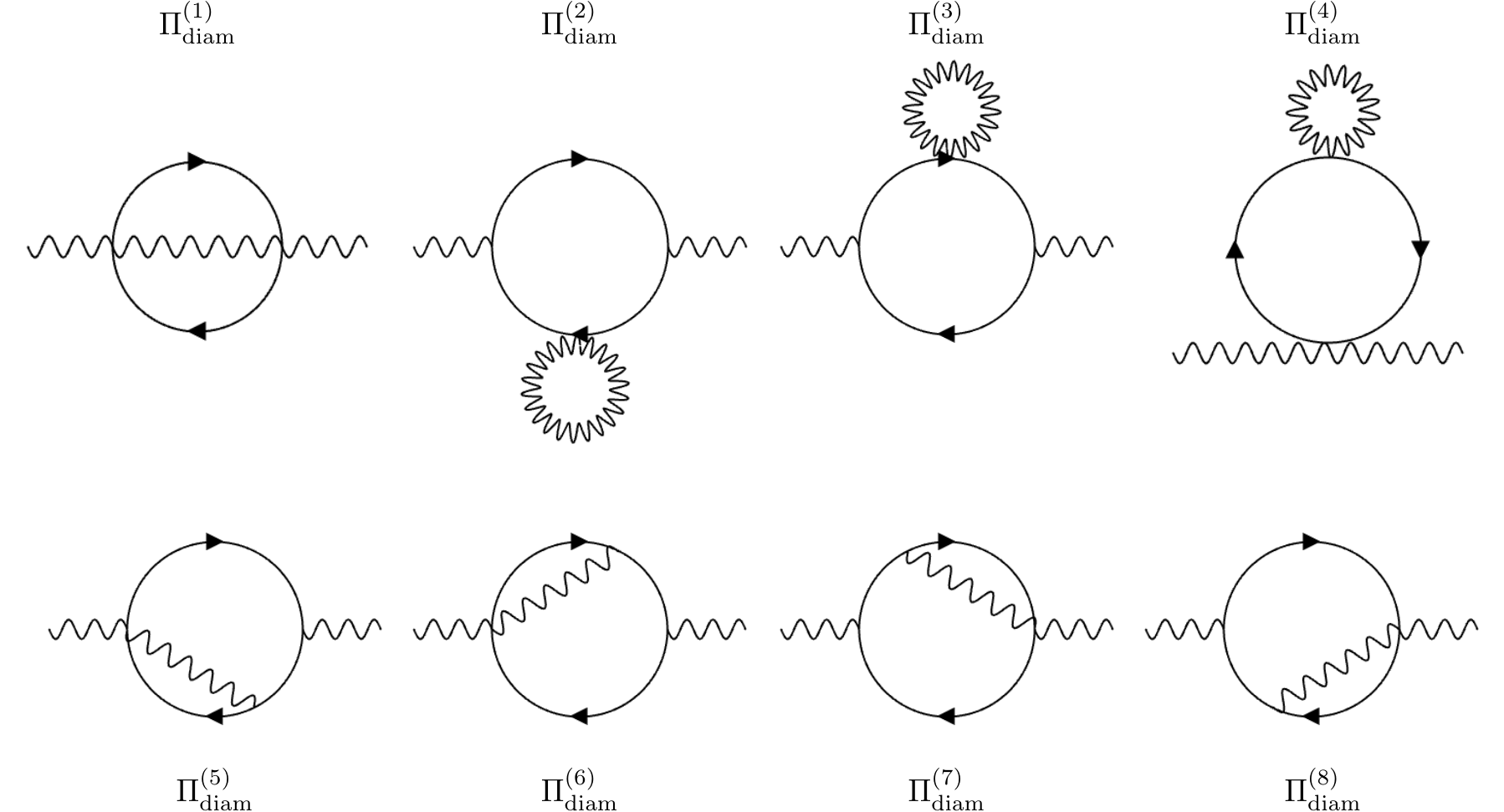} 
    \includegraphics[width = \textwidth]{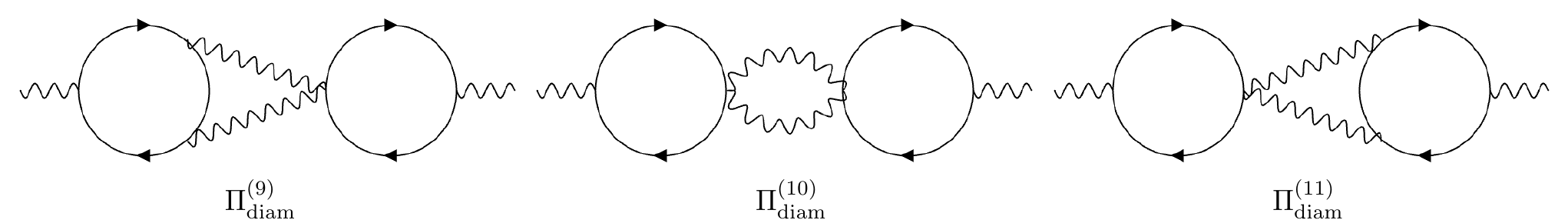}
    \caption{The set of diagrams contributing to the gauge field self energy at $\mathcal{O}(1/N)$ which involve at least one insertion of the diamagnetic current vertex.}
    \label{fig:Feynman_O(1:N)_diam}
\end{figure*}

To complete the argument, we finally turn to diagrams involving at least one insertion of the diamagnetic vertex. In total, there are eight such diagrams with a single fermion loop and three such diagrams with two fermion loops, as shown in Fig.~\ref{fig:Feynman_O(1:N)_diam}.
Let us evaluate these diagrams in turn. The first diagram evaluates to
\begin{equation}
    \Pi_{\rm diam}^{(1)} = \int_{\bs{k}, \omega} \int_{\bs{q}_1, \Omega_1} G(\bs{k}, \omega) G(\bs{k} - \bs{q}_1, \omega - \Omega_1 + \Omega) \mathcal{D}_{TT}(\bs{q}_1, \Omega_1) \left[V_T(\bs{k} - \bs{q}_1/2)\right]^2 \,.
\end{equation}
From a scaling analysis with dynamical critical exponent $z$, we see that the IR singular part of this diagram scales as $\Omega^2 \Omega^{6/z} \Omega^{-4/z} \Omega^{-(z-1)/z} = \Omega^{1 + 3/z}$, which is subleading relative to the $\Omega^{(4-z)/z}$ contribution coming from $\Pi^{(1)}_{\rm para}$. We thus conclude that $\Pi^{(1)}_{\rm diam}$ is negligible in the low energy limit. 

For diagrams (2) and (3), we have one fermion loop with three internal fermion propagators. These loops have the general structure 
\begin{equation}
    \int_{\bs{k}, \omega} F(\bs{k}) G(\bs{k}, \omega) G(\bs{k}, \omega) G(\bs{k}, \omega+\Omega) \,. 
\end{equation}
By a change of variables $\omega \rightarrow \omega - \Omega$, we can easily show that such loop integrals always vanish. For diagram (4), the fermion loop and the gauge field loop each contributes a decoupled $\Omega$-independent factor. The product is also $\Omega$-independent and does not lead to a non-analytic term in $\Pi(\bs{q}=0,\Omega)$. 

Diagrams (5)-(8) have the same scaling in the IR limit. Let us extract their common frequency scaling from a representative diagram (5)
\begin{equation}
    \begin{aligned}
    \Pi_{\rm diam}^{(5)} &= \int_{\bs{k}, \omega} \int_{\bs{q}_1, \Omega_1} G(\bs{k}, \omega) G(\bs{k} - \bs{q}_1, \omega - \Omega_1 + \Omega) \mathcal{D}_{TT}(\bs{q}_1, \Omega_1) G(\bs{k}, \omega + \Omega) \\
    &\times V_T(\bs{k} - \bs{q}_1/2,\hat{\bs{q}}_1, \hat{\bs{q}}) f(\bs{k} - \bs{q}_1/2, \hat{\bs{q}}_1) f(\bs{k}, \hat{\bs{q}})  \,. 
    \end{aligned}
\end{equation}
In a system with inversion symmetry, under a change of variables $\bs{k} \rightarrow - \bs{k}, \bs{q}_1 \rightarrow - \bs{q}_1$, every factor in the integrand is invariant except for $f(\bs{k}, \hat{\bs{q}})$ which changes sign. Therefore, the whole diagram is equal to the opposite of itself, which implies that it must vanish. When inversion is broken, this statement no longer holds. Using the fundamental identity in \eqref{eq:transport_fund_identities}, we can estimate the low frequency scaling as follows:
\begin{equation}
    \begin{aligned}
    \Pi_{\rm diam}^{(5)} &= \frac{1}{i\Omega} \int_{\bs{k}, \omega} \int_{\bs{q}_1, \Omega_1} \left[G(\bs{k}, \omega)  - G(\bs{k}, \omega + \Omega)\right] G(\bs{k} - \bs{q}_1, \omega - \Omega_1 + \Omega) \mathcal{D}_{TT}(\bs{q}_1, \Omega_1) \\
    &\cdot V_T(\bs{k} - \bs{q}_1/2,\hat{\bs{q}}_1, \hat{\bs{q}}) f(\bs{k} - \bs{q}_1/2, \hat{\bs{q}}_1) f(\bs{k}, \hat{\bs{q}})  \\
    &\sim \Omega \cdot \Omega^{6/z} \cdot \Omega^{-4/z} \cdot \Omega^{-(z-1)/z} = \Omega^{3/z} \,. 
    \end{aligned}
\end{equation}
This final scaling is again subleading relative to the dominant scaling $\Omega^{(4-z)/z}$ in $\Pi^{(1)}_{\rm para}$. 

Finally, we consider diagrams (9) - (11) with two fermion loops. While these diagrams are nontrivial for general external momentum $\bs{q}$, they all vanish at $\bs{q} = 0$ because each contains at least one isolated fermion loop with the structure
\begin{equation}
    \int_{\bs{k}, \omega} F(\bs{k}) G(\bs{k}, \omega) G(\bs{k}, \omega+\Omega)  = 0 \,. 
\end{equation}
Based on these arguments, we conclude that the total contribution from all diagrams involving diamagnetic vertices in Fig.~\ref{fig:Feynman_O(1:N)_diam} is upper bounded by $\mathcal{O}(\Omega^{3/z})$ and hence dominated by the contribution from paramagnetic diagrams in Fig.~\ref{fig:Feynman_O(1:N)_para}, which scales as $\Omega^{\frac{4-z}{z}}$.

\bibliography{C3_ACFL}

\end{document}